\documentclass[11pt,letterpaper]{article}

\usepackage{latexsym,multirow}
\usepackage{amssymb,amsmath, bm}
\usepackage{graphicx}
\usepackage[color,all,import,arrow]{xy}
\usepackage{enumerate}
\usepackage{booktabs}

\usepackage{natbib}
\usepackage{xcolor}
\usepackage[pdftex, bookmarksopen=true, bookmarksnumbered=true,
pdfstartview=FitH, breaklinks=true, urlbordercolor={0 1 0}, citebordercolor={0 0 1}]{hyperref}
\usepackage{colortbl}
\usepackage{subfigure}

\usepackage{tikz}
\usepackage{inputenc}
\usetikzlibrary{shapes,arrows,trees,fit,positioning,shapes.misc,calc}

\usepackage{dcolumn}
\newcolumntype{.}{D{.}{.}{-1}}
\newcolumntype{d}[1]{D{.}{.}{#1}}

\usepackage{theorem}
\theoremstyle{plain}
\theoremheaderfont{\scshape}
\newtheorem{assumption}{Assumption}

\newtheorem{corollary}{Corollary}

\newtheorem{proposition}{Proposition}
\newtheorem{theorem}{Theorem}

\newtheorem{lemma}{Lemma}

\newcommand{\ind}{\mbox{$\perp\!\!\!\perp$}}

\usepackage{rotating}

\usepackage[compact]{titlesec}

\allowdisplaybreaks

\newcommand\spacingset[1]{\renewcommand{\baselinestretch}%
{#1}\small\normalsize}

\newcommand{\blind}{0}

\newcommand*{\QEDB}{\hfill\ensuremath{\square}}

\newcommand{\s}{\textup{s}}

\newcommand{\hbx}{\bar{\bm{x}}}
\newcommand{\hbX}{\overline{\bm{X}}}
\newcommand{\hbz}{\bar{\bm{z}}}
\newcommand{\hbZ}{\overline{\bm{Z}}}

\newcommand{\bx}{\bm{x}}
\newcommand{\td}{\text{d}}
\newcommand{\bZ}{\bm{Z}}

\newcommand{\bz}{\bm{z}}

\newcommand{\E}{\mathbb{E}}
\newcommand{\bX}{\bm{X}}

\newcommand{\bV}{\bm{V}}

\begin{document} 

\newcommand{\tit}{Longitudinal Causal Inference with Selective
  Eligibility}
%
%
\spacingset{1.25}

\if0\blind

{\title{\bf\tit}

\author{Zhichao Jiang\thanks{Professor, School of Mathematics,
      Sun Yat-sen University, Guangzhou,  Guangdong 510275, China. Email:
      \href{mailto:jiangzhch7@mail.sysu.edu.cn}{jiangzhch7@mail.sysu.edu.cn} }
    \and Eli Ben-Michael\thanks{Assistant Professor, Department of Statistics \& Data Science and Heinz College of Information Systems \& Public Policy, Carnegie Mellon University, U.S.A. Email: \href{mailto:ebenmichael@cmu.edu}{ebenmichael@cmu.edu} URL:
    \href{https://ebenmichael.github.io}{ebenmichael.github.io}} 
  \and D. James Greiner\thanks{Honorable S.  William Green Professor
      of Public Law, Harvard Law School, 1525 Massachusetts Avenue,
      Griswold 504, Cambridge, MA 02138, U.S.A.} \and Ryan Halen\thanks{Data
      Analyst, Access to Justice Lab at Harvard Law School, 1607
      Massachusetts Avenue, Third Floor, Cambridge, MA 02138.}  \and Kosuke
  Imai\thanks{Professor, Department of Government and Department of
    Statistics, Harvard University.  1737 Cambridge Street,
    Institute for Quantitative Social Science, Cambridge MA 02138, U.S.A.
    Email: \href{mailto:imai@harvard.edu}{imai@harvard.edu} URL:
    \href{https://imai.fas.harvard.edu}{https://imai.fas.harvard.edu}}
}

\date{
\today
}

\maketitle

}\fi

\if1\blind
\title{\bf \tit}

\maketitle
\fi

\pdfbookmark[1]{Title Page}{Title Page}

\thispagestyle{empty}
\setcounter{page}{0}
         
\begin{abstract}
  Dropout poses a significant challenge to causal inference in
  longitudinal studies with time-varying treatments. However, existing
  research does not simultaneously address dropout and time-varying
  treatments.  We examine selective eligibility, an important yet
  overlooked source of non-ignorable dropout in such settings. This
  problem arises when a unit’s prior treatment history influences its
  eligibility for subsequent treatments, a common scenario in medical
  and other settings.  We propose a general methodological framework
  for longitudinal causal inference with selective eligibility.  By
  focusing on a subgroup of units who would become eligible for
  treatment given a specific past treatment sequence, we define the
  time-specific eligible treatment effect and expected number of
  outcome events under a treatment sequence of interest.  Under a
  generalized version of sequential ignorability, we derive two
  nonparametric identification formulae, each leveraging different
  parts of the observed data distribution. We then derive the
  efficient influence function of each causal estimand, yielding the
  corresponding doubly robust estimator.  Finally, we apply the
  proposed methodology to an impact evaluation of a pre-trial risk
  assessment instrument in the criminal justice system, in which
  selective eligibility arises due to recidivism.
  
\end{abstract}


\clearpage
\spacingset{1.5}

\section{Introduction}

Dropout significantly complicates longitudinal causal inference with
time-varying treatments. Much of existing research, however, does not
address time-varying treatments and dropout at the same time.  For
example, the causal inference literature has established
identification formulas and estimation strategies for time-varying
treatment effects but without dropout
\citep[e.g.,][]{robins1986new,richardson2013single,robins2008estimation,hernan2020causal}. In
addition, the survival analysis literature offers various model-based
methods to account for dropout --- often referred to as competing
events. However, treatment is typically time-invariant
\citep[e.g.,][]{comment2019survivor,xu2022bayesian,baer2023causal,janvin2024causal,rytgaard2024one}.

When treatments are time-varying {\it and} dropout is present,
existing methods are not directly applicable because standard causal
estimands are not well defined for those who have dropped out and are
no longer eligible for subsequent treatments.  In this paper, we
examine selective eligibility, an important yet overlooked source of
non-ignorable dropout. This problem occurs when a unit’s treatment
history influences its eligibility for subsequent
treatments. Selective eligibility is common in medical settings, in
which patients become eligible for another treatment only if they are
readmitted to a hospital.  Moreover, complete recovery or severe side
effects can render patients ineligible for additional
treatments. Recidivism in criminal justice systems also has the
selective eligibility problem because only those who are rearrested
are subject to another round of judicial decisions.

A key feature of selective eligibility is that it causes dropout {\it
  before} the administration of treatment.  This implies that both
treatment and outcome are defined only for treatment-eligible
units.\footnote{\cite{boruvka2018assessing} address the related issue
  of treatment availability in the context of mobile health
  applications.  They consider the setting where some units may not be
  able to receive the treatment even when assigned to the treatment
  condition.  Since these units remain in the study and take up the
  control condition, this is a problem of noncompliance rather than
  that of dropout.}
In contrast, existing literature primarily addresses a different type
of non-ignorable treatment-induced dropout, where dropout occurs
between treatment administration and outcome measurement.  In this
case, the outcome is observed only for a subset of units that received
treatment. This type of dropout is commonly handled using the
principal stratification framework
\citep[e.g.,][]{kurland2009longitudinal,lee2010causal,lee2013causal,shardell2015doubly,shardell2018joint,josefsson2016causal,
  comment2019survivor,josefsson2021bayesian,grossi2025bayesian}, which
extends the classic method for truncation by death problems
\citep[e.g.,][]{zhang2003estimation,rubin2006causal,imai2008sharp,ding2011identifiability,wang2017causal,yang2018using}.
In addition, we emphasize that among these works, only
\cite{josefsson2021bayesian} considers time-varying treatments while
the others focus on time-invariant treatments.


We develop a general methodological framework for selective
eligibility (Section~\ref{sec:methodology}). We define the average
treatment effect at each time period by focusing on a group of units
who become eligible for treatment given a specific treatment history.
This causal estimand, which we call the average eligible treatment
effect (ETE), characterizes treatment effect heterogeneity across time
periods and treatment histories.  We also introduce the expected
number of outcome events (EOE) under a given treatment
history. 
We consider stochastic interventions, for which deterministic
interventions are a special case.  Stochastic interventions can deal
with a relatively large number of time periods because they enable us
to specify high-dimensional treatment strategies with low-dimensional
parameters
\citep[e.g.,][]{kennedy2019nonparametric,papa:etal:22,schindl2024incremental,wu2024assessing}.

We develop a statistical methodology for estimating the ETE and EOE by
extending the sequential ignorability assumption \citep[see][for a
review]{hernan2020causal} to longitudinal studies with selective
eligibility.  Under this assumption, we first derive two
identification formulae based on the outcome regression and inverse
probability weighting (IPW).  The former leads to an estimator that
requires the correct specification of the outcome regression and
treatment eligibility models, while the latter implies an estimator
based on the correctly specified model for the propensity score.  The
existence of multiple estimators leads to the question of how to
derive a more efficient estimator by combining them.

We address this question by deriving the efficient influence function
\citep[EIF;][]{bickel1993efficient} for each causal estimand.  The
EIFs motivate new estimators that combine nonparametric models of
outcome regression, treatment eligibility probability, and propensity
score. The proposed estimators are doubly robust; they are consistent
so long as either the propensity score model or the outcome regression
and eligibility models are correctly specified. This result extends
the classic doubly robust (DR) estimator
\citep{bang2005doubly,hernan2020causal} to longitudinal studies with
selective eligibility.  Section~\ref{sec::simulation} presents
simulation studies to evaluate the finite sample performance of the
proposed estimators.

As an empirical application, we analyze recidivism in criminal justice
systems (Section~\ref{sec::application}).  We evaluate the impact of a
specific pre-trial risk assessment instrument, called the Public
Safety Assessment Instrument (PSA). We analyze the data from our own
randomized controlled trial (RCT), in which the provision of the PSA
is randomized across cases
\citep{greiner2020randomized,imai2020experimental}. A primary goal of
the PSA is to help judges make release decisions that address concerns
about public safety while avoiding unnecessary incarceration during a
pre-trial period. We investigate whether the use of this PSA reduces
subsequent negative outcomes such as new criminal activities committed
upon release.

The problem of selective eligibility arises in this context because an
arrestee would become eligible for the subsequent randomized PSA
provision only if rearrested. Moreover, rearrest can depend on the
judge's previous decisions, which in turn may be influenced by the
provision of the PSA. Previous analyses of this RCT focused on the
first arrest cases to avoid selective eligibility
\citep{imai2020experimental,ben2021safe,benmichael2024does}. In
contrast, we assess treatment effect heterogeneity over multiple
arrests.

\section{The Proposed Methodology}
\label{sec:methodology}

We now present the proposed methodology for longitudinal causal
inference with selective eligibility.
We first present assumptions and introduce causal estimands under this
setting.  We then derive identification formulae and develop efficient
estimators.

\subsection{Notation and assumptions}
\label{sec::setup}
We consider a simple random sample of $n$ units from a target
population.  Suppose that each unit can become eligible for the
treatment up to a total of $T$ times.  We use $S_{it}$ to indicate
whether unit $i$ is eligible for the $t$-th treatment where
$i=1,2,\ldots,n$ and $t=1,2,\ldots,T$. We assume that all units are
eligible for the first-period treatment, i.e., $S_{i1}=1$ for all $i$.
Furthermore, a unit can only become eligible for the $(t+1)$-th
treatment if it was previously eligible for the $t$-th treatment,
i.e., $S_{it} \geq S_{i,t+1}$ for all $i=1,2,\ldots,n$ and
$t=1,2,\ldots,T-1$.

We use $Z_{it}$ to represent the indicator variable for the $t$-th
treatment assignment for unit $i$.  This variable is defined only when
the unit becomes eligible for the $t$-th treatment, i.e.,
$S_{it} = 1$.  We use $\hbZ_{it}=(Z_{i1},\ldots,Z_{it})$ to denote the
treatment history up to the $t$-th treatment for unit $i$, and
$\hbz_t = (z_1,\ldots,z_t)$ to denote a treatment sequence from time 1
to time $t$.  The outcome is measured after each treatment.  We let
$Y_{it}$ denote the outcome for unit $i$ observed after the $t$-th
treatment.  Similar to the treatment, the outcome $Y_{it}$ is not
well-defined for units who do not become eligible for the $t$-th
treatment, i.e., $S_{it}=0$.

We adopt the potential outcomes framework under the stable unit
treatment value assumption (SUTVA; \cite{rubin1980randomization}).
This implies that the treatment status of one unit does not affect the
eligibility and outcome of another unit.  However, we allow for
within-unit carryover effects.  That is, for a given unit, the
treatment status at any time period may affect its own future
eligibility and outcome. Let $S_{it}(\hbz_T)$ and $Y_{it}(\hbz_T)$ be
the potential values of eligibility and outcome given a treatment
sequence $\hbz_T= (z_1,\ldots,z_T)$, respectively.  Then, the observed
eligibility and outcome are given by $S_{it} = S_{it}(\hbZ_{iT})$ and
$Y_{it} = Y_{it}(\hbZ_{iT})$.

We assume that future treatment does not affect current eligibility
and current outcome.
\begin{assumption}[No Anticipation] \spacingset{1}
\label{asm::noanticipation}
For any $i=1,2,\ldots,n$ and $t=1,2,\ldots,T$,
$Y_{it}(\hbz_T) = Y_{it}(\hbz'_T)$ if $\hbz_t = \hbz'_t$ and
$S_{it}(\hbz_T) = S_{it}(\hbz'_T)$ if $\hbz_{t-1} = \hbz'_{t-1}$.
\end{assumption}
Assumption~\ref{asm::noanticipation} is common in causal inference
with time series and panel data
\citep[e.g.,][]{bojinov2019time,ben2022synthetic,papa:etal:22,jiang2023instrumental}.
The assumption is plausible as long as units are not aware of their
future treatment. Under Assumption~\ref{asm::noanticipation}, we can
simplify $Y_{it}(\hbz_T)$ as $Y_{it}(\hbz_t)$ and $S_{it}(\hbz_T)$ as
$S_{it}(\hbz_{t-1})$.

\tikzstyle{level 1}=[level distance=3cm, sibling distance=2cm]
\tikzstyle{level 2}=[level distance=3cm, sibling distance=1cm]

\tikzstyle{bag} = [text width=4em, text centered]
\tikzstyle{end} = [circle, minimum width=3pt, fill, inner sep=0pt]

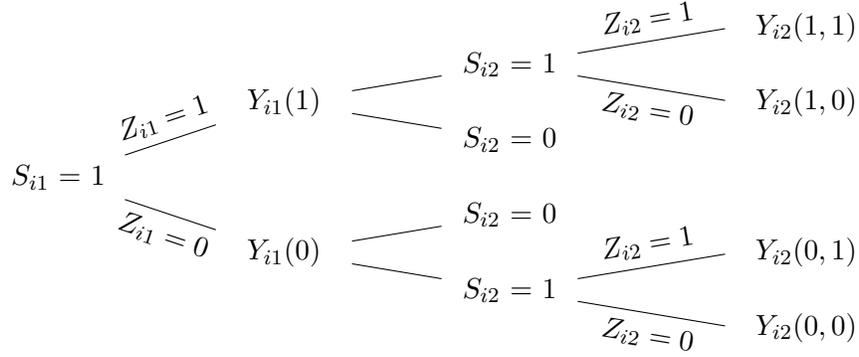
\begin{figure}[t] \spacingset{1}
  \centering
  \begin{tikzpicture}[grow=right, sloped]
  \node[bag] {$S_{i1}=1$}
  child {
    node[bag] {$Y_{i1}(0)$}
    child {
      node[bag] {$S_{i2}=1$}
      child{
        node[label=right: 
        {$Y_{i2}(0,0)$}]{}
        edge from parent 
        node[below]{$Z_{i2}=0$}
      }
      child{
        node[label=right: 
        {$Y_{i2}(0,1)$}]{}
        edge from parent 
        node[above]{$Z_{i2}=1$}
      }
      edge from parent 
    }
    child {
      node[bag] {$S_{i2}=0$}
      edge from parent
    }
    edge from parent 
    node[below] {$Z_{i1}=0$}
  }
  child {
    node[bag] {$Y_{i1}(1)$}        
    child {
      node[bag] {$S_{i2}=0$}
      edge from parent
    }
    child {
      node[bag] {$S_{i2}=1$}
      child{
        node[label=right: 
        {$Y_{i2}(1,0)$}]{}
        edge from parent 
        node[below]{$Z_{i2}=0$}
      }
      child{
        node[label=right: 
        {$Y_{i2}(1,1)$}]{}
        edge from parent 
        node[above]{$Z_{i2}=1$}
      }
      edge from parent 
    }
    edge from parent         
    node[above] {$Z_{i1}=1$}
  };
\end{tikzpicture}
\caption{The temporal ordering of the eligibility indicator $S_{it}$,
  treatment assignment $Z_{it}$, and potential outcome $Y_{it}(\hbz_t)$ in an
  example with two time periods.  At each time point, eligibility is
  measured first.  Treatment and outcome variables are only defined when a unit
  becomes eligible.}
\label{fig::order-eligibility}
\end{figure}

Figure~\ref{fig::order-eligibility} illustrates the temporal ordering
of the eligibility, treatment, and outcome variables with an example
of two time periods. In addition, we observe a set of covariates
$\bX_{it}$ for unit $i$ at time $t$. These covariates may include both
time-invariant and time-varying covariates, including past outcomes.
We use $\hbX_{it} = \{\bX_{1t},\ldots,\bX_{it}\}$ to denote the
covariate history up to the $t$-th treatment. Similar to the
treatment, the covariate history $\hbX_{it}$ is defined only for units
eligible at time $t$.

The proposed methodology is applicable to both RCTs and observational
studies.  We generalize the sequential ignorability assumption
to the settings with selective eligibility.
\begin{assumption}[Sequential ignorability with selective eligibility]
  \label{asm::ignorability} \spacingset{1} For any $i=1,2,\ldots,n$
  and $t=1,2,\ldots,T$,
$$
Z_{it} \ \ind \ \{Y_{it}(\hbz_t),S_{i,t+1}(\hbz_t),\ldots,S_{iT}(\hbz_{T-1})\} \mid \hbZ_{i,t-1}=\hbz_{t-1},S_{it}=1,\hbX_{it}=\hbx_t,
$$
for all $\hbz_{t-1},\hbz_t,\hbz_{t+1},\ldots,\hbz_{T-1}$, and $\hbx_t$.
\end{assumption}
The assumption implies that for eligible units, the treatment
assignment is conditionally independent of the current potential
outcome and the potential values of all future eligibility indicator
variables, given the treatment and covariate histories.
In addition, we generalize the standard overlap condition to the
current setting.
\begin{assumption}[Overlap condition with selective eligibility] \spacingset{1}
\label{asm::overlap}
For any $\hbz_{t-1}$ and $\hbx_t$, $0<\Pr(Z_{it}=1\mid S_{it}=1, \hbZ_{i,t-1}=\hbz_{t-1},\hbX_{it}=\hbx_t)<1$.
\end{assumption}
Assumption~\ref{asm::overlap} requires that eligible units at each
time $t$ have a non-zero probability of being assigned to both the
treatment and control conditions.

We do not impose sequential ignorability or overlap condition on the
eligibility indicator $S_{it}$. Thus, selective eligibility can either
be affected by unobserved variables or be a deterministic function of
past covariates, treatments, and outcomes --- a scenario commonly
encountered in clinical studies.  In general, observational studies
face additional difficulties in satisfying
Assumption~\ref{asm::overlap}, particularly when the covariate is
high-dimensional \citep[e.g.,][]{d2021overlap}. Longitudinal causal
inference with selective eligibility also leads to challenges in
meeting this assumption as the dimension of the treatment history
grows with time and the number of eligible units may become small for
certain treatment histories.


\subsection{Causal estimands}
\label{sec::estimand}

We first consider a causal effect of the treatment at each time $t$
given a treatment history $\hbz_{t-1}$. Recall that the outcome
$Y_{it}(\hbz_{t-1},z)$ is defined only for eligible units at time $t$.
Therefore, we introduce the following average eligible treatment
effect (ETE): \begin{equation}
  \label{eqn::def-tau}
  \tau_t(\hbz_{t-1})\ :=\ \E\{ Y_{it}(\hbz_{t-1},1) - Y_{it}(\hbz_{t-1},0) \mid S_{it}(\hbz_{t-1})=1 \}.
\end{equation}
The ETE represents the average effect of the $t$-th treatment among
units who are eligible at time $t$ when the treatment history is fixed
at $\hbZ_{i,t-1}=\hbz_{t-1}$.  Due to the potential carryover effects
resulting from past treatments, the ETE depends on the entire vector
of previous treatments $\hbz_{t-1}$, yielding a separate causal
quantity for each of $2^{t-1}$ different values of $\hbz_{t-1}$.

We also consider the expected number of outcome events (EOE) across
time periods:
\begin{equation*}
  \theta(\hbz_{T}) \ := \  \sum_{t=1}^T
  \E\{Y_{it}(\hbz_{t}) 
  S_{it}(\hbz_{t-1}) \},
\end{equation*}
which equals the sum of all outcome events averaged across units under
a treatment sequence $\hbZ_T=\hbz_T$. 

We can generalize the EOE for various treatment strategies. Let $\xi$
denote a treatment strategy, which is defined as a probability
distribution over the sequence of treatments. Thus, $\xi$ may
represent either a deterministic or stochastic intervention. We use
$\hbZ_{iT}^\xi = (Z_{i1}^\xi,\ldots, Z_{iT}^\xi)$ to denote a sequence
of interventions on unit $i$ that is generated according to the
treatment strategy $\xi$ for the duration of $T$ time periods. At each
time period $t$, the treatment strategy sequentially generates
$Z_{it}^\xi$ from a Bernoulli distribution conditional on the
treatment history
$\hbZ_{i,t-1}^\xi = (Z_{i1}^\xi,\ldots,Z_{i,t-1}^\xi)$ for
$t=2,\ldots,T$. Therefore, the treatment strategy can be characterized
by the parameters of these Bernoulli distributions, denoted by a
vector of probabilities
$(\xi_1(z_1),\xi_2(\hbz_2),\ldots,\xi_T(\hbz_T))$, where
$\xi_t(\hbz_t)$ denotes the probability that the stochastic
intervention assigns $Z^\xi_t=z_t$ conditional on
$\hbZ^\xi_{t-1}=\hbz_{t-1}$.

Due to selective eligibility, the stochastic intervention is
administered only for treatment-eligible units, i.e., the intervention
$Z_{it}^\xi$ is generated for unit $i$ only if
$S_{it}(\hbZ_{i,t-1}^\xi)=1$.  This implies that we can characterize a
treatment strategy as the following factorized probability:
\begin{equation*}
\xi_t(\hbz_t) := \Pr(Z^\xi_{i1}=z_1)\prod_{t'=1}^t \Pr(Z^\xi_{t'} = z_{t'}\mid
\hbZ_{i,t'-1}^\xi=\hbz_{t'-1},S_{it'}=1).
\end{equation*}
Note that a deterministic treatment strategy is a special case of this
stochastic intervention where the probability is one for a specific
sequence $\hbz_t$ and is zero otherwise.  Given the above definition,
we can define the EOE under a treatment strategy $\xi$: 
\begin{eqnarray}
  \theta(\xi) & = & \sum_{t=1}^T \E\{ Y_{it}(\hbZ_{it}^\xi)
  S_{it}(\hbZ_{i,t-1}^\xi)\} \nonumber \\
\label{eqn::def-theta}
 &=&   \sum_{t=1}^T  \sum_{\hbz_t} \E\{ Y_{it}(\hbz_{t})\mid  S_{it}(\hbz_{t-1})=1\} \Pr\{S_{it}(\hbz_{t-1})=1\}\xi_t(\hbz_t).
\end{eqnarray}

\subsection{Nonparametric identification}
\label{sec::identification}

We now turn to the nonparametric identification of the causal
estimands defined above.  We focus on the identification of
$\E\{Y_{it}(\hbz_{t})\mid S_{it}(\hbz_{t-1})=1\}$ and
$\Pr\{S_{it}(\hbz_{t-1})=1\}$, which in turn implies the
identification of $\tau_t(\hbz_{t-1})$ and $\theta(\xi)$ via
Equations~\eqref{eqn::def-tau}~and~\eqref{eqn::def-theta}.
 
We begin by introducing the following quantities to simplify the
presentation:
\begin{eqnarray*}
\mu_t(\hbz_t,\hbx_t) &:=&\E(Y_{it}\mid S_{it}=1,\hbZ_{it}=\hbz_t,\hbX_{it}=\hbx_t),\\
p_{t}(\hbz_{t-1},\hbx_{t-1})&:=&  \Pr(S_{it}=1 \mid S_{i,t-1}=1,\hbZ_{i,t-1}=\hbz_{t-1},\hbX_{i,t-1}=\hbx_{t-1}),\\
 w_{t}(\hbz_{t},\hbx_{t}) &:=& \Pr(Z_{it}=z_{t}\mid S_{it}=1,\hbZ_{i,t-1}=\hbz_{t-1},\hbX_{it}=\hbx_{t}),\\
 l_{t}(\hbz_{t-1},\hbx_{t}) & := & \Pr(\bX_{it}=\bx_{t}\mid S_{it}=1,\hbZ_{i,t-1}=\hbz_{t-1},\hbX_{i,t-1}=\hbx_{t-1}),
\end{eqnarray*}
which represent the conditional mean outcome, eligibility probability,
treatment probability, and covariate distribution, respectively, for
eligible units at each time $t$, given the treatment and covariate
histories up to time $t-1$.  When $t=1$, we have
$p_t(\hbz_{t-1},\hbx_{t-1})=\Pr(S_{i1}=1)=1$,
$ l(\hbz_{t-1},\hbx_{t}) =\Pr(\bX_{i1}=\bx_1)$, and
$w_{t}(\hbz_{t},\hbx_{t})=\Pr(Z_{i1}=z_1 \mid \bX_{i1}=\bx_1)$.

The following theorem provides two different identification formulae,
based on different components of the observed data distribution, for
the two key quantities, i.e.,
$\E\{Y_{it}(\hbz_{t-1},z_t)\mid S_{it}(\hbz_{t=1})=1\}$ and
$\Pr\{S_{it}(\hbz_{t-1})=1\}$.
\begin{theorem}[Nonparametric Identification] \spacingset{1}
\label{thm::identification-obs}
Under Assumptions~\ref{asm::noanticipation}--\ref{asm::overlap}, the
following identification formulae hold for
$\E\{Y_{it}(\hbz_{t-1},z_t)\mid S_{it}(\hbz_{t=1})=1\}$ and
$\Pr\{S_{it}(\hbz_{t-1})=1\}$.
\begin{enumerate}[(a)]
\item Formula based on the conditional mean outcome and eligibility
  probability:
\begin{eqnarray*}
\label{eqn::identification-tau-reg} \E\{Y_{it}(\hbz_{t})\mid S_{it}(\hbz_{t=1})=1\}
&=&\frac{\int \mu_t(\hbz_t,\hbx_t)  \prod_{t'=2}^t p_{t'}(\hbz_{t'-1},\hbx_{t'-1}) \prod_{t'=1}^tl_{t'}(\hbz_{t'-1},\hbx_{t’}) \text{d}\hbx_t}{\int \prod_{t'=2}^{t} p_{t'}(\hbz_{t'-1},\hbx_{t'-1}) \prod_{t'=1}^{t-1}l_{t'}(\hbz_{t'-1},\hbx_{t’}) \text{d}\hbx_{t-1}},\\
\nonumber \Pr\{S_{it}(\hbz_{t-1})=1\}
&=&\int \prod_{t'=2}^{t} p_{t'}(\hbz_{t'-1},\hbx_{t'-1}) \prod_{t'=1}^{t-1}l_{t'}(\hbz_{t'-1},\hbx_{t’}) \text{d}\hbx_{t-1},
\end{eqnarray*}
where integrals are replaced by summation for discrete covariates.
\item Formula based on the propensity score:
\begin{eqnarray*}
\label{eqn::identification-tau-ipw}\E\{Y_{it}(\hbz_{t-1},z_t)\mid S_{it}(\hbz_{t=1})=1\}
&=&\frac{\E\left[Y_{it}\bm{1}\{\hbZ_{it}=\hbz_t,S_{it}=1\}/\pi_t(\hbz_t,\hbX_{it})  \right]}{\E\left[\bm{1}\{\hbZ_{i,t-1}=\hbz_{t-1},S_{it}=1\}/\pi_{t-1}(\hbz_{t-1},\hbX_{i,t-1}) \right]},\\
\nonumber \Pr\{S_{it}(\hbz_{t-1})=1\}&=&\E\left[\frac{\bm{1}\{\hbZ_{i,t-1}=\hbz_{t-1},S_{it}=1\}}{\pi_{t-1}(\hbz_{t-1},\hbX_{i,t-1})} \right],
\end{eqnarray*}
where $\pi_t(\hbz_t,\hbX_{it}) = \prod_{t'=1}^t w_{t'}(\hbz_{t'},\hbX_{it'})$.
\end{enumerate}
\end{theorem}
We provide some intuition for the results of
Theorem~\ref{thm::identification-obs} by focusing on
$\E\{Y_{it}(\hbz_{t-1},z_t)\mid S_{it}(\hbz_{t-1})=1\}$.  The first
identification formula given in
Theorem~\ref{thm::identification-obs}(a) is based on the models for
the conditional mean outcome and eligibility probability. For example,
when $t=2$, this formula can be simplified as:
\begin{eqnarray*}
 \E\{Y_{i2}(\hbz_{2})\mid S_{i2}(z_1)=1\}&=& \frac{  \E \left[    \E\{ \mu_2(\hbz_2,\hbX_{i2})\mid z_1, \bX_{i1}\}p_2(z_1,\bX_{i1})   \right]  }{\E\{ p_2(z_1,\bX_{i1}) \}},
\end{eqnarray*}
where the covariates are integrated out. Therefore, we can view this
formula as a g-formula for longitudinal causal inference with
selective eligibility.  In contrast, the first equation given in
Theorem~\ref{thm::identification-obs}(b) presents the IPW formula,
relying only on the treatment probability at each time given the
treatment and covariate histories $w_{t}(\hbz_{t},\hbx_{t})$.  In both
formulae, the numerator equals
$\E\{Y_{it}(\hbz_{t-1},z_t) S_{it}(\hbz_{t-1})\}$ and the denominator
equals $\Pr\{ S_{it}(\hbz_{t-1})=1\}$.
 

As noted earlier, Theorem~\ref{thm::identification-obs} directly
implies the identification of $\tau_t(\hbz_{t-1})$ and $\theta(\xi)$.
The identification of $\tau_t(\hbz_{t-1})$ requires the overlap
condition (Assumption~\ref{asm::overlap}) to hold only for the
corresponding treatment history $\hbZ_{i,t-1}=\hbz_{t-1}$.  Similarly,
the identification of $\theta(\xi)$ only requires the overlap
condition to hold for all $\hbz_t$ with $\xi(\hbz_t)\neq 0$.


The identification formulae in Theorem~\ref{thm::identification-obs}
have simpler forms when all the covariates are time invariant.  We
present this special case as a corollary.
\begin{corollary}[Identification Formulae without Time-varying
  Covariates] \spacingset{1}
\label{cor::obs-inv}
Under Assumptions~\ref{asm::noanticipation}---\ref{asm::overlap} with
$\bX_{it}=\bX_i$ for all $i$ and $t$, the identification formulae in
Theorem~\ref{thm::identification-obs}(a) simplify as
\begin{eqnarray*}
\label{eqn::identification-tau-reg-inv} \E\{Y_{it}(\hbz_{t})\mid S_{it}(\hbz_{t-1})=1\}
&=&\frac{\E\left\{ \mu_t(\hbz_t,\bX_i)  \prod_{t'=2}^t p_{t'}(\hbz_{t'-1},\bX_i) \right\} }{\E\left\{\prod_{t'=2}^{t} p_{t'}(\hbz_{t'-1},\bX_i) \right\}},\\
\Pr\{S_{it}(\hbz_{t-1})=1\} &=&\E\left\{ \prod_{t'=2}^t p_{t'}(\hbz_{t'-1},\bX_i) \right\}.
\end{eqnarray*}
\end{corollary}
Unlike the formulae in Theorem~\ref{thm::identification-obs}(a),
without time-varying covariates, the formulae in
Corollary~\ref{cor::obs-inv} no longer include the covariate
distribution terms $l_t(\hbz_{t-1}, \hbx_t)$.  In addition, the formulae
in Theorem~\ref{thm::identification-obs}(b) do not change
except that $\hbX_{it}$ is replaced with $\bX_i$.

\subsection{Efficient influence functions}
\label{sec::eif}

The identification formulae in Theorem~\ref{thm::identification-obs}
imply the existence of multiple estimators for the causal estimands.
To combine them in a principled way, we derive the efficient influence
functions (EIFs) of the estimands, which in turn yield estimators with
certain robustness and optimality properties.  These estimators can
achieve parametric convergence rates even when nuisance functions are
estimated at slower rates through nonparametric estimation methods
\citep[see
e.g.,][]{bickel1993efficient,tsiatis2006semiparametric,kennedy2016}.

We begin by introducing the following two regression functions that
are building blocks of the EIFs and deriving their identification
formulae: \allowdisplaybreaks
\begin{align*}
  m_{Y_tS_t}(\hbz_{t'}, \hbX_{it'})  \ := \ & 
                                           \E\{Y_{it}(\hbz_t)S_{it}(\hbz_{t-1})\mid S_{it'}=1,\hbZ_{it'}=\hbz_{t'},
                                           \hbX_{it'}\},\quad t'\leq t, \\
  m_{S_t}(\hbz_{t'}, \hbX_{it'} ) \ := \ & \E\{S_{it}(\hbz_{t-1})\mid
                                        S_{it'}=1,\hbZ_{it'}=\hbz_{t'}, \hbX_{it'}\}, \quad t'\leq t-1.
\end{align*}
 When $t^\prime = 0$, we have
$m_{Y_tS_t} =\E\{Y_{it}(\hbz_t)S_{it}(\hbz_{t-1})\}$ and
$m_{S_t} =\E\{S_{it}(\hbz_{t-1})\} $.
\begin{proposition}[Identification of Regression Functions]
  \label{thm::identification-m} \spacingset{1} Under
  Assumptions~\ref{asm::noanticipation}--\ref{asm::overlap}, we have
  the following identification formulae:
\begin{eqnarray*}
  m_{Y_tS_t}(\hbz_{t'}, \hbX_{it'})&=& \begin{cases}
 \E(Y_{it} \mid S_{it}=1,\hbZ_{it}=\hbz_{t},\hbX_{it}), & \quad  t'=t,\\
 \E\left\{S_{i,t'+1}m_{Y_tS_t}(\hbz_{t'+1}, \hbX_{i,t'+1}) \mid S_{it'}=1,\hbZ_{it'}=\hbz_{t'}, \hbX_{i,t'}\right\}, & \quad t'< t.
\end{cases} \\
  m_{S_t}(\hbz_{t'}, \hbX_{it'})&=& \begin{cases}
\E(S_{it} \mid S_{i,t-1}=1,\hbZ_{i,t-1}=\hbz_{t-1},\hbX_{i,t-1}), & \quad  t'=t-1,\\
\E\left\{S_{i,t'+1}m_{S_t}(\hbz_{t'}, \hbX_{i,t'+1}) \mid S_{it'}=1,\hbZ_{it'}=\hbz_{t'}, \hbX_{it'}\right\}, & \quad t'< t-1.
\end{cases}
\end{eqnarray*}
\end{proposition}

Proposition~\ref{thm::identification-m} shows how to obtain the
identification formula for $m_{Y_tS_t}(\hbz_{t'}, \hbX_{it'})$ for any
$t' \leq t$.  Specifically, we first obtain the formula for
$m_{Y_tS_t}(\hbz_{t}, \hbX_{it})$ (i.e., $t^\prime = t$) and then
sequentially derive the formula for
$m_{Y_tS_t}(\hbz_{t'}, \hbX_{it'})$ for
$t'=t-1,t-2,\ldots,1,0$. Similarly, we can obtain the identification
formulae for $m_{S_t}(\hbz_{t'}, \hbX_{it'})$ for all $t' \leq
t-1$. When $t'=0$, the identification formula for $m_{S_t}$ coincides
with that of $\Pr\{S_{it}(\hbz_{t-1})=1\}$, while the identification
formula for the ratio of $m_{Y_tS_t}$ and $m_{S_t}$ equals the formula
for $ \E\{Y_{it}(\hbz_{t})\mid S_{it}(\hbz_{t-1})=1\}$.  Both formulae
are given in Theorem~\ref{thm::identification-obs}(a).

Given Proposition~\ref{thm::identification-m}, the following theorem
presents the EIFs of $\tau_t(\hbz_{t-1})$ and $\theta(\xi)$.
\begin{theorem}[Efficient Influence Functions]
\label{thm::eif} \spacingset{1}
Under Assumptions~\ref{asm::noanticipation}--\ref{asm::overlap}, the
EIFs of $\tau_t(\hbz_{t-1})$ and $\theta(\xi)$
are given by:
\begin{eqnarray}
\label{eqn::eif-tau} \phi_{\tau} &=&  \frac{\phi_{N}(\hbz_{t-1},1)-\phi_{N}(\hbz_{t-1},0) -  \tau_t(\hbz_{t-1}) \cdot \phi_{D}(\hbz_{t-1})  }{ \Pr\{ S_{it}(\hbz_{t-1})=1\}},\\
\label{eqn::eif-theta}  \phi_{\theta} &=&   \sum_{t=1}^T  \sum_{\hbz_t} \phi_{N}(\hbz_{t}) \xi_t(\hbz_t)- \theta(\xi),
\end{eqnarray}
respectively, where
\begin{eqnarray*}
\phi_{N}(\bz_t)
&=&  m_{Y_tS_t}(z_1,\bX_{i1}) + \frac{\bm{1}\{\hbZ_{it}=\hbz_t\} S_{it}\{Y_{it} -m_{Y_tS_t}(\hbz_{it},\hbX_{it})\}}{\pi_t(\hbz_t,\hbX_{it})}\\
&&+\sum_{t'=2}^{t}\frac{\bm{1}\{\hbZ_{i,t'-1}=\hbz_{t'-1}\} \{S_{it'}m_{Y_tS_t}(\hbz_{t'},\hbX_{it'})   -S_{i,t'-1}m_{Y_tS_t}(\hbz_{t'-1},\hbX_{i,t'-1})\}}{\pi_{t'-1}(\hbz_{t'-1},\hbX_{i,t'-1})},\\
\phi_{D}(\bz_{t-1})
&=& m_{S_t}(z_1,\bX_{i1}) + \frac{\bm{1}\{\hbZ_{i,t-1}=\hbz_{t-1}\} \{S_{it} -S_{i,t-1}m_{S_t}(\hbz_{t-1},\hbX_{i,t-1})\}}{\pi_{t-1}(\hbz_{t-1},\hbX_{i,t-1})}\\
&&+\sum_{t'=2}^{t-1}\frac{\bm{1}\{\hbZ_{i,t'-1}=\hbz_{t'-1}\}\{S_{it'} m_{S_t}(\hbz_{t'},\hbX_{it'})   -S_{i,t'-1}m_{S_t}(\hbz_{t'-1},\hbX_{i,t'-1})\}}{\pi_{t'-1}(\hbz_{t'-1},\hbX_{i,t'-1})}.
\end{eqnarray*}
The identification formulae of $m_{Y_tS_t}(\hbz_{t'}, \hbX_{it'})$ and
$m_{S_t}(\hbz_{t'}, \hbX_{it'} )$ are given in
Proposition~\ref{thm::identification-m}.
\end{theorem}
We can show that $\phi_N(\hbz_t)-m_{Y_tS_t}$ and
$\phi_D(\hbz_{t-1})-m_{S_t}$ are the EIFs of
$\E\{Y_{it}(\hbz_{t-1},z_t) S_{it}(\hbz_{t-1})\} $ and
$\Pr\{ S_{it}(\hbz_{t-1})=1\}$, respectively.  
 The EIFs do not include terms
corresponding to the perturbation of treatment
distribution. Therefore, like the case of the average treatment effect
estimation, knowing the propensity score does not affect the EIFs
\citep{hahn1998role}.

Theorem~\ref{thm::eif} generalizes the EIFs under the standard
longitudinal causal inference setting
\citep{van2012targeted,kennedy2019nonparametric,diaz2023nonparametric},
where all units are eligible for all treatments, i.e., $S_{it}=1$ for
all $i$ and $t$.  In this case, $\tau_t(\hbz_{t-1})$ reduces to the
average treatment effect of treatment trajectory $\hbz_{t-1}$ with
$z_t = 1$ versus $\hbz_{t-1}$ with $z_t=0$, while $\phi_\tau$ equals
the corresponding EIF in unconfounded longitudinal studies. In
addition, if there is only one time period $T=1$, $\tau_t(\hbz_{t-1})$
reduces to the average treatment effect in a single time point and
$\phi_\tau$ becomes the classic doubly robust formula
\citep{hahn1998role}. 

We provide additional intuition about these EIFs by focusing on
$\phi_N(\hbz_t)-m_{Y_tS_t}$, which consists of three terms.  First,
the difference term $m_{Y_tS_t}(z_1,\bX_{i1})- m_{Y_tS_t}$ corresponds
to the perturbation of the marginal distribution of $\bX_{i1}$.  The
second term corresponds to the perturbation of the conditional
distribution of $Y_{it}$ given $S_{it}=1$, $\hbZ_{it}$, and
$\hbX_{it}$.  Finally, the last term can be decomposed as:
\begin{eqnarray*}
&&\sum_{t'=2}^{t}  \frac{\bm{1}\{\hbZ_{i,t'-1}=\hbz_{t'-1}\} \{S_{it'}m_{Y_tS_t}(\hbz_{t'},\hbX_{it'})   -S_{i,t'-1}m_{Y_tS_t}(\hbz_{t'-1},\hbX_{i,t'-1})\}}{\pi_{t'-1}(\hbz_{t'-1},\hbX_{i,t'-1})}\\
&=&\sum_{t'=2}^{t} \left[\frac{\bm{1}\{\hbZ_{i,t'-1}=\hbz_{t'-1}\} S_{it'}}{\pi_{t'-1}(\hbz_{t'-1},\hbX_{i,t'-1})}\left\{ m_{Y_tS_t}(\hbz_{t'},\hbX_{it'}) - \frac{m_{Y_tS_t}(\hbz_{t'-1},\hbX_{i,t'-1})}{p_{t'}(\hbz_{t'-1},\hbX_{i,t'-1})} \right\}\right]\\
&& \hspace{.4in} +\sum_{t'=2}^{t} \left[\frac{\bm{1}\{\hbZ_{i,t'-1}=\hbz_{t'-1}\} \{S_{it'} -S_{i,t'-1}p_{t'}(\hbz_{t'-1},\hbX_{i,t'-1})\} m_{Y_tS_t}(\hbz_{t'-1},\hbX_{i,t'-1}) }{p_{t'}(\hbz_{t'-1},\hbX_{i,t'-1})\pi_{t'-1}(\hbz_{t'-1},\hbX_{i,t'-1})}\right],
\end{eqnarray*} 
where the equality follows in part from
$m_{Y_tS_t}(\hbz_{t'-1},\hbX_{i,t'-1}) = \E
\{m_{Y_tS_t}(\hbz_{t'},\hbX_{it'}) \mid
S_{it'}=1,\hbz_{t'-1},\hbX_{i,t'-1}\}$
$\times p_{t'}(\hbz_{t'-1},\hbX_{i,t'-1})$.  The first summation
corresponds to the perturbation of the conditional distribution of
$\hbX_{it'}$ given $S_{it'}=1$, $\hbZ_{i,t'-1}$, and $\hbX_{i,t'-1}$,
while the second represents that of the conditional distribution of
$S_{it'}$ given $S_{i,t'-1}=1$, $\hbZ_{i,t'-1}$, $\hbX_{i,t'-1}$.


Given these EIFs, we can solve $\E( \phi_\tau)=0$ and
$\E(\phi_\theta)=0$ to obtain the identification formulae for
$\tau_t(\hbz_{t-1})$ and $\theta(\xi)$, which are given in the
following corollary.
\begin{corollary}[Identification formulae based on the efficient
  influence functions]
\label{cor::eif-est} \spacingset{1}
Under  Assumptions~\ref{asm::noanticipation}--\ref{asm::overlap},
\begin{eqnarray*}
\tau_t(\hbz_{t-1}) &=& \frac{\E\left\{\phi_{N}(\hbz_{t-1},1)-\phi_{N}(\hbz_{t-1},0) \right\}}{ \E\left\{\phi_{D}(\hbz_{t-1})\right\}},\\
\theta(\xi) &=&   \sum_{t=1}^T  \sum_{\bz_t} \E\left\{\phi_{N}(\hbz_{t})\right\} \xi_t(\hbz_t).
\end{eqnarray*}
\end{corollary}
When all the covariates are time invariant, the EIFs in
Theorem~\ref{thm::eif} and identification formulae in
Corollary~\ref{cor::eif-est} can also be simplified by replacing
the formulae for $m_{Y_tS_t}(\hbz_{t'}, \hbX_{it'})$ and
$m_{S_t}(\hbz_{t'}, \hbX_{it'} )$ in Proposition~\ref{thm::identification-m} with the following:
\begin{eqnarray}
\label{eqn::mYS-invariant}m_{Y_tS_t}(\hbz_{t'}, \hbX_{it'})  &=&  \mu_t(\hbz_t,\bX_i) \prod_{s=t'}^t p_{s}(\hbz_{s-1},\bX_i),\\
\label{eqn::mS-invariant}m_{S_t}(\hbz_{t'}, \hbX_{it'})  &=&  \prod_{s=t'}^t p_{s}(\hbz_{s-1},\bX_i),
\end{eqnarray}
where $\bX_i$ is the set of time-invariant covariates.  Unlike those
shown in Theorem~\ref{thm::eif}, these formulae depend only on the
conditional mean of the outcome and eligibility probabilities. As a
result, they do not need to be iteratively defined.

\subsection{Estimation and inference}
\label{sec::est}

Using the EIFs derived above, we construct estimators that have a
double robustness property and attain parametric convergence rates
even when nuisance functions are estimated nonparametrically with
flexible machine learning algorithms.  We also introduce alternative
simple estimators that do not require models for the outcomes and
eligibility probabilities.

\paragraph{Estimators based on the EIFs.}
Corollary~\ref{cor::eif-est} motivates estimating the causal
quantities by replacing the nuisance functions with their estimates.
Therefore, we can estimate $\tau_t(\hbz_{t-1}) $ and $\theta(\xi)$ as:
\begin{eqnarray}
\label{eqn::mYS-eif}\hat\tau_t(\hbz_{t-1}) & = &  \frac{\hat \phi_{N}(\hbz_{t-1},1)-\hat \phi_{N}(\hbz_{t-1},0)}{\hat
                        \phi_{D}(\hbz_{t-1})}, \\
\label{eqn::mS-eif} \hat\theta(\xi) & = & \sum_{t=1}^T \sum_{\hbz_t} \hat
                  \phi_{N}(\hbz_{t})\xi(\hbz_t),
\end{eqnarray}
where
\begin{eqnarray*}
\nonumber \hat \phi_N(\hbz_t)
 &=& \frac{1}{n}\sum_{i=1}^n \left[\hat m_{Y_tS_t}(z_1,\bX_{i1}) + \frac{\bm{1}\{\bZ_{it}=\hbz_t\} S_{it}\{Y_{it} -\hat m_{Y_tS_t}(\hbz_t,\hbX_{it})\}}{ \hat{\pi}_t(\hbz_t,\hbX_{it})}\right.\\
\nonumber && \hspace{.5in} +\left. \sum_{t'=2}^{t}\frac{\bm{1}\{\hbZ_{i,t'-1}=\hbz_{t'-1}\} \{S_{it'}\hat m_{Y_tS_t}(\hbz_{t'},\hbX_{it'})   -S_{i,t'-1}\hat m_{Y_tS_t}(\hbz_{t'-1},\hbX_{i,t'-1})\}}{\hat{\pi}_{t'-1}(\hbz_{t'-1},\hbX_{i,t'-1})}\right],\\
\label{eqn::est-eif-tau}
\nonumber \hat \phi_D(\hbz_{t-1})
&=&\frac{1}{n}\sum_{i=1}^n \left[\hat
    m_{S_t}(z_1,\bX_{i1}) + \frac{\bm{1}\{\hbZ_{i,t-1}=\hbz_{t-1}\} \{S_{it} -S_{i,t-1}\hat m_{S_t}(\hbz_{t-1},\hbX_{i,t-1})\}}{\hat {\pi}_{t-1}(\hbz_{t-1},\hbX_{i,t-1})}\right.\\
\nonumber && \hspace{.5in} \left. +\sum_{t'=2}^{t-1}\frac{\bm{1}\{\hbZ_{i,t'-1}=\hbz_{t'-1}\}\{S_{it'} \hat m_{S_t}(\hbz_{t'},\hbX_{it'})   -S_{i,t'-1}\hat m_{S_t}(\hbz_{t'-1},\hbX_{i,t'-1})\}}{\hat{\pi}_{t'-1}(\hbz_{t'-1},\hbX_{i,t'-1})}\right].
\end{eqnarray*}

We discuss how to estimate the nuisance functions. Because
$\pi_{t}(\hbz_{t},\hbX_{it})= \prod_{s=1}^{t}
w_{s}(\hbz_{s},\hbX_{is})$, we estimate $w_{s}(\hbz_{s},\hbX_{is})$
for each $s$ by modeling $Z_{is}$ given $(\hbZ_{i,s-1}, \hbX_{is})$
for units with $S_{is}=1$ and then multiplying these estimates across
$s$ to obtain $\hat{\pi}_{t}(\hbz_{t},\hbX_{it})$.  For
$\hat m_{Y_tS_t}(\hbz_{t'},\hbX_{it'}) $, we propose the following
recursive modeling approach based on
Proposition~\ref{thm::identification-m}(a).
\begin{enumerate}
\item Fit a model of $Y_{it}$ given $\hbZ_{it}$ and $\hbX_{it}$ for
  units with $S_{it}=1$ to obtain an estimate of
  $m_{Y_tS_t}(\hbz_{t}, \hbX_{it})= \E(Y_{it} \mid
  S_{it}=1,\hbZ_{it}=\hbz_{t},\hbX_{it})$, denoted by
  $\hat m_{Y_tS_t}(\hbz_{t}, \hbX_{it})$.
\item Fit a model of
  $S_{i,t'+1}m_{Y_tS_t}(\hbz_{t'+1}, \hbX_{i,t'+1})$ given
  $\hbZ_{it'}$ and $\hbX_{it'}$ for units with $S_{it'}=1$, using
  $\hat m_{Y_tS_t}(\hbz_{t'+1}, \hbX_{i,t'+1})$, to recursively obtain
  an estimate of
  $m_{Y_tS_t}(\hbz_{t'}, \hbX_{it'})\ = \
  \E\left\{S_{i,t'+1}m_{Y_tS_t}(\hbz_{t'+1}, \hbX_{i,t'+1}) \mid
    S_{it'}=1,\hbZ_{it'}=\hbz_{t'}, \hbX_{it'}\right\}$ for
  $t'=t-1,t-2,\ldots,1$.
\end{enumerate}

The model in Step~2 may be difficult to specify since
$S_{i,t'+1}m_{Y_tS_t}(\hbz_{t'+1}, \hbX_{i,t'+1})=0$ for units with
$S_{i,t'+1}=0$.  We could alternatively write:
\begin{eqnarray}
\nonumber&& \E\left\{S_{i,t'+1}m_{Y_tS_t}(\hbz_{t'+1}, \hbX_{i,t'+1}) \mid S_{it'}=1,\hbZ_{it'}=\hbz_{t'}, \hbX_{it'}\right\}\\
\label{eqn::est-mYS-alternative}&=& \E\left\{m_{Y_tS_t}(\hbz_{t'+1}, \hbX_{i,t'+1}) \mid S_{i,t'+1}=1,\hbZ_{it'}=\hbz_{t'}, \hbX_{it'}\right\}p_{t'+1}(\hbz_{t'},\hbX_{it'}).
\end{eqnarray}
Then, the estimate of this term can be obtained by fitting a model of
$\hat m_{Y_tS_t}(\hbz_{t'+1}, \hbX_{i,t'+1})$ given $\hbZ_{it'}$ and
$\hbX_{it'}$ for units with $S_{it'}=1$, and a model of $S_{it'}$ given
$\hbZ_{i,t'-1},\hbX_{i,t'-1}$ for units with $S_{i,t'-1}=1$.

For $m_{S_t}(\hbz_{t'},\hbX_{it'}) $, we propose a similar recursive
approach based on Proposition~\ref{thm::identification-m}(b).
\begin{enumerate}
\item Fit a model of $S_{it}$ given $\hbZ_{i,t-1}$ and $\hbX_{i,t-1}$ for units with $S_{i,t-1}=1$ to obtain an estimate of $m_{S_t}(\hbz_{t-1}, \hbX_{i,t-1})= \E(S_{it} \mid S_{i,t-1}=1,\hbZ_{i,t-1}=\hbz_{t-1},\hbX_{i,t-1})$, denoted by $\hat m_{S_t}(\hbz_{t-1}, \hbX_{i,t-1})$.
\item Fit a model of $S_{i,t'+1}m_{S_t}(\hbz_{t'+1}, \hbX_{i,t'+1})$
  given $\hbZ_{it'},\hbX_{it'}$ for units with $S_{it'}$, using
  $\hat m_{S_t}(\hbz_{t'+1}, \hbX_{i,t'+1})$, to recursively obtain an
  estimate of \sloppy
  $m_{S_t}(\hbz_{t'},
  \hbX_{it'})=\E\left\{S_{i,t'+1}m_{S_t}(\hbz_{t'}, \hbX_{i,t'+1})
    \mid S_{it'}=1,\hbZ_{it'}=\hbz_{t'}, \hbX_{it'}\right\}$ for
  $t'=t-2,t-3,\ldots,1$.
\end{enumerate}
Similar to the above, for Step~2, an alternative estimation strategy
is to multiply the estimates of
$\E\left\{m_{S_t}(\hbz_{t'+1}, \hbX_{i,t'+1}) \mid
  S_{it'}=1,\hbZ_{it'}=\hbz_{t'}, \hbX_{it'}\right\}$ and
$p_{t'+1}(\hbz_{t'},\hbX_{it'})$.

The recursive model approach relies on the models for the conditional
mean outcome, eligibility probability, and treatment assignment.
However, we will show that the consistency of the proposed estimators
for $\tau_t(\hbz_{t-1})$ and $\theta(\xi)$ does not require the
correct specification of all models.  The following theorem shows that
the proposed estimators are consistent so long as one of the two
required models is correctly specified.
\begin{theorem}[Double Robustness]
  \label{thm::robustness} \spacingset{1} Let
  $\tilde m_{Y_tS_t}(\hbz_{t'},\hbx_{t'})$,
  $\tilde m_{S_t}(\hbz_{t'},\hbx_{t'})$, and
  $\tilde{\pi}_{t'}(\hbz_{t'},\hbx_{t'})$ denote the probability limit
  of $\hat m_{Y_tS_t}(\hbz_{t'},\hbx_{t'})$,
  $\hat m_{S_t}(\hbz_{t'},\hbx_{t'})$, and
  $\hat{\pi}_{t'}(\hbz_{t'},\hbx_{t'})$, respectively.  Then,   $\hat{\tau}_t(\hbz_{t-1})$ is consistent for $\tau_t(\hbz_{t-1})$ if either of the following two equalities hold:
\begin{eqnarray*}
(\tilde{m}_{Y_tS_t}(\hbz_{t'},\hbx_{t'}),\tilde{m}_{S_t}(\hbz_{t'},\hbx_{t'})) &=&(m_{Y_tS_t}(\hbz_{t'},\hbx_{t'}), m_{S_t}(\hbz_{t'},\hbx_{t'}) )\ \text{ for all} \ t^\prime,\\
  \tilde{\pi}_{t'}(\hbz_{t'},\hbx_{t'})&=&\pi_{t'}(\hbz_{t'},\hbx_{t'}) \ \text{ for all} \ t^\prime.
\end{eqnarray*}
  Similarly, $\hat\theta(\xi)$ is consistent for
  $\theta(\xi)$ if either of the following two equalities hold:
  \begin{eqnarray*}
\tilde{m}_{Y_tS_t}(\hbz_{t'},\hbx_{t'})=m_{Y_tS_t}(\hbz_{t'},\hbx_{t'}), \quad
\tilde{\pi}_{t'}(\hbz_{t'},\hbx_{t'})=\pi_{t'}(\hbz_{t'},\hbx_{t'}) \ \text{ for all} \ t^\prime.
\end{eqnarray*}
\end{theorem}
The proof shows that the bias of $\hat \phi_{N}(\hbz_{t})$ for
estimating $\E\{Y_{it}(\hbz_t)S_{it}(\hbz_{t-1})\}$ is given by:
\begin{align*}
 \sum_{t'=1}^t \int &  \{m_{Y_tS_t}(\hbz_{t'},\hbx_{t'})-\tilde
                         m_{Y_tS_t}(\hbz_{t'},\hbx_{t'})\}
                         \left\{\frac{
                         \pi_{t'}(\hbz_{t'},\hbx_{t'})}{\tilde
                         \pi_{t'}(\hbz_{t'},\hbx_{t'})} - \frac{
                         \pi_{t'-1}(\hbz_{t'-1},\hbx_{t'-1})}{\tilde
                         \pi_{t'-1}(\hbz_{t'-1},\hbx_{t'-1})}
                         \right\} \\
  & \times \prod_{s=2}^{t'} p_s (\hbz_{s-1},\hbx_{s-1}) \prod_{s=1}^{t'}l_s  (\hbz_{s-1},\hbx_s)  \text{d} \hbx_{t'},
\end{align*}
while the bias of $\hat \phi_{D}(\bz_{t-1})$ for estimating
$\Pr\{S_{it}(\bz_{t-1})=1\}$ has the following expression:
\begin{align*}
 \sum_{t'=1}^{t-1} \int & \{m_{S_t}(\hbz_{t'},\hbx_{t'})-\tilde
                           m_{S_t}(\hbz_{t'},\hbx_{t'})\} \left\{\frac{
                           \pi_{t'}(\hbz_{t'},\hbx_{t'})}{\tilde
                           \pi_{t'}(\hbz_{t'},\hbx_{t'})} - \frac{
                           \pi_{t'-1}(\hbz_{t'-1},\hbx_{t'-1})}{\tilde
                           \pi_{t'-1}(\hbz_{t'-1},\hbx_{t'-1})}
                           \right\}  \\
  & \times \prod_{s=2}^{t'} p_s (\hbz_{s-1},\hbx_{s-1}) \prod_{s=1}^{t'}l_s  (\hbz_{s-1},\hbx_s)   \text{d} \hbx_{t'},
\end{align*}
The double robustness follows immediately from these bias expressions.

Finally, if all models are correctly specified, then the estimators
are asymptotically Normal and semiparametrically efficient under
certain regularity conditions.
\begin{theorem}[Asymptotic Normality and Efficiency]
  \label{thm::efficiency} \spacingset{1} Suppose that in addition to
  Assumptions~\ref{asm::noanticipation}--\ref{asm::overlap}, the
  following conditions hold:
\begin{enumerate}[(a)]
\item
  $\tilde
  m_{Y_tS_t}(\hbz_{t'},\hbx_{t'})=m_{Y_tS_t}(\hbz_{t'},\hbx_{t'})$,
  $\tilde m_{S_t}(\hbz_{t'},\hbx_{t'})=m_{S_t}(\hbz_{t'},\hbx_{t'})$, and
  $\tilde{\pi}_{t'}(\hbz_{t'},\hbx_{t'})=\pi_{t'}(\bz_{t'},\bx_{t'})$
  for all $t^\prime$;
\item The estimated functions $\hat m_{Y_tS_t}(\hbz_{t'},\hbx_{t'})$,
  $\hat m_{S_t}(\hbz_{t'},\hbx_{t'}) $, and
  $\hat \pi_{t'}(\hbz_{t'},\hbx_{t'})$ are in a Donsker class;
\item $\delta<\pi_{t}(\hbz_{t},\hbx_{t})<1-\delta$ and
  $\delta<\hat{\pi}_{t}(\hbz_{t},\hbx_{t})<1-\delta$ for all $\hbx_{t}$
  and some $\delta \in (0,\frac{1}{2})$;
\item For all $t^\prime$,
\begin{eqnarray*}
|| m_{Y_tS_t}(\hbz_{t'},\hbX_{it'})-\tilde
  m_{Y_tS_t}(\hbz_{t'},\hbX_{it'}) ||_{2,s}\cdot ||\hat
  w_{t'}(\hbz_{t'},\hbX_{it'})-w_t(\hbz_{t'},\hbX_{it'}) ||_{2,s} &=&
  o_P(n^{-1/2}),\\
  || m_{S_t}(\hbz_{t'},\hbX_{it'})-\tilde
  m_{S_t}(\hbz_{t'},\hbX_{it'}) ||_{2,s} \cdot ||\hat
  w_{t'}(\hbz_{t'},\hbX_{it'})-w_t(\hbz_{t'},\hbX_{it'}) ||_{2,s} &=&
  o_P(n^{-1/2}),
\end{eqnarray*}
where $||\cdot||_{2,s}$ is defined as
  follows for any function of $\hbX_{it'}$, $f(\hbX_{it'})$:
 \begin{eqnarray*}
||f(\hbX_{it'})||_{2,s} &=& \int f^2(\hbx_{t'})  \prod_{s=2}^{t'} p_s
                            (\hbz_{s-1},\hbx_{s-1}) \prod_{s=1}^{t'}l_s  (\hbz_{s-1},\hbx_s)   \text{d} \hbx_{t'}.
 \end{eqnarray*}
\end{enumerate}
Then, $ \{\hat \phi_{N}(\hbz_{t-1},1)-\hat \phi_{N}(\hbz_{t-1},0) \}/\hat \phi_{D}(\hbz_{t-1})$ and  $\sum_{t=1}^T  \sum_{\hbz_t} \hat \phi_{N}(\hbz_{t})\xi(\hbz_t)$ are asymptotically Normal and achieve the semiparametric efficiency bound.
\end{theorem}
Theorem~\ref{thm::efficiency} extends double machine learning
\citep[e.g.,][]{chernozhukov2018double} to longitudinal settings with
selective eligibility.  Condition~(a) requires the consistency of all
nuisance models; Condition~(b) imposes constraints on the complexity
of the spaces of their estimators, which can be relaxed by
cross-fitting; Condition~(c) requires the overlap condition to hold
for both the true and estimated treatment probabilities, enabling
bounding the estimation errors; Condition~(d) requires the nuisance
estimators to converge at a fast enough rate, though it can be slower
than the $n^{1/2}$ parametric convergence rates.

\paragraph{Alternative simple estimators.}
The above estimators are efficient but require complex modeling.  In
practice, researchers may prefer a simpler estimation strategy even at
the expense of some efficiency.  We consider two alternative
estimation strategies for
$\E\{Y_{it}(\hbz_t)\mid S_{it}(\hbz_{t-1})=1\} $ and
$\Pr\{ S_{it}(\hbz_{t-1})=1\} $ using the identification results shown
in Theorem~\ref{thm::identification-obs}.

First, based on the identification formula of
Theorem~\ref{thm::identification-obs}(a), we obtain the estimates
$\hat m_{Y_tS_t}(\hbz_t, \hbX_{it})$ and
$\hat m_{S_t}(\hbz_t, \hbX_{it})$ and then compute the following
estimators:
\begin{eqnarray}
\label{eqn::mYS-outreg}\widehat \E\{Y_{it}(\hbz_t)\mid S_{it}(\hbz_{t-1})=1\} &=& \frac{\sum_{i=1}^n\hat m_{Y_tS_t}(\hbz_t, \hbX_{it})  }{\sum_{i=1}^n\hat m_{S_t}(\hbz_t, \hbX_{it})  },\\
\label{eqn::mS-outreg}\widehat \Pr\{ S_{it}(\hbz_{t-1})=1\} &=& \frac{1}{n} \sum_{i=1}^n\hat m_{S_t}(\hbz_t, \hbX_{it}).
\end{eqnarray}
This estimation strategy avoids modeling the covariate distribution, relying only on the outcome mean and eligibility probability models.

Next, based on Theorem~\ref{thm::identification-obs}(b), we estimate
$\hat{\pi}_{t-1}(\hbz_{t-1},\hbX_{i,t-1})$ by modeling the treatment
assignment.  Using these estimates, we obtain the following
estimators:
\begin{eqnarray}
\label{eqn::mYS-ipw}\widehat \E\{Y_{it}(\hbz_t)\mid S_{it}(\hbz_{t-1})=1\} & = & \frac{\sum_{i=1}^nY_{it}\bm{1}\{\hbZ_{it}=\hbz_t,S_{it}=1\}/\hat{\pi}_{t-1}(\hbz_{t-1},\hbX_{i,t-1}) }{\sum_{i=1}^n\bm{1}\{\hbZ_{i,t-1}=\hbz_{t-1},S_{it}=1\}/\hat{\pi}_{t-1}(\hbz_{t-1},\hbX_{i,t-1}) },\\
\label{eqn::mS-ipw}\widehat \Pr\{ S_{it}(\hbz_{t-1})=1\} &=& \sum_{i=1}^n\frac{\bm{1}\{\hbZ_{i,t-1}=\hbz_{t-1},S_{it}=1\}}{\hat{\pi}_{t-1}(\hbz_{t-1},\hbX_{i,t-1})}.
\end{eqnarray}
This estimation strategy relies only on the modeling of the treatment probability. However, it may be unstable if some of the estimated treatment probabilities are close to zero or one. 

Finally, the recursive modeling approach becomes unnecessary when the
covariates are time invariant. From
Equations~\eqref{eqn::mYS-invariant}~and~\eqref{eqn::mS-invariant}, we
can plug in the estimated outcome mean and the eligibility
probabilities in the identification formula and replace the
expectation with the empirical average to obtain the estimators of
$\hat m_{Y_tS_t}(\hbz_{t'},\hbX_{it'}) $ and
$\hat m_{S_t}(\hbz_{t'},\hbX_{it'}) $.

\section{Simulation Study}
\label{sec::simulation}


\subsection{Setup}

We first generate the potential values of outcome and eligibility
indicators, and then compute their observed values after generating
the realized treatment:
\begin{enumerate}
\item Generate four independent standard Normal random variables, $\bX_i=(X_{i1},\ldots,X_{i4})$;
\item Generate $Y_{i1}(z_1)$ and $S_{i2}(z_1)$ from Normal linear and
  logistic models:
  $\E\{Y_{i1}(z_1)\mid \bX_i\}= -1+z_1+0.5X_{i1}-X_{i3}$, and
  $\Pr\{S_{i2}(z_1)=1\mid \bX_i\} =
  \text{logit}^{-1}(1+z_1+0.5X_{i2}-0.5X_{i3}-X_{i4})$.
\item Generate $Y_{i2}(z_1,z_2)$ and $S_{i3}(z_1,z_2)$ from Normal
  linear and logistic models conditional on $S_{i2}(z_1)=1$ while
  setting $S_{i3}(z_1,z_2)=0$ for units with $S_{i2}(z_1)=0$:
  $\E\{Y_{i2}(z_1,z_2)\mid S_{i2}(z_1)=1,\bX_i,Y_{i1}(z_1)\}
  =-0.5-0.5z_1-0.5z_1z_2+X_{i2}-0.5X_{i4}+\delta Y_{i1}(z_1)$, and
  $\Pr\{S_{i3}(z_1,z_2)=1\mid S_{i2}(z_1)=1,\bX_i\} =
  \text{logit}^{-1}(1-0.5z_1-z_2+0.5X_{i2}-X_{i3})$.
\item Generate $Y_{i3}(z_1,z_2,z_3)$ from a Normal linear model
  conditional on $S_{i3}(z_1,z_2)=1$:
  $\E\{Y_{i3}(z_1,z_2,z_3)\mid
  S_{i3}(z_1,z_2)=1,\bX_i,Y_{i1}(z_1),Y_{i2}(z_1,z_2)\} =
  -1-0.5z_2-z_3+0.5z_2z_3+X_{i1}-0.5X_{i3}-\delta Y_{i2}(z_1,z_2)$.
\item Generate $Z_{i1}$ from a logistic model and compute the
  observed data $Y_{i1}$ and $S_{i2}$:
  $\Pr\{Z_{i1}=1\mid \bX_i\}=
  \text{logit}^{-1}(0.2+0.2X_{i1}-0.4X_{i2})$.

\item Generate $Z_{i2}$ from a logistic model conditional on
  $S_{i2}=1$ and compute the observed data $Y_{i2}$ and $S_{i3}$:
  $\Pr\{Z_{i2}=1\mid S_{i2}=1,Z_{i1}=z_1,\bX_i,Y_{i1}\} =
  \text{logit}^{-1}(0.5-0.5z_1+0.5X_{i2}-0.5X_{i4}+\delta Y_{i1})$.

\item Generate $Z_{i3}$ from a logistic model conditional on
  $S_{i3}=1$ and compute $Y_{i3}$:
  $\Pr\{Z_{i3}=1\mid
  S_{i3}=1,Z_{i1}=z_1,Z_{i2}=z_2,\bX_i,Y_{i1},Y_{i2}\} =
  \text{logit}^{-1}(1- 0.2z_1-0.5z_2+0.5X_{i1}+0.5X_{i3}+\delta Y_2)$.

\end{enumerate}
In this data generating process, $\delta$ determines the degree to which the outcome in the preceding period affects the current treatment assignment and outcome; we consider two scenarios: $\delta=0$ and $\delta=0.5$.  When
$\delta=0$, Assumption~\ref{asm::ignorability} holds with time
invariant confounders $\bX_{it}=\bX_i$ for $t=1,2,3$. When
$\delta=0.5$, Assumption~\ref{asm::ignorability} holds with time
varying covariates $\bX_{i1}=\bX_i$, $\bX_{i2}=(\bX_i,Y_{i1})$, and
$\bX_{i2}=(\bX_i,Y_{i2})$.

Under each scenario, we consider the outcome regression estimators
based on Equations~\eqref{eqn::mYS-outreg}~and~\eqref{eqn::mS-outreg},
the IPW estimators based on
Equations~\eqref{eqn::mYS-ipw}~and~\eqref{eqn::mS-ipw}, and the DR
estimators based on
Equations~\eqref{eqn::mYS-eif}~and~\eqref{eqn::mS-eif}. The causal
quantities of interest include $\tau_1$, $\tau_2(z_1)$, and
$\tau_3(z_1,z_2)$ for $z_1,z_2=0,1$, and the expected cumulated
outcomes under the treat-nobody strategy $\theta_0$ and treat-everyone
strategy $\theta_1$.

Following \citet{kang2007demystifying} and
\citet{kennedy2019nonparametric}, we use both parametric and
nonparametric methods to estimate the nuisance functions with both
correctly specified covariates $\bX_i$ and mis-specified covariates
$\bX_i^\ast$, which are constructed by transforming the true
covariates as follows:
$X_1^\ast = \exp(X_1/2), \quad X_2^\ast\ =\ \frac{X_2}{1+\exp(X_1)}+10, \quad
X_3^\ast =  \left(\frac{X_1X_3}{25}+0.6\right)^3, \quad X_4^\ast\ = \ (X_1+X_4+20)^2$.

In total, we have four different versions of simulation.  For
parametric estimation, we use linear models for the continuous
variables and logistic models for binary variables.  For nonparametric
estimation, we use the Super Learner ensemble \citep{van2007super}
that combines parametric models, random forest, and generalized
additive models.

\subsection{Results}

\begin{figure}[!t]
 \centering \spacingset{1}
 \includegraphics[width=\textwidth]{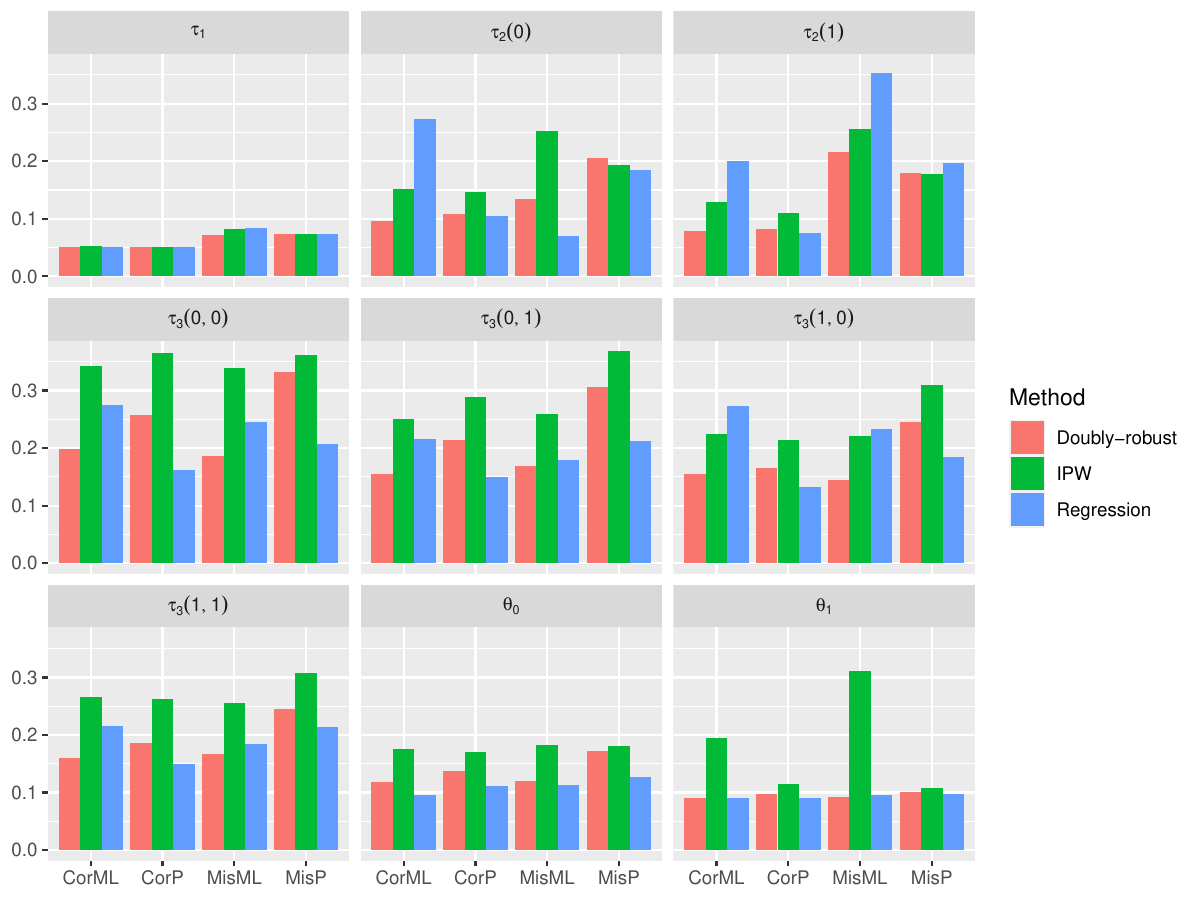}
 \caption{Absolute bias of each estimator across $500$ replications
   with $\delta=0$ and $n=1,000$. Each plot presents the results for
   one target causal quantity. The labels on the horizontal axis ---
   ``CorML,'' ``CorP,'' ``MisML,'' and ``MisP'' --- stand for the
   nonparametric estimation with correctly specified covariates,
   parametric estimation with correctly specified covariates,
   nonparametric estimation with mis-specified covariates, and
   parametric estimation with mis-specified covariates,
   respectively. The height of each bar represents the absolute bias
   of the regression (blue), IPW (green), or doubly robust estimator
   (red).  }
\label{fig::sim_inv_n1_con}
\end{figure}

\begin{figure}[!t]
 \centering \spacingset{1}
 \includegraphics[width=\textwidth]{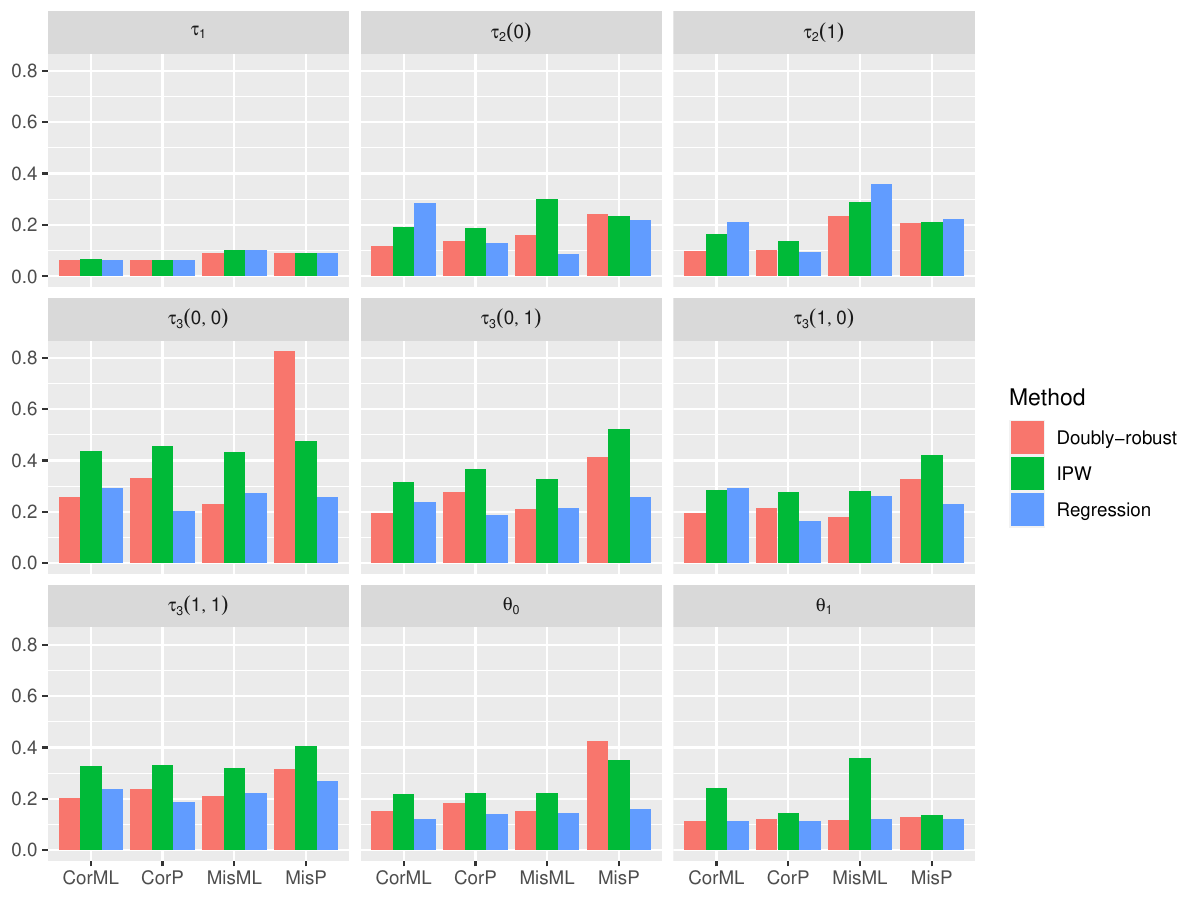}
 \caption{Root mean squared error (RMSE) of each estimator with
   $\delta=0$ and $n=1,000$. Each subplot presents the results for one
   target causal quantity. Within each subplot, the $x$-axis labels
   `CorML,'' ``CorP,'' ``MisML,'' and ``MisP'' stand for the
   nonparametric estimation with correctly specified covariates,
   parametric estimation with correctly specified covariates,
   nonparametric estimation with mis-specified covariates, and
   parametric estimation with mis-specified covariates, respectively.
   The height of each bar represents the RMSE of the regression
   (blue), IPW (green), or doubly robust estimator (red).}
\label{fig::sim_inv_n1_con_rmse}
\end{figure}

Figure~\ref{fig::sim_inv_n1_con} presents the absolute bias of each
estimator based on $500$ Monte Carlo simulations under the scenario
with time-invariant confounders and a sample size of $n=1,000$.
Figure~\ref{fig::sim_inv_n1_con_rmse} shows the root mean square error
(RMSE) under the same settings, while
Figures~\ref{fig::sim_var_n1_con_bias}~and~\ref{fig::sim_var_n1_con_rmse}
in the appendix present the corresponding results with the time
varying confounders.

Across all scenarios other than the parametric models with
misspecification (``MisP''), the DR estimator (red) consistently
demonstrates comparable or superior performance when compared to the
outcome regression (blue) and IPW estimators (green).  All three
estimators yield small absolute biases and RMSEs when the nuisance
functions are estimated using parametric models with correctly
specified covariates (``CorP'').  In contrast, the outcome regression
and IPW estimators exhibit large biases and RMSEs when nuisance
functions are estimated nonparametrically (``CorML'').  This aligns
with the semiparametric theory that these estimators may converge at a
much slower rate than the DR estimator.

With mis-specified covariates, biases and RMSEs tend to increase
across all estimators. Notably, the DR estimator, when nuisance
functions are nonparametrically estimated (``MisML''), outperforms
other estimators, possibly due to a more accurate approximation to the
true models. In contrast, the IPW and outcome regression estimators
tend to have greater bias and RMSE.  This is primarily attributable to
extreme treatment probabilities estimated using mis-specified
covariates.

Additional simulation results in the appendix include scenarios with
an increased sample size and settings with binary outcomes generated
from logistic models with both time-invariant and time-varying
confounders.  These results are qualitatively similar. 

\section{Empirical Analysis}
\label{sec::application}

In this section, we apply the proposed methodology to analyze our own RCT of the PSA used in the U.S. criminal justice system. We briefly describe the background of this original RCT.  \citet{greiner2020randomized} and 
\cite{imai2020experimental} provide additional details of this study. 

\subsection{Background}

In many U.S. jurisdictions, individuals who have been arrested attend
first appearance hearings. During this hearing, a judge decides
whether to release an arrestee pending the disposition of any criminal
charges and the conditions to impose for their release (e.g., cash
bail and supervision). A judge has to weigh the costs of incarceration
against the risk of the following negative outcomes if the arrestee is
released: (1) failing to appear (FTA) at subsequent court hearings,
(2) engaging in a new criminal activity (NCA), and (3) engaging in a
new violent criminal activity (NVCA).\footnote{ Note that because it
  is possible for individuals to make bail to secure their release,
  assigning cash bail does not necessarily preclude the negative
  outcomes from occurring.}

The PSA is an algorithmic recommendation system designed to help
judges make pretrial release decisions. It uses nine factors,
including an arrestee's age and criminal history, while excluding
other demographic factors such as race and gender. The PSA consists of
three separate risk scores for FTA, NCA, and NVCA. The information
about the formulae used to calculate these scores is publicly
available (\url{https://advancingpretrial.org/psa/factors/}). Finally,
the PSA combines the risk scores and other factors to produce overall
release and monitoring recommendations for the judge.

We conducted a randomized experiment in Dane County, Wisconsin to
examine how the PSA affects a judge's pretrial release decision.  We
randomized the provision of PSA across cases, observed the judge's
decisions, and collected the outcome variables, i.e., the occurrence
of FTA, NCA, and NVCA.  For each case assigned to the treatment
condition, the judge received a PSA report that contained the three
risk scores, the values of the nine factors supporting those scores,
and the results of the overall recommendation for that case. For cases
randomized to the control condition, the judge received no such
report.

An important complication in this analysis is the issue of recidivism.
Because treatment is assigned only to arrestees when they attend their
first appearance hearing, arrestees who recidivate receive the
treatment assignment multiple times while non-recidivating arrestees
receive it only once. Moreover, the treatment status of a previous
case can affect \emph{whether} an individual recidivates, and so the
initial treatment assignment can affect how frequently arrestees
appear in the data. Therefore, assessing the criminal justice impacts
of providing the PSA requires accounting for selective eligibility.

Previous analyses have avoided this selective eligibility problem by
focusing on only each individual's first case, limiting their scope to
first-time arrests
\citep{imai2020experimental,ben2021safe,benmichael2024does}. However,
in our extended dataset of $4,445$ arrestees from Dane County, $968$
have at least two cases and $315$ have at least three cases.  Note
that individuals may be arrested multiple times for different reasons,
each varying in how they relate to our three main outcomes of
interest.  For example, an arrestee who fails to appear in court (FTA)
after the first arrest could be rearrested for committing a new crime.

\begin{table}[!t] \centering \spacingset{1} \setlength{\tabcolsep}{3.5pt}
\begin{tabular}{@{\extracolsep{5pt}} lrrrrrrrrr}
\\[-1.8ex]\hline
		\hline \\[-1.8ex]	
		& \multicolumn{4}{c}{PSA (Treatment)}  & 	\multicolumn{4}{c}{No PSA (Control)} &\\
		\cmidrule(lr){2-5}\cmidrule(lr){6-9}
		& FTA & NCA & NVCA &Group total & FTA & NCA & NVCA &Group total  & Total\\ \hline
First	&  462   & 700    & 188 &2210 & 471 &653 &184&   2235    & 4445\\
                 & (10.4) &  (15.7) & (4.2) & (49.7)& (10.6)& (14.7) & (4.1) &(50.3) & (100) \\     
Second	&  143   & 210    & 63& 488& 124 &205 &57       &480& 968\\
                 & (14.8) &  (21.7) & (6.5)& (50.4) & (12.8)& (21.2) & (5.9) & (49.6)&(100) \\     
Third		&  39   & 87    & 23&164 & 43 &68 &20       &151& 315\\
                 & (12.4) &  (27.6) & (7.3) &(52.1)& (13.7)& (21.6) & (6.3) &(47.9)& (100) \\     \hline        
\end{tabular}
\caption{The distribution of negative outcomes for the first, second,
  and third cases in the treatment and control groups; failure to
  appear (FTA), new criminal activity (NCA), and new violent criminal
  activity (NVCA). The table shows the number
     of cases in each category with the corresponding percentage
     in parentheses. The percentages are computed with respect to the
     total number of cases for each arrest.}
\label{tab::summary}
\end{table}


Table~\ref{tab::summary} shows the distribution of the three outcomes
for the first, second, and third arrests with the corresponding
percentage in parentheses. We find that the outcome distribution
varies substantially across subgroups. Therefore, restricting the
analysis to the first cases jeopardizes the external validity of the
experimental results to the overall population by overlooking the
heterogeneity across different cases of the arrestees. The issue of
recidivism also motivates an exploration of arrestees' behavior over
time under specific PSA provision or bail decision strategies. For
instance, one might be interested in the total number of crimes an
arrestee would commit if a PSA report were provided to the judge for
each of their cases, relative to the number of crimes if a PSA report
were never provided.

\subsection{Setup}
We limit our analyses to estimating the
causal effects up to the third case $T=3$, as the number of cases
decreases to only 113 when considering the fourth case.  We first
examine whether the judge
becomes more likely to be influenced by the PSA recommendation in
later arrest cases.
We also evaluate the impact of PSA provision on the occurrence of each negative
outcome (FTA, NCA, or NVCA).   Finally, we estimate the expected number of
negative outcomes under two policies: one that provides PSA
recommendations to all cases and the other that never provides PSA
recommendations to any case.

Let $Z_{it}$ represent the randomized treatment assignment variable
with $Z_{it}=1$ implying that the judge received the PSA
recommendation for arrestee $i$ after the $t$th arrest for
$i=1,2,\ldots,n$ and $t=1,2,3$.  We consider two outcomes --- the
agreement between the judge's decision and PSA recommendation and the
occurrence of each negative outcome (FTA, NCA, and NVCA).  For each of
these two outcomes, we estimate several causal quantities.  They
include the average effect of the first PSA provision $\tau_1$, the
average effects of the second PSA provision $\tau_2(z_1)$ for
$z_1=0,1$, and the average effects of the third PSA provision
$\tau_3(z_1,z_2)$ for $z_1,z_2=0,1$.  In addition, we also estimate
the expected number of negative outcomes under two policy-relevant
scenarios: if the judge were to receive PSA recommendation for every
case, $\theta_1$, and if the judge were to receive PSA recommendation
for no cases, $\theta_0$. Finally, we estimate the difference between
these two scenarios, i.e., $\theta_1 - \theta_0$.

Because the PSA provision is randomized,
Assumption~\ref{asm::ignorability} holds without conditioning on
covariates. Nevertheless, for potential efficiency gains, we include
several covariates in our models.  Because the treatment probability
model is always correctly specified, the DR estimator will be
consistent regardless of whether the other nuisance functions are
correctly specified. First, we use demographic variables: age, gender,
race, and the interaction between gender and race.  The second set of
covariates is the PSA recommendation: three risk scores (for FTA, NCA,
and NVCA), and the binary overall release recommendation.  The final
set include those used to compute the PSA recommendation: a binary
variable for the presence of pending charge (felony, misdemeanor, or
both) at the time of offense, two binary variables for prior
convictions (misdemeanor and felony), a four-level variable for prior
violent convictions, and a three-level variable for FTAs within the
past two years.

\subsection{Findings}

\begin{figure}[!t]
 \centering \spacingset{1}
 \includegraphics[width=\textwidth]{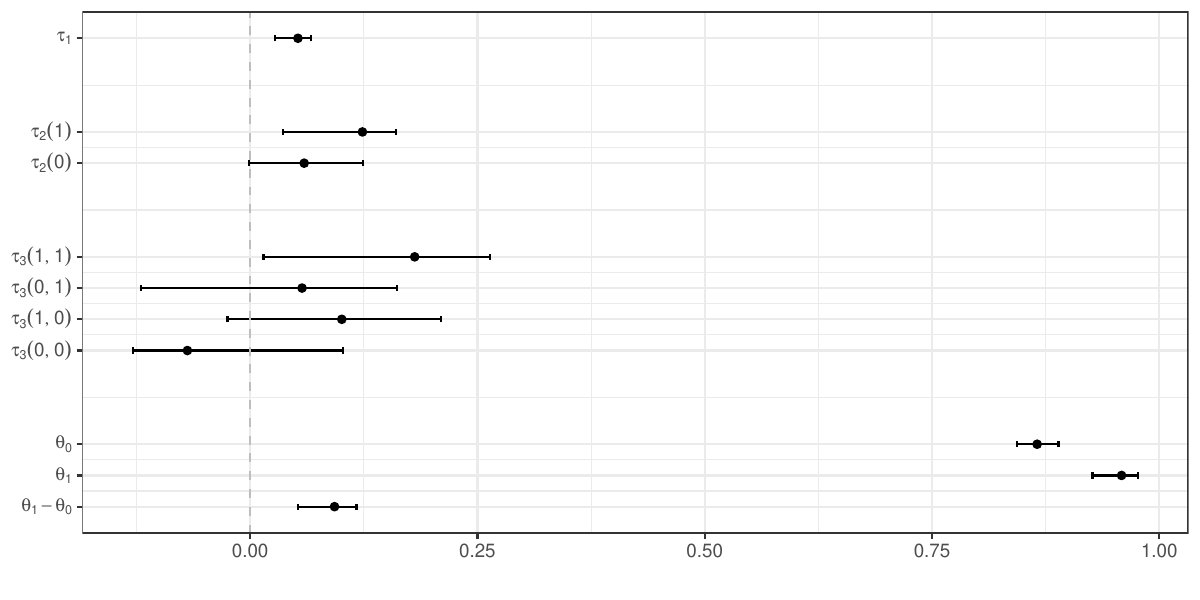}
 \caption{Estimated average effects of PSA provision on whether the
   judge's decision agrees with the PSA recommendation using
   nonparametric models.  The figure shows the estimated effect at
   time $t=1$ on the top ($\tau_1$), the two estimated effects at time
   $t=2$ in the middle ($\tau_2(z_1)$), the four estimated effects at
   time $t=3$ in the bottom ($\tau_3(z_2,z_1)$), and the expected
   number of each negative outcome under the never-treat strategy
   ($\theta_0$) and always-treat strategy ($\theta_1$). The horizontal
   lines represent the 95\% bootstrap confidence intervals.}
\label{fig::rct-agree}
\end{figure}

We first estimate the effect of PSA provision on whether the judge's
decision agrees with the PSA recommendation at time $t=1,2,3$. We set
$Y_{it}=1$ if they agree and $Y_{it}=0$
otherwise. Figure~\ref{fig::rct-agree} presents the results. We find
that PSA provision generally increases the probability of agreement
between the judge's decision and the PSA recommendation.  The
estimated causal effects are statistically significant for the first
two arrest cases.  For the third arrest case, which has a limited
sample size, the estimated effects are not statistically significant
with the exception of the cases, in which the judge receives the PSA
recommendation for the first two arrest cases.  This finding, along
with the fact that $\tau_2(1)$ is slightly greater than $\tau_2(0)$
(though the difference is statistically insignificant), may suggest
that the judge is more likely to be influenced by the PSA
recommendation in the second and third arrest cases if the
recommendation were given to them in the earlier arrest cases.

\begin{figure}[!t]
 \centering \spacingset{1}
 \includegraphics[width=\textwidth]{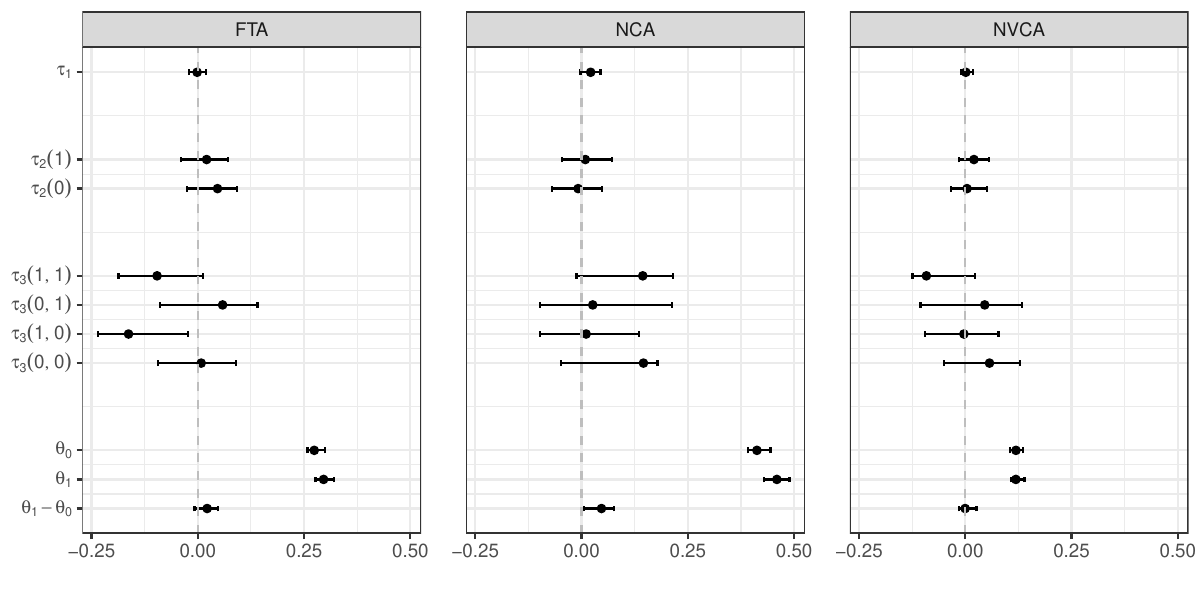}
 \caption{Estimated average effects of PSA provision on the three
   negative outcomes (FTA, NCA, and NVCA) based on nonparametric
   models.  Each panel includes the estimated effect at time $t=1$ on
   the top ($\tau_1$), the two estimated effects at time $t=2$ in the
   middle ($\tau_2(z_1)$), the four estimated effects at time $t=3$ in
   the bottom ($\tau_3(z_2,z_1)$), and the expected number of each
   negative outcome under the never-treat strategy ($\theta_0$) and
   the always-treat strategy ($\theta_1$). The horizontal lines
   represent 95\% bootstrap confidence intervals.}
\label{fig::rct}
\end{figure}

Furthermore, Figure~\ref{fig::rct} shows the estimated effects of PSA
provision on the occurrence of each negative outcome with their 95\%
bootstrap confidence intervals. We find that none of the estimated
effects is statistically significant with the exception of the
estimated difference in the expected number of NCAs under the
always-provide and never-provide strategies (middle panel).  The
result suggests that the provision of the PSA may increase the number of
NCAs.  The fact that most estimates are statistically insignificant is
consistent with an earlier analysis of first arrest cases
\citep{imai2020experimental}.

\section{Concluding remarks}

In this paper, we proposed a general causal framework for defining and
estimating treatment effects in longitudinal studies with selective
eligibility.
We established identification results and developed estimation
approaches for the causal quantities.  The proposed methodology,
therefore, generalizes existing methods for standard longitudinal
studies to this new setting.


In our randomized evaluation of the PSA, we find that providing the
PSA increases the rate of agreement between the algorithm's
recommendation and the judge's decision, with potentially larger
impacts on agreement when the PSA is repeatedly provided for the same
arrestee across re-arrests.  We mostly find null impacts of the
provision of the PSA on the three negative outcomes, though there is
some evidence that the total number of new criminal activities would
increase if the PSA were to always be provided to the judge relative
to cases where it were never provided to the judge.

There are several avenues for future work. While we focused on
nonparametric identification and estimation, with a large value of
$t$, the causal effect $\tau_t(\bz_{t-1})$ conditions on a long
treatment history, increasing the likelihood of violating the overlap
condition required for identification. In addition, the number of
causal estimands grows exponentially with $t$.  In such cases,
incorporating modeling assumptions becomes necessary for addressing
identification challenges and improving estimation.  One possible
approach is to apply structural nested mean models under selective
eligibility \citep{robins1992estimation,hernan2020causal}.

It is also of interest to extend the proposed framework to continuous
time longitudinal studies with selective eligibility. The survival
analysis literature has formulated the problem as semi-competing
risks, in which the primary outcome is the nonterminal event and the
process of selective eligibility acts as a competing risk
\citep{fine2001semi}.  However, most
existing research does not involve time-varying treatments \citep{comment2019survivor,xu2022bayesian}.  Our framework may be generalized to assess the effect of a time-varying treatment in the presence of semi-competing risks.

\newpage

\pdfbookmark[1]{References}{References}
\spacingset{1.45}
\bibliographystyle{Chicago}
\bibliography{longitudinaldropout-ref}

\begin{thebibliography}{}

\bibitem[\protect\citeauthoryear{Baer, Strawderman, and Ertefaie}{Baer
  et~al.}{2023}]{baer2023causal}
Baer, B.~R., R.~L. Strawderman, and A.~Ertefaie (2023).
\newblock Causal inference for the expected number of recurrent events in the
  presence of a terminal event.
\newblock Technical report, arXiv preprint arXiv:2306.16571.

\bibitem[\protect\citeauthoryear{Bang and Robins}{Bang and
  Robins}{2005}]{bang2005doubly}
Bang, H. and J.~M. Robins (2005).
\newblock Doubly robust estimation in missing data and causal inference models.
\newblock {\em Biometrics\/}~{\em 61\/}(4), 962--973.

\bibitem[\protect\citeauthoryear{Ben-Michael, Feller, and
  Rothstein}{Ben-Michael et~al.}{2022}]{ben2022synthetic}
Ben-Michael, E., A.~Feller, and J.~Rothstein (2022).
\newblock Synthetic controls with staggered adoption.
\newblock {\em Journal of the Royal Statistical Society Series B: Statistical
  Methodology\/}~{\em 84\/}(2), 351--381.

\bibitem[\protect\citeauthoryear{Ben-Michael, Greiner, Huang, Imai, Jiang, and
  Shin}{Ben-Michael et~al.}{2024}]{benmichael2024does}
Ben-Michael, E., D.~J. Greiner, M.~Huang, K.~Imai, Z.~Jiang, and S.~Shin
  (2024).
\newblock Does {AI} help humans make better decisions? {A} methodological
  framework for experimental evaluation.

\bibitem[\protect\citeauthoryear{Ben-Michael, Greiner, Imai, and
  Jiang}{Ben-Michael et~al.}{2021}]{ben2021safe}
Ben-Michael, E., D.~J. Greiner, K.~Imai, and Z.~Jiang (2021).
\newblock Safe policy learning through extrapolation: Application to pre-trial
  risk assessment.
\newblock Technical report, arXiv:2109.11679.

\bibitem[\protect\citeauthoryear{Bickel, Klaassen, Bickel, Ritov, Klaassen,
  Wellner, and Ritov}{Bickel et~al.}{1993}]{bickel1993efficient}
Bickel, P.~J., C.~A. Klaassen, P.~J. Bickel, Y.~Ritov, J.~Klaassen, J.~A.
  Wellner, and Y.~Ritov (1993).
\newblock {\em Efficient and adaptive estimation for semiparametric models},
  Volume~4.
\newblock Springer.

\bibitem[\protect\citeauthoryear{Bojinov and Shephard}{Bojinov and
  Shephard}{2019}]{bojinov2019time}
Bojinov, I. and N.~Shephard (2019).
\newblock Time series experiments and causal estimands: exact randomization
  tests and trading.
\newblock {\em Journal of the American Statistical Association\/}~{\em
  114\/}(528), 1665--1682.

\bibitem[\protect\citeauthoryear{Boruvka, Almirall, Witkiewitz, and
  Murphy}{Boruvka et~al.}{2018}]{boruvka2018assessing}
Boruvka, A., D.~Almirall, K.~Witkiewitz, and S.~A. Murphy (2018).
\newblock Assessing time-varying causal effect moderation in mobile health.
\newblock {\em Journal of the American Statistical Association\/}~{\em
  113\/}(523), 1112--1121.

\bibitem[\protect\citeauthoryear{Chernozhukov, Chetverikov, Demirer, Duflo,
  Hansen, Newey, and Robins}{Chernozhukov
  et~al.}{2018}]{chernozhukov2018double}
Chernozhukov, V., D.~Chetverikov, M.~Demirer, E.~Duflo, C.~Hansen, W.~Newey,
  and J.~Robins (2018).
\newblock Double/debiased machine learning for treatment and structural
  parameters.

\bibitem[\protect\citeauthoryear{Comment, Mealli, Haneuse, and Zigler}{Comment
  et~al.}{2019}]{comment2019survivor}
Comment, L., F.~Mealli, S.~Haneuse, and C.~Zigler (2019).
\newblock Survivor average causal effects for continuous time: a principal
  stratification approach to causal inference with semicompeting risks.
\newblock Technical report, arXiv:1902.09304.

\bibitem[\protect\citeauthoryear{D{\'\i}az, Williams, Hoffman, and
  Schenck}{D{\'\i}az et~al.}{2023}]{diaz2023nonparametric}
D{\'\i}az, I., N.~Williams, K.~L. Hoffman, and E.~J. Schenck (2023).
\newblock Nonparametric causal effects based on longitudinal modified treatment
  policies.
\newblock {\em Journal of the American Statistical Association\/}~{\em
  118\/}(542), 846--857.

\bibitem[\protect\citeauthoryear{Ding, Geng, Yan, and Zhou}{Ding
  et~al.}{2011}]{ding2011identifiability}
Ding, P., Z.~Geng, W.~Yan, and X.-H. Zhou (2011).
\newblock Identifiability and estimation of causal effects by principal
  stratification with outcomes truncated by death.
\newblock {\em Journal of the American Statistical Association\/}~{\em 106},
  1578--1591.

\bibitem[\protect\citeauthoryear{D’Amour, Ding, Feller, Lei, and
  Sekhon}{D’Amour et~al.}{2021}]{d2021overlap}
D’Amour, A., P.~Ding, A.~Feller, L.~Lei, and J.~Sekhon (2021).
\newblock Overlap in observational studies with high-dimensional covariates.
\newblock {\em Journal of Econometrics\/}~{\em 221\/}(2), 644--654.

\bibitem[\protect\citeauthoryear{Fine, Jiang, and Chappell}{Fine
  et~al.}{2001}]{fine2001semi}
Fine, J.~P., H.~Jiang, and R.~Chappell (2001).
\newblock On semi-competing risks data.
\newblock {\em Biometrika\/}~{\em 88\/}(4), 907--919.

\bibitem[\protect\citeauthoryear{Greiner, Halen, Stubenberg, and
  Griffin}{Greiner et~al.}{2020}]{greiner2020randomized}
Greiner, J., R.~Halen, M.~Stubenberg, and C.~L. Griffin (2020).
\newblock Randomized control trial evaluation of the implementation of the
  psa-dmf system in dane county, wi.

\bibitem[\protect\citeauthoryear{Grossi, Mariani, Mattei, and Mealli}{Grossi
  et~al.}{2025}]{grossi2025bayesian}
Grossi, G., M.~Mariani, A.~Mattei, and F.~Mealli (2025).
\newblock Bayesian principal stratification with longitudinal data and
  truncation by death.
\newblock {\em Econometrics and Statistics\/}.

\bibitem[\protect\citeauthoryear{Hahn}{Hahn}{1998}]{hahn1998role}
Hahn, J. (1998).
\newblock On the role of the propensity score in efficient semiparametric
  estimation of average treatment effects.
\newblock {\em Econometrica\/}~{\em 66\/}(2), 315--331.

\bibitem[\protect\citeauthoryear{Hern{\'a}n and Robins}{Hern{\'a}n and
  Robins}{2020}]{hernan2020causal}
Hern{\'a}n, M.~A. and J.~M. Robins (2020).
\newblock {\em Causal inference}.
\newblock Boca Raton: Chapman \& Hall/CRC.

\bibitem[\protect\citeauthoryear{Imai}{Imai}{2008}]{imai2008sharp}
Imai, K. (2008).
\newblock Sharp bounds on the causal effects in randomized experiments with
  “truncation-by-death”.
\newblock {\em Statistics \& probability letters\/}~{\em 78\/}(2), 144--149.

\bibitem[\protect\citeauthoryear{Imai, Jiang, Greiner, Halen, and Shin}{Imai
  et~al.}{2023}]{imai2020experimental}
Imai, K., Z.~Jiang, D.~J. Greiner, R.~Halen, and S.~Shin (2023).
\newblock Experimental evaluation of computer-assisted human decision-making:
  Application to pretrial risk assessment instrument (with discussion).
\newblock {\em Journal of the Royal Statistical Society, Series A (Statistics
  in Society)\/}~{\em 186\/}(2), 167--189.

\bibitem[\protect\citeauthoryear{Janvin, Young, Ryalen, and Stensrud}{Janvin
  et~al.}{2024}]{janvin2024causal}
Janvin, M., J.~G. Young, P.~C. Ryalen, and M.~J. Stensrud (2024).
\newblock Causal inference with recurrent and competing events.
\newblock {\em Lifetime data analysis\/}~{\em 30\/}(1), 59--118.

\bibitem[\protect\citeauthoryear{Jiang, Chen, and Ding}{Jiang
  et~al.}{2023}]{jiang2023instrumental}
Jiang, Z., S.~Chen, and P.~Ding (2023).
\newblock An instrumental variable method for point processes: generalised wald
  estimation based on deconvolution.
\newblock {\em Biometrika\/}~{\em 110\/}(4), 989--1008.

\bibitem[\protect\citeauthoryear{Jiang, Yang, and Ding}{Jiang
  et~al.}{2022}]{jiang2022multiply}
Jiang, Z., S.~Yang, and P.~Ding (2022).
\newblock Multiply robust estimation of causal effects under principal
  ignorability.
\newblock {\em Journal of the Royal Statistical Society Series B: Statistical
  Methodology\/}~{\em 84\/}(4), 1423--1445.

\bibitem[\protect\citeauthoryear{Josefsson and Daniels}{Josefsson and
  Daniels}{2021}]{josefsson2021bayesian}
Josefsson, M. and M.~J. Daniels (2021).
\newblock Bayesian semi-parametric {G}-computation for causal inference in a
  cohort study with mnar dropout and death.
\newblock {\em Journal of the Royal Statistical Society. Series C, Applied
  statistics\/}~{\em 70\/}(2), 398.

\bibitem[\protect\citeauthoryear{Josefsson, de~Luna, Daniels, and
  Nyberg}{Josefsson et~al.}{2016}]{josefsson2016causal}
Josefsson, M., X.~de~Luna, M.~J. Daniels, and L.~Nyberg (2016).
\newblock Causal inference with longitudinal outcomes and non-ignorable
  drop-out: Estimating the effect of living alone on cognitive decline.
\newblock {\em Journal of the Royal Statistical Society. Series C, Applied
  Statistics\/}~{\em 65\/}(1), 131.

\bibitem[\protect\citeauthoryear{Kang and Schafer}{Kang and
  Schafer}{2007}]{kang2007demystifying}
Kang, J.~D. and J.~L. Schafer (2007).
\newblock Demystifying double robustness: A comparison of alternative
  strategies for estimating a population mean from incomplete data.
\newblock {\em Statistical Science\/}~{\em 22\/}(4), 523--539.

\bibitem[\protect\citeauthoryear{Kennedy}{Kennedy}{2016}]{kennedy2016}
Kennedy, E.~H. (2016).
\newblock {\em Semiparametric Theory and Empirical Processes in Causal
  Inference}, pp.\  141--167.
\newblock Springer International Publishing.

\bibitem[\protect\citeauthoryear{Kennedy}{Kennedy}{2019}]{kennedy2019nonparametric}
Kennedy, E.~H. (2019).
\newblock Nonparametric causal effects based on incremental propensity score
  interventions.
\newblock {\em Journal of the American Statistical Association\/}~{\em
  114\/}(526), 645--656.

\bibitem[\protect\citeauthoryear{Kurland, Johnson, Egleston, and Diehr}{Kurland
  et~al.}{2009}]{kurland2009longitudinal}
Kurland, B.~F., L.~L. Johnson, B.~L. Egleston, and P.~H. Diehr (2009).
\newblock Longitudinal data with follow-up truncated by death: match the
  analysis method to research aims.
\newblock {\em Statistical science\/}~{\em 24\/}(2), 211.

\bibitem[\protect\citeauthoryear{Lee and Daniels}{Lee and
  Daniels}{2013}]{lee2013causal}
Lee, K. and M.~J. Daniels (2013).
\newblock Causal inference for bivariate longitudinal quality of life data in
  presence of death by using global odds ratios.
\newblock {\em Statistics in medicine\/}~{\em 32\/}(24), 4275--4284.

\bibitem[\protect\citeauthoryear{Lee, Daniels, and Sargent}{Lee
  et~al.}{2010}]{lee2010causal}
Lee, K., M.~J. Daniels, and D.~J. Sargent (2010).
\newblock Causal effects of treatments for informative missing data due to
  progression/death.
\newblock {\em Journal of the American Statistical Association\/}~{\em
  105\/}(491), 912--929.

\bibitem[\protect\citeauthoryear{Papadogeorgou, Imia, Lyall, and
  Li}{Papadogeorgou et~al.}{2022}]{papa:etal:22}
Papadogeorgou, G., K.~Imia, J.~Lyall, and F.~Li (2022).
\newblock Causal inference with spatio-temporal data: Estimating the effects of
  airstrikes on insurgent violence in {Iraq}.
\newblock {\em Journal of the Royal Statistical Society, Series B (Statistical
  Methodology)\/}~{\em 84\/}(5), 1969--1999.

\bibitem[\protect\citeauthoryear{Richardson and Robins}{Richardson and
  Robins}{2013}]{richardson2013single}
Richardson, T.~S. and J.~M. Robins (2013).
\newblock Single world intervention graphs (swigs): A unification of the
  counterfactual and graphical approaches to causality.
\newblock {\em Center for the Statistics and the Social Sciences, University of
  Washington Series. Working Paper\/}~{\em 128\/}(30), 2013.

\bibitem[\protect\citeauthoryear{Robins}{Robins}{1986}]{robins1986new}
Robins, J. (1986).
\newblock A new approach to causal inference in mortality studies with a
  sustained exposure period—application to control of the healthy worker
  survivor effect.
\newblock {\em Mathematical modelling\/}~{\em 7\/}(9-12), 1393--1512.

\bibitem[\protect\citeauthoryear{Robins}{Robins}{1992}]{robins1992estimation}
Robins, J. (1992).
\newblock Estimation of the time-dependent accelerated failure time model in
  the presence of confounding factors.
\newblock {\em Biometrika\/}~{\em 79\/}(2), 321--334.

\bibitem[\protect\citeauthoryear{Robins and Hernan}{Robins and
  Hernan}{2008}]{robins2008estimation}
Robins, J. and M.~Hernan (2008).
\newblock Estimation of the causal effects of time-varying exposures.
\newblock In {\em Longitudinal Data Analysis}, pp.\  553--599. Chapman and
  Hall/CRC.

\bibitem[\protect\citeauthoryear{Rubin}{Rubin}{1980}]{rubin1980randomization}
Rubin, D.~B. (1980).
\newblock Comment on ``{R}andomization analysis of experimental data: The
  {F}isher randomization test''.
\newblock {\em J. Am. Statist. Ass.\/}~{\em 75}, 591--593.

\bibitem[\protect\citeauthoryear{Rubin}{Rubin}{2006}]{rubin2006causal}
Rubin, D.~B. (2006).
\newblock Causal inference through potential outcomes and principal
  stratification: application to studies with ``censoring'' due to death.
\newblock {\em Statistical Science\/}~{\em 21}, 299--309.

\bibitem[\protect\citeauthoryear{Rytgaard and van~der Laan}{Rytgaard and
  van~der Laan}{2024}]{rytgaard2024one}
Rytgaard, H.~C. and M.~J. van~der Laan (2024).
\newblock One-step targeted maximum likelihood estimation for targeting
  cause-specific absolute risks and survival curves.
\newblock {\em Biometrika\/}~{\em 111\/}(1), 129--145.

\bibitem[\protect\citeauthoryear{Schindl, Shen, and Kennedy}{Schindl
  et~al.}{2024}]{schindl2024incremental}
Schindl, K., S.~Shen, and E.~H. Kennedy (2024).
\newblock Incremental effects for continuous exposures.
\newblock Technical report, arXiv preprint arXiv:2409.11967.

\bibitem[\protect\citeauthoryear{Shardell and Ferrucci}{Shardell and
  Ferrucci}{2018}]{shardell2018joint}
Shardell, M. and L.~Ferrucci (2018).
\newblock Joint mixed-effects models for causal inference with longitudinal
  data.
\newblock {\em Statistics in medicine\/}~{\em 37\/}(5), 829--846.

\bibitem[\protect\citeauthoryear{Shardell, Hicks, and Ferrucci}{Shardell
  et~al.}{2015}]{shardell2015doubly}
Shardell, M., G.~E. Hicks, and L.~Ferrucci (2015).
\newblock Doubly robust estimation and causal inference in longitudinal studies
  with dropout and truncation by death.
\newblock {\em Biostatistics\/}~{\em 16\/}(1), 155--168.

\bibitem[\protect\citeauthoryear{Tsiatis}{Tsiatis}{2006}]{tsiatis2006semiparametric}
Tsiatis, A.~A. (2006).
\newblock {\em Semiparametric theory and missing data}.
\newblock Springer.

\bibitem[\protect\citeauthoryear{van~der Laan and Gruber}{van~der Laan and
  Gruber}{2012}]{van2012targeted}
van~der Laan, M.~J. and S.~Gruber (2012).
\newblock Targeted minimum loss based estimation of causal effects of multiple
  time point interventions.
\newblock {\em The International Journal of Biostatistics\/}~{\em 8\/}(1).

\bibitem[\protect\citeauthoryear{van~der Laan, Polley, and Hubbard}{van~der
  Laan et~al.}{2007}]{van2007super}
van~der Laan, M.~J., E.~C. Polley, and A.~E. Hubbard (2007).
\newblock Super learner.
\newblock {\em Statistical Applications in Genetics and Molecular
  Biology\/}~{\em 6\/}(1), 1--21.

\bibitem[\protect\citeauthoryear{Wang, Richardson, and Zhou}{Wang
  et~al.}{2017}]{wang2017causal}
Wang, L., T.~S. Richardson, and X.-H. Zhou (2017).
\newblock Causal analysis of ordinal treatments and binary outcomes under
  truncation by death.
\newblock {\em Journal of the Royal Statistical Society: Series B (Statistical
  Methodology)\/}~{\em 79}, 719--735.

\bibitem[\protect\citeauthoryear{Wu, Weinberger, Wellenius, Dominici, and
  Braun}{Wu et~al.}{2024}]{wu2024assessing}
Wu, X., K.~R. Weinberger, G.~A. Wellenius, F.~Dominici, and D.~Braun (2024).
\newblock Assessing the causal effects of a stochastic intervention in time
  series data: are heat alerts effective in preventing deaths and
  hospitalizations?
\newblock {\em Biostatistics\/}~{\em 25\/}(1), 57--79.

\bibitem[\protect\citeauthoryear{Xu, Scharfstein, M{\"u}ller, and Daniels}{Xu
  et~al.}{2022}]{xu2022bayesian}
Xu, Y., D.~Scharfstein, P.~M{\"u}ller, and M.~Daniels (2022).
\newblock A bayesian nonparametric approach for evaluating the causal effect of
  treatment in randomized trials with semi-competing risks.
\newblock {\em Biostatistics\/}~{\em 23\/}(1), 34--49.

\bibitem[\protect\citeauthoryear{Yang and Ding}{Yang and
  Ding}{2018}]{yang2018using}
Yang, F. and P.~Ding (2018).
\newblock Using survival information in truncation by death problems without
  the monotonicity assumption.
\newblock {\em Biometrics\/}~{\em 74\/}(4), 1232--1239.

\bibitem[\protect\citeauthoryear{Zhang and Rubin}{Zhang and
  Rubin}{2003}]{zhang2003estimation}
Zhang, J.~L. and D.~B. Rubin (2003).
\newblock Estimation of causal effects via principal stratification when some
  outcomes are truncated by “death”.
\newblock {\em Journal of Educational and Behavioral Statistics\/}~{\em
  28\/}(4), 353--368.

\end{thebibliography}

\newpage

\appendix

\setcounter{equation}{0}
\setcounter{figure}{0}
\setcounter{theorem}{0}
\setcounter{lemma}{0}
\setcounter{section}{0}
\renewcommand {\theequation} {S\arabic{equation}}
\renewcommand {\thefigure} {S\arabic{figure}}
\renewcommand {\thetheorem} {S\arabic{theorem}}
\renewcommand {\thelemma} {S\arabic{lemma}}
\renewcommand {\thesection} {S\arabic{section}}

\begin{center}
  \LARGE {\bf Supplementary Appendix}
\end{center}
Section~\ref{app::proof} provides the proofs of the results and Section~\ref{app::sim} presents additional simulation results.

\section{Proofs}
\label{app::proof}

\subsection{Proof of Theorem~\ref{thm::identification-obs}}
We first prove Theorem~\ref{thm::identification-obs}(a).
 We can write
 $$
 \E\{Y_{it}(\hbz_{t})\mid S_{it}(\hbz_{t-1})=1\} \ = \ \frac{\E\{Y_{it}(\hbz_{t})S_{it}(\hbz_{t-1})\}}{\Pr\{S_{it}(\hbz_{t-1})=1\}}.
 $$
We consider the numerator and the denominator separately. For the denominator, 
because $S_{it}=1$ implies $S_{it'}=1$ for any $t\geq t'$, we have,
\begin{eqnarray*}
  &&\Pr\{S_{it}(\hbz_{t-1})=1\}\\
  &=&\Pr\{S_{it}(\hbz_{t-1})=1,\ldots,S_{i2}(z_1)=1\}\\
  &=&\int \Pr\{S_{it}(\hbz_{t-1})=1,\ldots,S_{i2}(z_1)=1\mid \bX_{i1}=\bx_1\}l_1(\bx_1)  \text{d} \bx_1\\
  &=&\int \Pr\{S_{it}(\hbz_{t-1})=1,\ldots,S_{i2}(z_1)=1\mid
      Z_{i1}=z_1, \bX_{i1}= \bx_1\}l_1(\bx_1)   \text{d} \bx_1\\
  &=&\int \Pr\{S_{it}(\hbz_{t-1})=1,\ldots,S_{i3}(\hbz_2)=1\mid
      S_{i2}=1,Z_{i1}=z_1, \bX_{i1}= \bx_1\} p_2(z_1, \bx_1)l_1(\bx_1)  \text{d} \bx_1,
\end{eqnarray*}
where the third equality follows from
Assumption~\ref{asm::ignorability}. We can further write:
\begin{eqnarray*}
  &&\Pr\{S_{it}(\hbz_{t-1})=1,\ldots,S_{i3}(\hbz_2)=1\mid S_{i2}=1,Z_{i1}=z_1,\bX_{i1}=\bx_1\}\\
  &=&\int \Pr\{S_{it}(\hbz_{t-1})=1,\ldots,S_{i3}(\hbz_2)=1\mid S_{i2}=1,Z_{i1}=z_1,\hbX_{i2}=\hbx_2\}  l_2(z_1,\hbx_2) \text{d} \bx_2\\
  &=&\int \Pr\{S_{it}(\hbz_{t-1})=1,\ldots,S_{i3}=1\mid S_{i2}=1,\hbZ_{i2}=\hbz_2,\hbX_{i2}=\hbx_2\} l_2(z_1,\hbx_2) \text{d} \bx_2\\
  &=&\int \Pr\{S_{it}(\hbz_{t-1})=1,\ldots,S_{i4}(\hbz_3)=1\mid S_{i3}=1,\hbZ_{i2}=\hbz_2,\hbX_{i2}=\hbx_2\}p_3(\hbz_2,\hbx_2)  l_2(z_1,\hbx_2) \text{d} \bx_2,
\end{eqnarray*}
where the third equality again follows from
Assumption~\ref{asm::ignorability}.  Applying this step iteratively,
we obtain:
\begin{eqnarray}
\label{eqn::proof-tau-reg1}\Pr\{S_{it}(\hbz_{t-1})=1\}&=& \int \prod_{t'=2}^t p_{t'}(\hbz_{t'-1},\hbx_{t'-1})  \prod_{t'=1}^{t-1}  l_{t'}(\hbz_{t'-1},\hbx_{t'}) \text{d} \hbx_{t-1}.
\end{eqnarray}
For the numerator, we use a similar procedure to obtain:
\begin{eqnarray}
\nonumber & & \E\{Y_{it}(\hbz_{t})S_{it}(\hbz_{t-1})\} \\
\nonumber&=&\int   \E\{Y_{it}(\hbz_{t}) \mid S_{it}=1, \hbZ_{i,t-1}=\hbz_{t-1}, \hbX_{i,t-1}=\hbx_{t-1} \}\prod_{t'=2}^t p_{t'}(\hbz_{t'-1},\hbx_{t'-1})  \prod_{t'=1}^{t-1}  l_{t'}(\hbz_{t'-1},\hbx_{t'}) \text{d} \hbx_{t-1}\\
\nonumber&=&\int   \E\{Y_{it}(\hbz_{t}) \mid S_{it}=1, \hbZ_{i,t-1}=\hbz_{t-1}, \hbX_{it}=\hbx_{t} \}\prod_{t'=2}^t p_{t'}(\hbz_{t'-1},\hbx_{t'-1})  \prod_{t'=1}^{t}  l_{t'}(\hbz_{t'-1},\hbx_{t'}) \text{d} \hbx_{t}\\
\label{eqn::proof-tau-reg2}&=&\int   \mu_t(\hbz_t,\hbx_t)\prod_{t'=2}^t p_{t'}(\hbz_{t'-1},\hbx_{t'-1})  \prod_{t'=1}^{t}  l_{t'}(\hbz_{t'-1},\hbx_{t'}) \text{d} \hbx_{t},
\end{eqnarray}
where the third equality follows from
Assumption~\ref{asm::ignorability}.  Combining
Equations~\eqref{eqn::proof-tau-reg1}~and~\eqref{eqn::proof-tau-reg2}
leads to the desired result.

We next prove Theorem~\ref{thm::identification-obs}(b).  Using the law
of total expectation, we have:
\begin{eqnarray*}
&&\E\left[\frac{\bm{1}\{\hbZ_{i,t-1}=\hbz_{t-1},S_{it}=1\}}{\prod_{t'=1}^{t-1}\Pr(Z_{it'}=z_{t'}\mid S_{it'}=1,\hbZ_{i,t'-1}=\hbz_{t'-1},\hbX_{it'}=\hbx_{t'})} \right]\\
&=&\int
    \E\left[\frac{\bm{1}\{\hbZ_{i,t-1}=\hbz_{t-1},S_{it}=1\}}{\prod_{t'=1}^{t-1}\Pr(Z_{it'}=z_{t'}\mid
    S_{it'}=1,\hbZ_{i,t'-1}=\hbz_{t'-1},\hbX_{it'}=\hbx_{t'})} \ \Bigr | \  \bX_{i1}=\bx_1 \right] l_1(\bx_1) \text{d} \bx_1\\
&=&\int   \E\left[\frac{\bm{1}\{\hbZ_{i,t-1}=\hbz_{t-1},S_{it}=1\}}{\prod_{t'=2}^{t-1}\Pr(Z_{it'}=z_{t'}\mid S_{it'}=1,\hbZ_{i,t'-1}=\hbz_{t'-1},\hbX_{it'}=\hbx_{t'})} \ \Bigr | \  S_{i2}=1,Z_{i1}=z_1,\bX_{i1}=\bx_1 \right] \\
&&\hspace{0.5cm}\times p_2(z_1,\bx_1) l_1(\bx_1) \text{d} \bx_1\\
&=&\int   \E\left[\frac{\bm{1}\{\hbZ_{i,t-1}=\hbz_{t-1},S_{it}=1\}}{\prod_{t'=2}^{t-1}\Pr(Z_{it'}=z_{t'}\mid S_{it'}=1,\hbZ_{i,t'-1}=\hbz_{t'-1},\hbX_{it'}=\hbx_{t'})} \ \Bigr | \ S_{i2}=1,Z_{i1}=z_1,\hbX_{i2}=\hbx_2 \right] \\
&&\hspace{0.5cm}\times p_2(z_1,\bx_1) l_2(z_1,\hbx_2)l_1(\bx_1) \text{d} \hbx_2\\
&=&\int   \E\left[\frac{\bm{1}\{\hbZ_{i,t-1}=\hbz_{t-1},S_{it}=1\}}{\prod_{t'=3}^{t-1}\Pr(Z_{it'}=z_{t'}\mid S_{it'}=1,\hbZ_{i,t'-1}=\hbz_{t'-1},\hbX_{it'}=\hbx_{t'})} \ \Bigr | \ S_{i3}=1,\hbZ_{i2}=\hbz_2,\hbX_{i2}=\hbx_2 \right] \\
&&\hspace{0.5cm}\times  p_3(\hbz_2,\hbx_2)p_2(z_1,\bx_1) l_2(z_1,\hbx_2)l_1(\bx_1) \text{d} \hbx_2.
\end{eqnarray*}
Applying the procedure iteratively, we obtain:
\begin{eqnarray*}
&&\E\left[\frac{\bm{1}\{\bZ_{i,t-1}=\hbz_{t-1},S_{it}=1\}}{\prod_{t'=1}^{t-1}\Pr(Z_{it'}=z_{t'}\mid S_{it'}=1,\hbZ_{i,t'-1}=\hbz_{t'-1},\hbX_{it'}=\hbx_{t'})} \right]\\
&=&\int \prod_{t'=2}^t p_{t'}(\hbz_{t'-1},\hbx_{t'-1})  \prod_{t'=1}^{t-1}  l_{t'}(\hbz_{t'-1},\hbx_{t'}) \text{d} \hbx_{t-1},
\end{eqnarray*}
which, together with Equation~\eqref{eqn::proof-tau-reg1}, yields the
inverse probability weighting formula for
$\Pr\{S_{it}(\hbz_{t-1})=1\}$.  Similarly, we can derive:
\begin{eqnarray*}
&&\E\left[\frac{Y_{it}\bm{1}\{\hbz_{it}=\hbz_t,S_{it}=1\}}{\prod_{t'=1}^t\Pr(Z_{it'}=z_{t'}\mid S_{it'}=1,\hbZ_{i,t'-1}=\hbz_{t'-1},\hbX_{it'}=\hbx_{t'})} \right]\\
  &=&\int \E\left[\frac{Y_{it}\bm{1}\{\hbz_{it}=\hbz_t,S_{it}=1\}}{\Pr(Z_{it}=z_{t}\mid S_{it}=1,\hbz_{i,t-1}=\hbz_{t-1},\hbX_{i,t'-1}=\hbx_{t'-1})} \ \Bigr | \ S_{it}=1,\hbZ_{i,t-1}=\hbz_{t-1},\hbX_{i,t-1}=\hbx_{t-1} \right]\\
   &&\hspace{0.5cm}\times  \prod_{t'=2}^t p_{t'}(\hbz_{t'-1},\hbx_{t'-1})  \prod_{t'=1}^{t-1}  l_{t'}(\hbz_{t'-1},\hbx_{t'}) \text{d} \hbx_{t-1}\\
   &=&\int   \mu_t(\hbz_t,\hbx_t)\prod_{t'=2}^t p_{t'}(\hbz_{t'-1},\hbx_{t'-1})  \prod_{t'=1}^{t}  l_{t'}(\hbz_{t'-1},\hbx_{t'}) \text{d} \hbx_{t},
\end{eqnarray*}
which, combined with Equation~\eqref{eqn::proof-tau-reg2}, yields the
inverse probability weighting formula for
$\E\{Y_{it}(\hbz_{t})S_{it}(\hbz_{t-1})\}$.  \QEDB

\subsection{Proof of Proposition~\ref{thm::identification-m}}

For  $m_{Y_tS_t}(\hbz_{t'}, \hbX_{it'})$ with $t'<t$, we have: 
\begin{eqnarray*}
&&m_{Y_tS_t}(\hbz_{t'}, \hbX_{it'})\\
&=& \E\{Y_{it}(\hbz_t)S_{it}(\hbz_{t-1})\mid S_{i,t'+1}=1,\hbZ_{it'}=\hbz_{t'}, \hbX_{it'}\}\Pr(S_{i,t'+1}=1\mid S_{it'}=1,\hbZ_{it'}=\hbz_{t'}, \hbX_{it'})\\
&=& \E\left[ \E\{Y_{it}(\hbz_t)S_{it}(\hbz_{t-1})\mid S_{i,t'+1}=1,\hbZ_{it'}=\hbz_{t'}, \hbX_{i,t'+1}\} \mid S_{i,t'+1}=1,\hbZ_{it'}=\hbz_{t'}, \hbX_{it'}\right]\\
&&\times \Pr(S_{i,t'+1}=1\mid S_{it'}=1,\hbZ_{it'}=\hbz_{t'}, \hbX_{it'})  \\
&=& \E\left\{m_{Y_tS_t}(\hbz_{t'+1}, \hbX_{i,t'+1}) \mid S_{i,t'+1}=1,\hbZ_{it'}=\hbz_{t'}, \hbX_{it'} \right\}p_{t'+1}(\hbz_{t'},\hbX_{it'})\\
&=&\E\left\{S_{i,t'+1}m_{Y_tS_t}(\hbz_{t'+1}, \hbX_{i,t'+1}) \mid S_{it'}=1,\hbZ_{it'}=\hbz_{t'}, \hbX_{it'}\right\},
\end{eqnarray*}
where the third equality follows from
Assumption~\ref{asm::ignorability}. 

For  $m_{S_t}(\hbz_{t'}, \hbX_{it'})$ with $t'<t-1$, we have:
\begin{eqnarray*}
&&m_{S_t}(\hbz_{t'}, \hbX_{it'})\\
&=& \E\{S_{it}(\hbz_{t-1})\mid S_{i,t'+1}=1,\hbZ_{it'}=\hbz_{t'}, \hbX_{it'}\}\Pr(S_{i,t'+1}=1\mid S_{it'}=1,\hbZ_{it'}=\hbz_{t'}, \hbX_{it'})\\
&=& \E\left[ \E\{S_{it}(\hbz_{t-1})\mid S_{i,t'+1}=1,\hbZ_{it'}=\hbz_{t'}, \hbX_{i,t'+1}\} \mid S_{i,t'+1}=1,\hbZ_{it'}=\hbz_{t'}, \hbX_{it'}\right]\\
&&\times \Pr(S_{i,t'+1}=1\mid S_{it'}=1,\hbZ_{it'}=\hbz_{t'}, \hbX_{it'})  \\
&=& \E\left\{m_{S_t}(\hbz_{t'+1}, \hbX_{i,t'+1}) \mid S_{i,t'+1}=1,\hbZ_{it'}=\hbz_{t'}, \hbX_{it'} \right\}p_{t'+1}(\hbz_{t'},\hbX_{it'})\\
&=&\E\left\{S_{i,t'+1}m_{S_t}(\hbz_{t'+1}, \hbX_{i,t'+1}) \mid S_{it'}=1,\hbZ_{it'}=\hbz_{t'}, \hbX_{it'}\right\},
\end{eqnarray*}
where the third equality again follows from Assumption~\ref{asm::ignorability}. \QEDB

\subsection{Proof of Theorem~\ref{thm::eif}}

We derive the EIFs for the case with $T=3$. The derivation with
general $T$ is similar and thus omitted.  We first introduce the
following lemma for the EIF of a ratio-type parameter.  The proof of
this lemma appears in \citet{jiang2022multiply}.
\begin{lemma}\label{lem::ratio} Consider a ratio-type parameter $R=N/D$, where $N$ and $D$ are defined as functions of the distribution of variables $V$.
If $\dot{N}_{\theta}|_{\theta=0}=\E\{\varphi_{N}(V)\textup{s}(V)\}$ and $\dot{D}_{\theta}|_{\theta=0}=\E\{\varphi_{D}(V)\textup{s}(V)\}$ where $\theta$ is the parameter of a submodel and $s(\cdot)$ is the score function, 
then $\dot{R}_{\theta}|_{\theta=0}=\E\{\varphi_{R}(V)\textup{s}(V)\}$ where
\begin{equation}
\varphi_{R}(V)=\frac{1}{D}\varphi_{N}(V)-\frac{R}{D}\varphi_{D}(V).\label{eq:ratio-ses}
\end{equation}
In particular, if $\varphi_{N}(V)$ and $\varphi_{D}(V)$ are the
EIFs for $N$ and $D$, then $\varphi_{R}(V)$ is the EIF for $R.$
\end{lemma}

This lemma implies that because
\begin{eqnarray*}
\tau_3(\hbz_2)&= & \frac{\E\{Y_{i3}(\hbz_2,1)S_{i3}(\hbz_2)\}-\E\{Y_{i3}(\hbz_2,0)S_{i3}(\hbz_2)\}}{\E\{S_{i3}(\hbz_2)\}},\\
\theta(\xi)&=&\sum_{t=1}^3  \sum_{\hbz_t} \E\{ Y_{it}(\hbz_{t}) S_{it}(\hbz_{t-1})\} \Pr\{S_{it}(\hbz_{t-1})=1\}\xi_t(\hbz_t).
\end{eqnarray*}
it suffices to derive the EIFs for $\E\{Y_{i3}(\hbz_3)S_{i3}(\hbz_2)\}$ and $\E\{S_{i3}(\hbz_2)\}$.
We will use the semiparametric theory of \citet{bickel1993efficient} to derive the EIFs.  Denote 
 $$\bV_i=(\bX_{i1},Z_{i1},S_{i2}, S_{i2}\bX_{i2},S_{i2}Z_{i2},S_{i3},S_{i3}\bX_{i3},S_{i3}Z_{i3},S_{i3}Y_{i3}),$$ which consists of all the observed variables up to time $t=3$.
 The likelihood of $\bV_i$ factorizes as 
\begin{eqnarray}
\nonumber
\nonumber f(\bV_i) 
 &=& f(\bX_{i1})f (Z_{i1} \mid \bX_{i1}) f(S_{i2}\mid Z_{i1}, \bX_{i1})\\
  \nonumber     && \times \{f(\bX_{i2}\mid S_{i2}=1,Z_{i1}, \bX_{i1})f(Z_{i2}\mid S_{i2}=1,Z_{i1},\hbX_{i2}) f(S_{i3}\mid S_{i2}=1,\hbZ_{i2},\hbX_{i2})\}^{S_{i2}}\\
  \label{eqn::eif-fv}      && \times \{f(\bX_{i3}\mid S_{i3}=1,\hbZ_{i2},\hbX_{i2})f(Z_{i3}\mid S_{i2}=1,\hbZ_{i2},\hbX_{i3}) f(Y_{i3}\mid S_{i3}=1,\hbZ_{i3},\hbX_{i3})\}^{S_{i3}}.
\end{eqnarray}
To derive the EIF, we consider a one-dimensional parametric submodel $f_\theta(\bV_i)$, which contains the true model at $\theta=0$, i.e., $f_\theta(\bV_i)\vert_{\theta=0}=f(\bV_i)$. We use  $\theta$ in the subscript to denote the quantities with respect to the submodel, e.g., $\E_\theta\{Y_{i3}(\hbz_3)S_{i3}(\hbz_2)\}$ is the value of $\E\{Y_{i3}(\hbz_3)S_{i3}(\hbz_2)\}$ in the submodel, and $m_{\theta,Y_3S_3}(z_1,\bx_1)$ is the value of $m_{Y_3S_3}(z_1,\bx_1)$ in the submodel. We use dot to denote the partial derivative with respect to $\theta$, e.g., $\dot \E_\theta\{Y_{i3}(\hbz_3)S_{i3}(\hbz_2)\}= \partial \E_\theta\{Y_{i3}(\hbz_3)S_{i3}(\hbz_2)\}/\partial \theta$, and use $\s_\theta(\cdot)$ to denote the score function of the submodel.
From Equation~\eqref{eqn::eif-fv}, the score function under the submodel decomposes as
\begin{eqnarray*}
\s_\theta(\bV_i)&=& \sum_{t=1}^3 S_{it} \cdot \s_\theta(\bX_{it}\mid S_{it}=1,\hbZ_{i,t-1},\hbX_{i,t-1})+ \sum_{t=1}^3 S_{it}\cdot \s_\theta(Z_{it}\mid S_{it}=1,\hbZ_{i,t-1},\hbX_{it})\\
    &&+\sum_{t=2}^3 S_{i,t-1}\cdot \s_\theta(S_{it}\mid S_{i,t-1}=1,\hbZ_{i,t-1},\hbX_{i,t-1})
    +S_{i3}\cdot\s_\theta(Y_{i3}\mid S_{i3}=1,\hbZ_{i3},\hbX_{i3}),
\end{eqnarray*}
where $\s(\cdot)$ represents the score function with $\s_\theta(\cdot)\vert_{\theta=0} = \s(\cdot)$.

From the semiparametric theory, the tangent space is 
\begin{eqnarray*}
\Lambda &=& H^X_1 \oplus H^X_2\oplus H^X_3 \oplus H^Z_1 \oplus H^Z_2\oplus H^Z_3\oplus H^S_2 \oplus H^S_3 \oplus H^Y_3,
\end{eqnarray*}
which is the direct sum of orthogonal subspaces
\begin{eqnarray*}
H^{\bX}_t &=& \{ S_{it}h(\hbZ_{i,t-1},\hbX_{it}): \E \{h(\hbZ_{i,t-1},\hbX_{it}) \mid  S_{it}=1, \hbZ_{i,t-1},\hbX_{i,t-1} \}=0 \},\\
H^Z_t &=& \{ S_{it}h(\hbZ_{it},\hbX_{it}): \E \{h(\hbZ_{it},\hbX_{it}) \mid  S_{it}=1, \hbZ_{i,t-1},\hbX_{it} \}=0 \},\\
H_2^S &=& \{ h(S_{i2},Z_{i1},\bX_{i1}) :  \E\{h(S_{i2},Z_{i1},\bX_{i1})\mid  Z_{i1},\bX_{i1} \}=0 \},\\
H_3^S &=& \{ S_{i2} h(S_{i3},\hbZ_{i2},\hbX_{i2}) :  \E\{h(S_{i3},\hbZ_{i2},\hbX_{i2})\mid  S_{i2}=1,\hbZ_{i2},\hbX_{i2} \}=0 \},\\
H^Y_3&=& \{ S_{i3} h(Y_{i3},\hbZ_{i3},\hbX_{i3}): \E\{h(Y_{i3},\hbZ_{i3},\hbX_{i3})\mid S_{i3}=1,\hbZ_{i3},\hbX_{i3}\} \}.
\end{eqnarray*}

The EIF for $m_{Y_3S_3}=  \E\{Y_{i3}(\hbz_3)S_{i3}(\hbz_2)\}$, denoted by $\phi_1(\bV_i)$, must satisfy
\begin{eqnarray*}
\dot m_{\theta,Y_3S_3}\vert_{\theta=0}\ = \ \E\{ \phi_1(\bV_i)  \s(\bV_i)\}
\end{eqnarray*}
and $\phi_1(\bV_i) \in \Lambda$.
From the chain rule, we have 
\begin{eqnarray*}
\dot m_{\theta,Y_3S_3}\vert_{\theta=0} &=&B_1+ B_2,
\end{eqnarray*}
where
\begin{eqnarray*}
B_1&=&\int m_{Y_3S_3}(z_1,\bx_1) \dot l_{1\theta}(\bx_1)\vert_{\theta=0} \td \bx_1
\ =\ \int m_{Y_3S_3}(z_1,\bx_1)  \s(x_1) l_1(\bx_1) \td \bx_1\\
&=&\E\{m_{Y_3S_3}(z_1,\bX_{i1})\s(\bX_{i1})\},\\
B_2&=&  \int  \dot m_{\theta,Y_3S_3}(z_1,\bx_1)\vert_{\theta=0} l_1(\bx_1)\td \bx_1.
\end{eqnarray*}
For $B_2$, we can further decompose it as $B_2= C_1+ C_2$, where
\begin{eqnarray*}
C_1 &=&  \int  \E\{Y_{i3}(\hbz_3)S_{i3}(\hbz_2)\mid S_{i2}=1,Z_{i1}=z_1,\bX_{i1}=\bx_1\} \dot p_{2\theta}(z_1,\bx_1)\vert_{\theta=0} l_1(\bx_1)\td \bx_1,\\
C_2 &=&  \int  \dot\E_\theta\{Y_{i3}(\hbz_3)S_{i3}(\hbz_2)\mid S_{i2}=1,Z_{i1}=z_1,\bX_{i1}=\bx_1\}\vert_{\theta=0} p_{2}(z_1,\bx_1) l_1(\bx_1)\td \bx_1.
\end{eqnarray*}
For $C_1$, we have
\begin{eqnarray*}
C_1&=& \int  \E\{Y_{i3}(\hbz_3)S_{i3}(\hbz_2)\mid S_{i2}=1,Z_{i1}=z_1,\bX_{i1}=\bx_1\} \s(S_{i2}=1\mid z_1,\bx_1) p_2(z_1,\bx_1) l_1(\bx_1)\td \bx_1\\
&=& \int  \E\left[\frac{\bm{1}\{Z_{i1}=z_1\}S_{i2}
    \E\{Y_{i3}(\hbz_3)S_{i3}(\hbz_2)\mid S_{i2}=1,Z_{i1},\bX_{i1}=\bx_1\}
    }{\pi_1(z_1, \bX_{i1})}\cdot \s(S_{i2}\mid Z_{i1},\bX_{i1}) \ \bigr | \ \bX_{i1}=\bx_1\right]\\
&&\hspace{0.5cm}\times l_1(\bx_1)\td \bx_1\\
&=& \E\left[\frac{\bm{1}\{Z_{i1}=z_1\}S_{i2} \E\{Y_{i3}(\hbz_3)S_{i3}(\hbz_2)\mid S_{i2}=1,Z_{i1},\bX_{i1}\} }{\pi_1(z_1, \bX_{i1})}\cdot \s(S_{i2}\mid Z_{i1},\bX_{i1})\right],
\end{eqnarray*}
where we simplify $\s(S_{i2}=1\mid Z_{i1}=z_1,\bX_{i1}=\bx_1)$ as $\s(S_{i2}=1\mid z_1,\bx_1)$.
For $C_2$, we can decompose $C_2 = D_1+D_2$, where
\begin{eqnarray*}
D_1&=& \int m_{Y_3S_3}(\hbz_2,\hbx_2)   \dot  l_{2\theta}(z_1,\hbx_2)\vert_{\theta=0} p_{2}(z_1,\bx_1) l_1(\bx_1)\td \hbx_2,\\
D_2&=&\int  \dot m_{\theta,Y_3S_3}(\hbz_2,\hbx_2)\vert_{\theta=0} l_{2}(z_1,\hbx_2)p_{2}(z_1,\bx_1) l_1(\bx_1)\td \hbx_2.
\end{eqnarray*}
For $D_1$, we have
\begin{eqnarray*}
D_1&=&  \int m_{Y_3S_3}(\hbz_2,\hbx_2)   \s(\bX_{i2}  = \bx_2\mid S_{i2}=1,z_1,\bx_1)l_2(z_1,\hbx_2) p_{2}(z_1,\bx_1) l_1(\bx_1)\td \hbx_2, \\
&=& \int  \E \left\{ m_{Y_3S_3}(\hbz_2,\bx_1,\bX_{i2})   \s(\bX_{i2}\mid S_{i2}=1,z_1,\bx_1)\mid S_{i2}=1,z_1,\bx_1 \right\}p_{2}(z_1,\bx_1) l_1(\bx_1) \td \bx_1 \\
&=&\int     \E\left[
    \frac{\bm{1}\{Z_{i1}=z_1\}S_{i2}m_{Y_3S_3}(\hbz_2,\bx_1,\bX_{i2})
    }{\pi_1(z_1, \bX_{i1})}  \cdot \s(\bX_{i2}\mid S_{i2}=1,Z_{i1},\bx_1) \ \bigr
    | \ \bX_{i1} = \bx_1 \right]   l_1(\bx_1)\td \bx_1\\
&=&   \E\left[  \frac{\bm{1}\{Z_{i1}=z_1\}S_{i2}m_{Y_3S_3}(\bz_2,\bX_{i2}) }{\pi_1(z_1, \bX_{i1})} \cdot  \s(\bX_{i2}\mid S_{i2}=1,Z_{i1},\bX_{i1})  \right],
\end{eqnarray*}
For $D_2$, we have $D_2 =E_1+E_2$, where
\begin{eqnarray*}
E_1&=& \int \E\{Y_{i3}(\hbz_3)\mid S_{i3}=1,\hbZ_{i2}=\hbz_2,\hbX_{i2}=\hbx_2\}\dot p_{3\theta}(\hbz_2,\hbx_2)\vert_{\theta=0} l_{2}(z_1,\hbx_2)p_{2}(z_1,\bx_1) l_1(\bx_1)\td \hbx_2,\\
E_2&=&  \int \dot \E_\theta\{Y_{i3}(\hbz_3)\mid S_{i3}=1,\hbZ_{i2}=\hbz_2,\hbX_{i2}=\hbx_2\}\vert_{\theta=0}  p_{3}(\hbz_2,\hbx_2)l_{2}(z_1,\hbx_2)p_{2}(z_1,\bx_1) l_1(\bx_1)\td \hbx_2.
\end{eqnarray*}
For $E_1$, we have 
\begin{eqnarray*}
E_1&=&\int \E\{Y_{i3}(\hbz_3)\mid S_{i3}=1,\hbZ_{i1}=\hbz_2,\hbX_{i2}=\hbx_2\} \s(S_{i3}=1\mid S_{i2}=1,\hbz_2,\hbx_2 )\\
&&\hspace{1cm} \times p_{3}(\hbz_2,\hbx_2) l_{2}(z_1,\hbx_2)p_{2}(z_1,\bx_1) l_1(\bx_1)\td \hbx_2\\
&=& \E\left[\frac{\bm{1}\{\hbZ_{i2}=\hbz_2\} S_{i3}  \E\{Y_{i3}(\hbz_3)S_{i3}(\hbz_2)\mid S_{i3}=1,\hbZ_{i2},\hbX_{i2}\} }{\pi_2(\hbz_2,\hbX_{i2})}\cdot \s(S_{i3}=1\mid S_{i2}=1,\hbZ_{i2},\hbX_{i2})\right].
\end{eqnarray*}
For $E_2$, we have $E_2=F_1+F_2$, where we use the double integral to
denote multiple integrals,
\begin{eqnarray*}
F_1&=& \int  m_{Y_3S_3}(\hbz_3,\hbx_3)  \dot l_{3\theta}(\hbz_2,\hbx_3)\vert_{\theta=0}  p_{3}(\hbz_2,\hbx_2)l_{2}(z_1,\hbx_2)p_{2}(z_1,\bx_1) l_1(\bx_1)\td \hbx_3\\
&=& \int m_{Y_3S_3}(\hbz_3,\hbx_3)  \s(\bX_{i3} = \bx_3\mid S_{i3}=1,\hbz_2,\hbx_2) l_{3}(\hbz_2,\hbx_3)  p_{3}(\hbz_2,\hbx_2)l_{2}(z_1,\hbx_2)p_{2}(z_1,\bx_1) l_1(\bx_1)\td \hbx_3\\
&=& \E\left[\frac{\bm{1}\{\hbZ_{i2}=\hbz_2\} S_{i3} m_{Y_3S_3}(\hbz_3,\hbX_{i3})  }{\pi_2(\hbz_2,\hbX_{i2})}\cdot \s(\bX_{i3}\mid S_{i3}=1,\hbZ_{i2},\hbX_{i2})\right],\\
F_2&=& \int  \dot m_{\theta,Y_3S_3}(\hbz_3,\hbx_3)\vert_{\theta=0} l_{3}(\hbz_2,\hbx_3) p_{3}(\hbz_2,\hbx_2)l_{2}(z_1,\hbx_2)p_{2}(z_1,\bx_1) l_1(\bx_1)\td \hbx_3.
\end{eqnarray*}
For $F_2$, we have
\begin{eqnarray*}
 \dot m_{\theta,Y_3S_3}(\hbz_3,\hbx_3)\vert_{\theta=0} 
&=&  \dot \E_\theta(Y_{i3}\mid S_{i3}=1,\hbZ_{i3}=\hbz_3,\hbX_{i3}=\hbx_3)\vert_{\theta=0} \\
&=&  \E  \{  Y_{i3}\cdot \s(Y_{i3}\mid S_{i3}=1,\hbZ_{i3}=\hbz_3,\hbX_{i3}=\hbx_3 )  \mid S_{i3}=1,\hbZ_{i3}=\hbz_3,\hbX_{i3}=\hbx_3\}.
\end{eqnarray*}
Therefore, we obtain
\begin{eqnarray*}
F_2  
&=& \E\left[\frac{\bm{1}\{\hbZ_{i3}=\hbz_3\} S_{i3}  Y_{i3} }{\pi_3(\hbz_3,\hbX_{i3})}\cdot \s(Y_{i3}\mid S_{i3}=1,\hbZ_{i3},\hbX_{i3})\right].
\end{eqnarray*}
Combining all the results, we obtain
\begin{eqnarray*}
&&\dot \E_\theta\{Y_{i3}(\hbz_3)S_{i3}(\hbz_2)\}\vert_{\theta=0} \\
&=& \E\{ h_1(\bX_{i1})\s(\bX_{i1})\}+ \E\{ h_2(S_{i2},Z_{i1},\bX_{i1})\s(S_{i2}\mid Z_{i1},\bX_{i1})\}\\
&&+ \E\{ h_3(S_{i2},Z_{i1},\hbX_{i2})\s(\bX_{i2}\mid S_{i2}=1,Z_{i1},\bX_{i1})\}+\E\{ h_4(S_{i3},\hbZ_{i2},\hbX_{i2})\s(S_{i3}\mid S_{i2}=1,\hbZ_{i2},\hbX_{i2})\}\\
&&+\E\{ h_5(S_{i3},\hbZ_{i2},\hbX_{i3})\s(\bX_{i3}\mid S_{i3}=1,\hbZ_{i2},\hbX_{i2})\}+\E\{ h_6(Y_{i3},S_{i3},\hbZ_{i3},\hbX_{i3})\s(Y_{i3}\mid S_{i3}=1,\hbZ_{i3},\hbX_{i3})\},
\end{eqnarray*}
where 
\begin{eqnarray*}
h_1(\bX_{i1}) &=& m_{Y_3S_3}(z_1,\bX_{i1}),\\
h_2(S_{i2},Z_{i1},\bX_{i1})
&=&\frac{\bm{1}\{Z_{i1}=z_1\}S_{i2} \E\{Y_{i3}(\hbz_3)S_{i3}(\hbz_2)\mid S_{i2}=1,Z_{i1},\bX_{i1}\} }{\pi_1(z_1, \bX_{i1})},\\
h_3(S_{i2},Z_{i1},\hbX_{i2})&=&\frac{\bm{1}\{Z_{i1}=z_1\}S_{i2}m_{Y_3S_3}(\hbz_2,\hbX_{i2}) }{\pi_1(z_1, \bX_{i1})},\\
h_4(S_{i3},\hbZ_{i2},\hbX_{i2}) &=& \frac{\bm{1}\{\hbZ_{i2}=\hbz_2\} S_{i3}  \E\{Y_{i3}(\hbz_3)S_{i3}(\hbz_2)\mid S_{i3}=1,\hbZ_{i2},\hbX_{i2}\} }{\pi_2(\hbz_2,\hbX_{i2})},\\
h_5(S_{i3},\hbZ_{i2},\hbX_{i3})&=&\frac{\bm{1}\{\hbZ_{i2}=\hbz_2\} S_{i3} m_{Y_3S_3}(\hbz_3,\hbX_{i3})  }{\pi_2(\hbz_2,\hbX_{i2})},\\
h_6(Y_{i3},S_{i3},\hbZ_{i3},\hbX_{i3})&=&\frac{\bm{1}\{\hbZ_{i3}=\hbz_3\}
                                          S_{i3}  Y_{i3} }{\pi_3(\hbz_3,\hbX_{i3})}.
\end{eqnarray*}
Denote 
\begin{eqnarray*}
\phi_{N1}(\bX_{i1})&=&h_1(\bX_{i1})  - \E\{h_1(\bX_{i1}) \}, \\
\phi_{N2}(S_{i2},Z_{i1},\bX_{i1})&=&h_2(S_{i2},Z_{i1},\bX_{i1})-\E\{h_2(S_{i2},Z_{i1},\bX_{i1})\mid Z_{i1},\bX_{i1}\},\\
\phi_{N3}(S_{i2},Z_{i1},\hbX_{i2})&=&h_3(S_{i2},Z_{i1},\hbX_{i2})-\E\{h_3(S_{i2},Z_{i1},\hbX_{i2})\mid S_{i2}=1,Z_{i1},\bX_{i1} \},\\
\phi_{N4}(S_{i3},\hbZ_{i2},\hbX_{i2})&=&h_4(S_{i3},\hbZ_{i2},\hbX_{i2}) - \E\{h_4(S_{i3},\hbZ_{i2},\hbX_{i2})\mid S_{i2}=1,\hbZ_{i2},\hbX_{i2} \},\\
\phi_{N5}(S_{i3},\hbZ_{i2},\hbX_{i3})&=&h_5(S_{i3},\hbZ_{i2},\hbX_{i3})- \E\{h_5(S_{i3},\hbZ_{i2},\hbX_{i3})\mid  S_{i3}=1,\hbZ_{i2},\hbX_{i2}  \},\\
\phi_{N6}(Y_{i3},S_{i3},\hbZ_{i3},\hbX_{i3})&=&h_6(Y_{i3},S_{i3},\hbZ_{i3},\hbX_{i3})-\E\{h_6(Y_{i3},S_{i3},\hbZ_{i3},\hbX_{i3})\mid S_{i3}=1,\hbZ_{i3},\hbX_{i3} \}.
\end{eqnarray*}
We can verify that $\phi_{N1}(X_{i1}) \in H_1^{\bX}$,  $\phi_{N2}(S_{i2},Z_{i1},\bX_{i1}) \in H_2^S$, $\phi_{N3}(S_{i2},Z_{i1},\hbX_{i2}) \in H_2^{\bX}$, $\phi_{N4}(S_{i3},\hbZ_{i2},\hbX_{i2})\in H_3^S$, $\phi_{N5}(S_{i3},\hbZ_{i2},\hbX_{i3}) \in H_3^{\bX}$, and $\phi_{N6}(Y_{i3},S_{i3},\hbZ_{i3},\hbX_{i3})\in H_3^Y$. Therefore,
\begin{eqnarray*}
\dot \E_\theta\{Y_{i3}(\hbz_3)S_{i3}(\hbz_2)\}\vert_{\theta=0} &=& \E\{\phi_N(\bV_i) \s(\bV_i)\},
\end{eqnarray*}
where 
\begin{eqnarray*}
&&\phi_N(\hbz_t)\\
 &=& \phi_{N1}(\bX_{i1})+\phi_{N2}(S_{i2},Z_{i1},\bX_{i1})+\phi_{N3}(S_{i2},Z_{i1},\hbX_{i2})+\phi_{N4}(S_{i3},\hbZ_{i2},\hbX_{i2})\\
&&+\phi_{N5}(S_{i3},\hbZ_{i2},\hbX_{i3})+\phi_{N6}(Y_{i3},S_{i3},\hbZ_{i3},\hbX_{i3})\\
&=&\frac{\bm{1}\{\hbZ_{i3}=\hbz_3\}S_{i3}\cdot \{Y_{i3} -m_{Y_3S_3}(\hbz_3,\hbX_{i3})\} }{\pi_3(\hbz_3,\hbX_{i3})}
+\frac{\bm{1}\{\hbZ_{i2}=\hbz_2\}S_{i3}}{\pi_2(\hbz_2,\hbX_{i2}) }\cdot \left\{m_{Y_3S_3}(\hbz_3,\hbX_{i3}) -\frac{m_{Y_3S_3}(\hbz_2,\hbX_{i2})}{p_3(\hbz_2,\hbX_{i2})}\right \}
\\
&&+ \frac{\bm{1}\{\hbZ_{i2}=\hbz_2\}\{S_{i3}-S_{i2}p_3(\hbz_2,\hbX_{i2})\}\cdot m_{Y_3S_3}(\hbz_2,\hbX_{i2}) }{ p_3(\hbz_2,\hbX_{i2})\pi_2(\hbz_2,\hbX_{i2})}\\
&&+ \frac{\bm{1}\{Z_{i1}=z_1\}S_{i2} }{\pi_1(z_1,\bX_{i1})}\cdot \left\{m_{Y_3S_3}(\hbz_2,\hbX_{i2})-\frac{m_{Y_3S_3}(z_1,\bX_{i1})}{p_2(z_1,\bX_{i1}) } \right\}\\
&&+\frac{\bm{1}\{Z_{i1}=z_1\}\{S_{i2}-p_2(z_1,\bX_{i1})\} \cdot m_{Y_3S_3}(z_1,\bX_{i1}) }{p_2(z_1,\bX_{i1}) \pi_1(\bX_{i1})}+m_{Y_3S_3}(z_1,\bX_{i1})-m_{Y_3S_3}\\
&=&\frac{\bm{1}\{\hbZ_{i3}=\hbz_3\}S_{i3}\cdot \{Y_{i3} -m_{Y_3S_3}(\hbz_3,\hbX_{i3})\} }{\pi_3(\hbz_3,\hbX_{i3})}+\frac{\bm{1}\{\hbZ_{i2}=\hbz_2\}\left\{S_{i3}m_{Y_3S_3}(\hbz_3,\hbX_{i3}) -S_{i2}m_{Y_3S_3}(\hbz_2,\hbX_{i2})\right \} }{\pi_2(\hbz_2,\hbX_{i2}) }\\
&&+ \frac{\bm{1}\{Z_{i1}=z_1\}\cdot \left\{S_{i2}m_{Y_3S_3}(\hbz_2,\hbX_{i2})-  m_{Y_3S_3}(z_1,\bX_{i1}) \right\} }{\pi_1(z_1,\bX_{i1})}+m_{Y_3S_3}(z_1,\bX_{i1})-m_{Y_3S_3}.
\end{eqnarray*}
Similarly, we can show that 
\begin{eqnarray*}
\dot \E_\theta\{S_{i3}(\hbz_2)\}\vert_{\theta=0} &=& \E\{\phi_D(\bV_i) \s(\bV_i)\},
\end{eqnarray*}
where
\begin{eqnarray*}
\phi_D(\hbz_{t-1})
&=& \frac{\bm{1}\{\hbZ_{i2}=\hbz_2\}\{S_{i3}-S_{i2}m_{S_3}(\hbz_2,\hbX_{i2})\}}{\pi_2(\hbz_2,\hbX_{i2})}+ \frac{\bm{1}\{Z_{i1}=z_1\}S_{i2}}{\pi_1(z_1,\bX_{i1})}\cdot \left\{m_{S_3}(\hbz_2,\hbX_{i2})-\frac{m_{S_3}(z_1,\bX_{i1})}{p_2(z_1,\bX_{i1})} \right\} \\
&&+\frac{\bm{1}\{Z_{i1}=z_1\}\{S_{i2}-p_2(z_1,\bX_{i1})\} \cdot m_{S_3}(z_1,\bX_{i1}) }{p_2(z_1,\bX_{i1})\pi_1(z_1,\bX_{i1})}+m_{S_3}(z_1,\bX_{i1})-m_{S_3}\\
&=& \frac{\bm{1}\{\hbZ_{i2}=\hbz_2\}\{S_{i3}-S_{i2}m_{S_3}(\hbz_2,\hbX_{i2})\}}{\pi_2(\hbz_2,\hbX_{i2})}+ \frac{\bm{1}\{Z_{i1}=z_1\}\{S_{i2}m_{S_3}(\hbz_2,\hbX_{i2})-m_{S_3}(z_1,\bX_{i1})\}}{\pi_1(z_1,\bX_{i1})}\\
&&+m_{S_3}(z_1,\bX_{i1})-m_{S_3}.
\end{eqnarray*}
Therefore,  the EIF for $\tau_3(\hbz_2)$ is
\begin{eqnarray*}
 \frac{\phi_{N}(\hbz_2,1)-\phi_{N}(\hbz_2,0) -  \tau_t(\hbz_2) \cdot \phi_{D}(\hbz_2)  }{ \Pr\{ S_{it}(\hbz_2)=1\}}
\end{eqnarray*}
and the EIF for $\theta(\xi)$ is 
\begin{eqnarray*}
\sum_{t=1}^T \sum_{\hbz_t}\phi_N(\hbz_t)\xi_t(\hbz_t)-\theta(\xi).
\end{eqnarray*}
\QEDB

\subsection{Proof of Theorem~\ref{thm::robustness}}

The estimator $\hat \phi_N(\hbz_t)$ is consistent for
\begin{eqnarray}
\nonumber &&\E\left[\frac{\bm{1}\{\hbZ_{it}=\hbz_t\} S_{it}\{Y_{it} -\tilde m_{Y_tS_t}(\hbz_{it},\hbX_{it})\}}{\tilde{\pi}_t(\hbz_t,\hbX_{it})}\right]\\
 \nonumber&&+\sum_{t'=2}^{t}\E\left[\frac{\bm{1}\{\hbZ_{i,t'-1}=\hbz_{t'-1}\} \{S_{it'}\tilde m_{Y_tS_t}(\hbz_{t'},\hbX_{it'})   -S_{i,t'-1}\tilde m_{Y_tS_t}(\hbz_{t'-1},\hbX_{i,t'-1})\}}{\tilde{\pi}_{t'-1}(\hbz_{t'-1},\hbX_{i,t'-1})}\right]\\
\label{eqn::phi-N-robustness} && 
\end{eqnarray}
We compute the terms in Equation~\eqref{eqn::phi-N-robustness} separately. For any $t'$, we have
\begin{eqnarray*}
&&\E\left[\frac{\bm{1}\{\hbZ_{it'}=\hbz_{t'}\} S_{it'}\tilde m_{Y_tS_t}(\hbz_{t'},\hbX_{it'}) }{\tilde{\pi}_{t'}(\hbz_{t'},\hbX_{it'})}\right]\\
&=&\int \E\left[\frac{\bm{1}\{\hbZ_{it'}=\hbz_{t'}\} S_{it'}\tilde
    m_{Y_tS_t}(\hbz_{t'},\hbX_{it'})
    }{\tilde{\pi}_{t'-1}(\hbz_{t'},\hbX_{it'})} \ \bigr | \  \bX_{i1}
    = \bx_1\right]l_1(\bx_1) \text{d} \bx_1\\
&=&\int \E\left[\frac{\bm{1}\{\hbZ_{it'}=\hbz_{t'}\} S_{it'}\tilde
    m_{Y_tS_t}(\hbz_{t'},\hbX_{it'}) w_1(z_1,\bx_1)
    }{\tilde{\pi}_{t'}(\hbz_{t'},\hbX_{it'})} \ \bigr | \ Z_{i1}= z_1,
    \bX_{i1} = x_1\right]l_1(\bx_1) \text{d} \bx_1\\
&=&\int \E\left[\frac{\bm{1}\{\hbZ_{it'}=\hbz_{t'}\} S_{it'}\tilde
    m_{Y_tS_t}(\hbz_{t'},\hbX_{it'}) w_1(z_1,\bx_1)
    }{\tilde{\pi}_{t'}(\hbz_{t'},\hbX_{it'})} \ \bigr | \
    S_{i2}=1, Z_{i1} = z_1,\bX_{i1} = \bx_1\right]p_2(z_1,\bx_1)l_1(\bx_1) \text{d} \bx_1,
\end{eqnarray*}
where the first equality follows from the law of total expectation. By
applying this calculation iteratively, we obtain
\begin{eqnarray*}
&&\E\left[\frac{\bm{1}\{\hbZ_{it'}=\hbz_{t'}\} S_{it'}\tilde m_{Y_tS_t}(\hbz_{t'},\hbX_{it'}) }{\tilde{\pi}_{t'}(\hbz_{t'},\hbX_{it'})}\right]\\
&=&\int \frac{ \tilde m_{Y_tS_t}(\hbz_{t'},\hbx_{t'})  \pi_{t'}(\hbz_{t'},\hbx_{t'}) }{\tilde{\pi}_{t'}(\hbz_{t'},\hbx_{t'})}  \prod_{s=2}^{t'} p_s(\hbz_{s-1},\hbx_{s-1}) \prod_{s=1}^{t'}l_s(\hbz_{s-1},\hbx_s) \text{d} \hbx_{t'},
\end{eqnarray*}
Similarly, we have 
\begin{eqnarray*}
&&\E\left[\frac{\bm{1}\{\hbZ_{i,t'-1}=\hbz_{t'-1}\} S_{i,t'}\tilde m_{Y_tS_t}(\hbz_{t'},\hbX_{it'}) }{\tilde{\pi}_{t'-1}(\hbz_{t'-1},\hbX_{it'-1})}\right]\\
&=&\int \frac{ \tilde m_{Y_tS_t}(\hbz_{t'},\hbx_{t'})  \pi_{t'-1}(\hbz_{t'-1},\hbx_{t'-1}) }{\tilde{\pi}_{t'-1}(\hbz_{t'-1},\hbx_{t'-1})}  \prod_{s=2}^{t'} p_s(\hbz_{s-1},\hbx_{s-1}) \prod_{s=1}^{t'}l_s(\hbz_{s-1},\hbx_s) \text{d} \hbx_{t'},\\
&&\E\left[\frac{\bm{1}\{\hbZ_{it}=\hbz_t\} S_{it}Y_{it} }{\tilde{\pi}_t(\hbz_t,\hbX_{it})}\right]\\
&=&\int \frac{ m_{Y_tS_t}(\hbz_{t},\hbx_{t})  \pi_{t}(\hbz_{t},\hbx_{t}) }{\tilde{\pi}_{t}(\hbz_{t},\hbx_{t})}  \prod_{s=2}^{t} p_s(\hbz_{s-1},\hbx_{s-1}) \prod_{s=1}^{t}l_s(\hbz_{s-1},\hbx_s) \text{d} \hbx_{t}.
\end{eqnarray*}
For simplicity, we omit the dependency of $p_s(\hbz_{s-1},\hbx_{s-1}) $,
$l_s(\hbz_{s-1},\hbx_s)$, and $ \pi_{t}(\hbz_{t},\hbx_{t}) $ on $\hbz$ and
$\hbx$ and write them as $p_s$, $l_s$, and $\pi_t$, respectively. We
can then write Equation~\eqref{eqn::phi-N-robustness} as 
\begin{eqnarray}
\nonumber &&\int  \frac{ \{m_{Y_tS_t}(\hbz_{t},\hbx_{t}) -\tilde m_{Y_tS_t}(\hbz_{t},\hbx_{t})\}  \pi_{t} }{\tilde{\pi}_{t}}  \prod_{s=2}^{t} p_s \prod_{s=1}^{t}l_s \text{d} \hbx_{t}\\
\nonumber &&+\sum_{t'=2}^{t} \left[ \int  \frac{\tilde m_{Y_tS_t}(\hbz_{t'},\hbx_{t'})\pi_{t'-1} }{\tilde \pi_{t'-1}}   \prod_{s=2}^{t'} p_s \prod_{s=1}^{t'}l_s \text{d} \hbx_{t'}- \int  \frac{\tilde m_{Y_tS_t}(\hbz_{t'-1},\hbx_{t'-1})\pi_{t'-1} }{\tilde \pi_{t'-1}}   \prod_{s=2}^{t'-1} p_s \prod_{s=1}^{t'-1}l_s \text{d} \hbx_{t'-1}\right]\\
\nonumber&& +\int  \tilde m_{Y_tS_t}(z_1,\bx_1) l_1 \text{d} \bx_1\\
\label{eqn::phi-N-robustness-2}&=& \int  \frac{ m_{Y_tS_t}(\hbz_{t},\hbx_{t})  \pi_{t} }{\tilde{\pi}_{t}}  \prod_{s=2}^{t} p_s \prod_{s=1}^{t}l_s \text{d} \hbx_{t}-  \sum_{t'=1}^t  \iint \tilde m_{Y_tS_t}(\hbz_{t'},\hbx_{t'}) \left(\frac{ \pi_{t'}}{\tilde  \pi_{t'}} - \frac{ \pi_{t'-1}}{\tilde  \pi_{t'-1}}  \right)  \prod_{s=2}^{t'} p_s \prod_{s=1}^{t'}l_s    \text{d} \hbx_{t'}.
\end{eqnarray}
For the term within the summation in Equation~\eqref{eqn::phi-N-robustness-2}, we have 
\begin{eqnarray}
\nonumber && \int \tilde m_{Y_tS_t}(\hbz_{t'},\hbx_{t'}) \left(\frac{ \pi_{t'}}{\tilde  \pi_{t'}} - \frac{ \pi_{t'-1}}{\tilde  \pi_{t'-1}}  \right)  \prod_{s=2}^{t'} p_s \prod_{s=1}^{t'}l_s    \text{d} \hbx_{t'}\\
\nonumber &=& \int \{\tilde m_{Y_tS_t}(\hbz_{t'},\hbx_{t'})-m_{Y_tS_t}(\hbz_{t'},\hbx_{t'})\} \left(\frac{ \pi_{t'}}{\tilde  \pi_{t'}} - \frac{ \pi_{t'-1}}{\tilde  \pi_{t'-1}}  \right)  \prod_{s=2}^{t'} p_s \prod_{s=1}^{t'}l_s    \text{d} \hbx_{t'}\\
\nonumber &&+\int   m_{Y_tS_t}(\hbz_{t'},\hbx_{t'}) \left(\frac{ \pi_{t'}}{\tilde  \pi_{t'}} - \frac{ \pi_{t'-1}}{\tilde  \pi_{t'-1}}  \right)  \prod_{s=2}^{t'} p_s \prod_{s=1}^{t'}l_s    \text{d} \hbx_{t'}\\
\nonumber &=& \int \{\tilde m_{Y_tS_t}(\hbz_{t'},\hbx_{t'})-m_{Y_tS_t}(\hbz_{t'},\hbx_{t'})\} \left(\frac{ \pi_{t'}}{\tilde  \pi_{t'}} - \frac{ \pi_{t'-1}}{\tilde  \pi_{t'-1}}  \right)  \prod_{s=2}^{t'} p_s \prod_{s=1}^{t'}l_s    \text{d} \hbx_{t'}\\
\nonumber &&+\int   m_{Y_tS_t}(\hbz_{t'},\hbx_{t'}) \frac{ \pi_{t'}}{\tilde  \pi_{t'}}  \prod_{s=2}^{t'} p_s \prod_{s=1}^{t'}l_s    \text{d} \hbx_{t'}-\int   m_{Y_tS_t}(\hbz_{t'-1},\hbx_{t'-1}) \frac{ \pi_{t'-1}}{\tilde  \pi_{t'-1}}  \prod_{s=2}^{t'-1} p_s \prod_{s=1}^{t'-1}l_s    \text{d} \hbx_{t'-1},\\
\label{eqn::phi-N-robustness-3}
\end{eqnarray}
where the last equality follows from the fact that
\begin{eqnarray*}
\int  m_{Y_tS_t}(\hbz_{t'},\hbx_{t'}) p_{t'}l_{t'} \text{d} \hbx_{t'} &=&m_{Y_tS_t}(\hbz_{t'-1},\hbx_{t'-1}).
\end{eqnarray*}
Substituting Equation~\eqref{eqn::phi-N-robustness-3} into
Equation~\eqref{eqn::phi-N-robustness-2}, we can write Equation~\eqref{eqn::phi-N-robustness-2} as
\begin{eqnarray*}
&& \int  \frac{ m_{Y_tS_t}(\hbz_{t},\hbx_{t})  \pi_{t} }{\tilde{\pi}_{t}}  \prod_{s=2}^{t} p_s \prod_{s=1}^{t}l_s \text{d} \hbx_{t}\\
 &&- \sum_{t'=1}^t \int \{\tilde m_{Y_tS_t}(\hbz_{t'},\hbx_{t'})-m_{Y_tS_t}(\hbz_{t'},\hbx_{t'})\} \left(\frac{ \pi_{t'}}{\tilde  \pi_{t'}} - \frac{ \pi_{t'-1}}{\tilde  \pi_{t'-1}}  \right)  \prod_{s=2}^{t'} p_s \prod_{s=1}^{t'}l_s    \text{d} \hbx_{t'}\\
 &&- \sum_{t'=1}^t\int  \left[ m_{Y_tS_t}(\hbz_{t'},\hbx_{t'}) \frac{ \pi_{t'}}{\tilde  \pi_{t'}}  \prod_{s=2}^{t'} p_s \prod_{s=1}^{t'}l_s    \text{d} \hbx_{t'}-\iint   m_{Y_tS_t}(\hbz_{t'-1},\hbx_{t'-1}) \frac{ \pi_{t'-1}}{\tilde  \pi_{t'-1}}  \prod_{s=2}^{t'-1} p_s \prod_{s=1}^{t'-1}l_s    \text{d} \hbx_{t'-1}\right]\\
 &=&m_{Y_tS_t}+ \sum_{t'=1}^t \int \{m_{Y_tS_t}(\hbz_{t'},\hbx_{t'})-\tilde  m_{Y_tS_t}(\hbz_{t'},\hbx_{t'})\} \left(\frac{ \pi_{t'}}{\tilde  \pi_{t'}} - \frac{ \pi_{t'-1}}{\tilde  \pi_{t'-1}}  \right)  \prod_{s=2}^{t'} p_s \prod_{s=1}^{t'}l_s    \text{d} \hbx_{t'}.
\end{eqnarray*}
Therefore, the bias of the estimator is 
\begin{eqnarray*}
 \sum_{t'=1}^t \int \{m_{Y_tS_t}(\hbz_{t'},\hbx_{t'})-\tilde  m_{Y_tS_t}(\hbz_{t'},\hbx_{t'})\} \left(\frac{ \pi_{t'}}{\tilde  \pi_{t'}} - \frac{ \pi_{t'-1}}{\tilde  \pi_{t'-1}}  \right)  \prod_{s=2}^{t'} p_s \prod_{s=1}^{t'}l_s    \text{d} \hbx_{t'},
\end{eqnarray*}
which equals zero when either $(\tilde m_{Y_tS_t}, \tilde m_{S_t})=( m_{Y_tS_t},  m_{S_t})$ or $\tilde{\pi}=\pi$. \QEDB

\subsection{Proof of Theorem~\ref{thm::efficiency}}

Let $\mathcal{P}_n$ denote the empirical average and $\mathcal{P} $ denote the expectation. For example, $\mathcal{P}_n f(\bV) = n^{-1} \sum_{i=1}^n f(\bV_i)$ and $\mathcal{P}  f(\bV)= \E\{f(\bV_i)\}$.
By the empirical process theory, we have 
\begin{eqnarray*}
\hat \phi_N -  \E(\phi_N) &=&  (\mathcal{P}_n-\mathcal{P} ) \hat \phi_N + \mathcal{P}(\hat \phi_N-\phi_N)\\
&=& (\mathcal{P}_n-\mathcal{P} ) \phi_N + \mathcal{P}(\hat \phi_N-\phi_N)+ o_p(n^{-1/2}),
\end{eqnarray*}
where the second equality follows from Condition 2.
From the derivation of the bias in the proof of Theorem~\ref{thm::robustness}, we have 
\begin{eqnarray*}
&&\mathcal{P}(\hat \phi_N-\phi_N)\\
&=&  \sum_{t'=1}^t \int \{\hat m_{Y_tS_t}(\hbz_{t'},\hbx_{t'})-\tilde  m_{Y_tS_t}(\hbz_{t'},\hbx_{t'})\} \left(\frac{ \hat \pi_{t'}}{\tilde  \pi_{t'}} - \frac{ \hat \pi_{t'-1}}{\tilde  \pi_{t'-1}}  \right)  \prod_{s=2}^{t'} p_s \prod_{s=1}^{t'}l_s    \text{d} \hbx_{t'}\\
&=&  \sum_{t'=1}^t \int \{\hat m_{Y_tS_t}(\hbz_{t'},\hbx_{t'})-  m_{Y_tS_t}(\hbz_{t'},\hbx_{t'})\} \left(\frac{ \hat \pi_{t'}}{  \pi_{t'}} - \frac{ \hat \pi_{t'-1}}{  \pi_{t'-1}}  \right)  \prod_{s=2}^{t'} p_s \prod_{s=1}^{t'}l_s    \text{d} \hbx_{t'}\\
&\leq &\sum_{t'=1}^t   \int \{\hat m_{Y_tS_t}(\hbz_{t'},\hbx_{t'})-  m_{Y_tS_t}(\hbz_{t'},\hbx_{t'})\}^2  \prod_{s=2}^{t'} p_s \prod_{s=1}^{t'}l_s    \text{d} \hbx_{t'} \int \left(\frac{ \hat \pi_{t'}}{  \pi_{t'}} - \frac{ \hat \pi_{t'-1}}{  \pi_{t'-1}}  \right)^2  \prod_{s=2}^{t'} p_s \prod_{s=1}^{t'}l_s    \text{d} \hbx_{t'}\\
&\leq& C \sum_{t'=1}^t   \int \{\hat m_{Y_tS_t}(\hbz_{t'},\hbx_{t'})-  m_{Y_tS_t}(\hbz_{t'},\hbx_{t'})\}^2  \prod_{s=2}^{t'} p_s \prod_{s=1}^{t'}l_s    \text{d} \hbx_{t'} \int \left( \hat w_{t'} - w_{t'} \right)^2  \prod_{s=2}^{t'} p_s \prod_{s=1}^{t'}l_s    \text{d} \hbx_{t'}\\
&=& o_p(n^{-1/2}).
\end{eqnarray*}
where $C$ is a constant, the second equality follows from Condition 1, the first inequality follows from the Cauchy-Schwarz inequality, the second inequality follows from Condition 3, and the last equality follows from Condition 4. As a result, $\hat \phi_N(\hbz_t)$ is a semiparametrically efficient for $m_{Y_tS_t}$. 
Similarly, we can show that  $\hat \phi_D(\hbz_t)$ is a semiparametrically efficient for $m_{S_t}$.    \QEDB

\section{Additional  simulation results}
\label{app::sim}
\subsection{Results with continuous outcomes}
\begin{figure}[htbp]
 \centering \spacingset{1}
 \includegraphics[width=\textwidth]{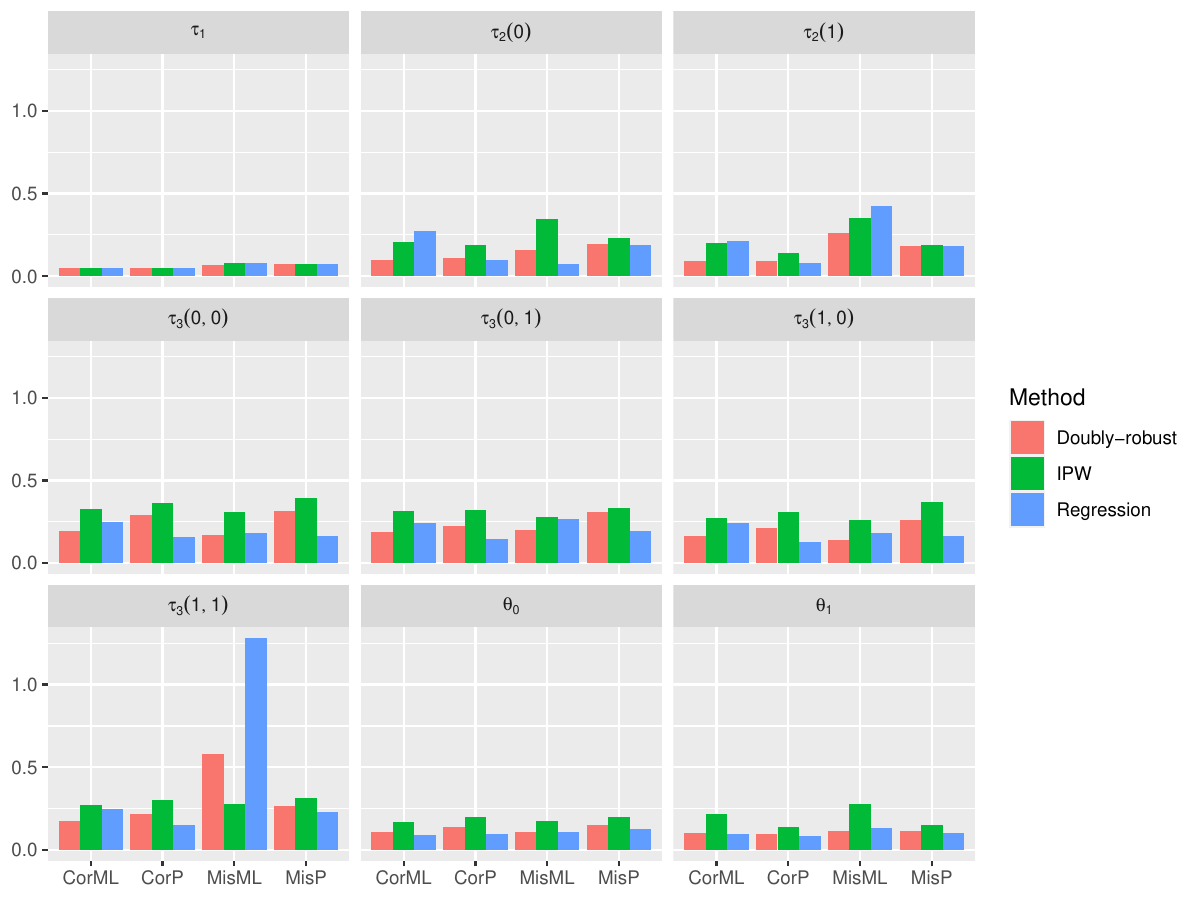}
 \caption{Biases of the estimators across $500$ replications with continuous outcomes, $\delta=0.5$, and $n=1,000$. Each row presents the results for one target causal quantity. The column labels ``CorML,'' ``CorP,'' ``MisML,'' and ``MisP'' stand for the nonparametric estimation with correctly specified covariates, parametric estimation with correctly specified covariates,  nonparametric estimation with mis-specified covariates, and parametric estimation with mis-specified covariates, respectively.}
\label{fig::sim_var_n1_con_bias}
\end{figure}

\begin{figure}[htbp]
 \centering \spacingset{1}
 \includegraphics[width=\textwidth]{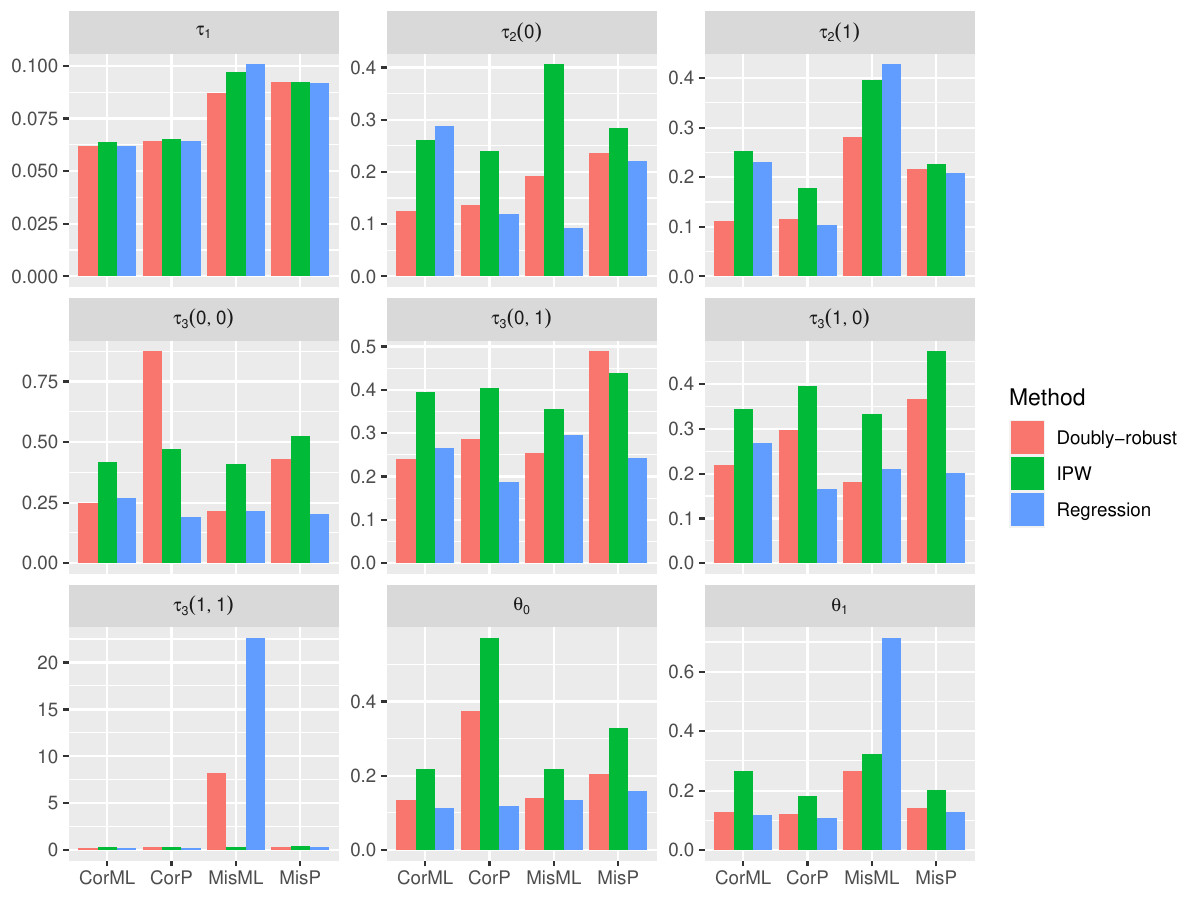}
 \caption{Root mean squared errors for the estimators with continuous outcomes, $\delta=0.5$, and $n=1,000$. Each subplot presents the results for one target causal quantity. Within each subplot, the $x$-axis labels `CorML,'' ``CorP,'' ``MisML,'' and ``MisP'' stand for the nonparametric estimation with correctly specified covariates, parametric estimation with correctly specified covariates,  nonparametric estimation with mis-specified covariates, and parametric estimation with  mis-specified covariates, respectively.  The heights of the bars represent the root mean squared errors of the regression (blue), inverse probability weighting (red), and doubly robust estimators (green).}
\label{fig::sim_var_n1_con_rmse}
\end{figure}

Figure~\ref{fig::sim_var_n1_con_bias} shows the absolute biases of the estimators with time varying confounders and sample size $n=1,000$ and Figure~\ref{fig::sim_var_n1_con_rmse} presents the corresponding RMSEs. 

\begin{figure}[htbp]
 \centering \spacingset{1}
 \includegraphics[width=\textwidth]{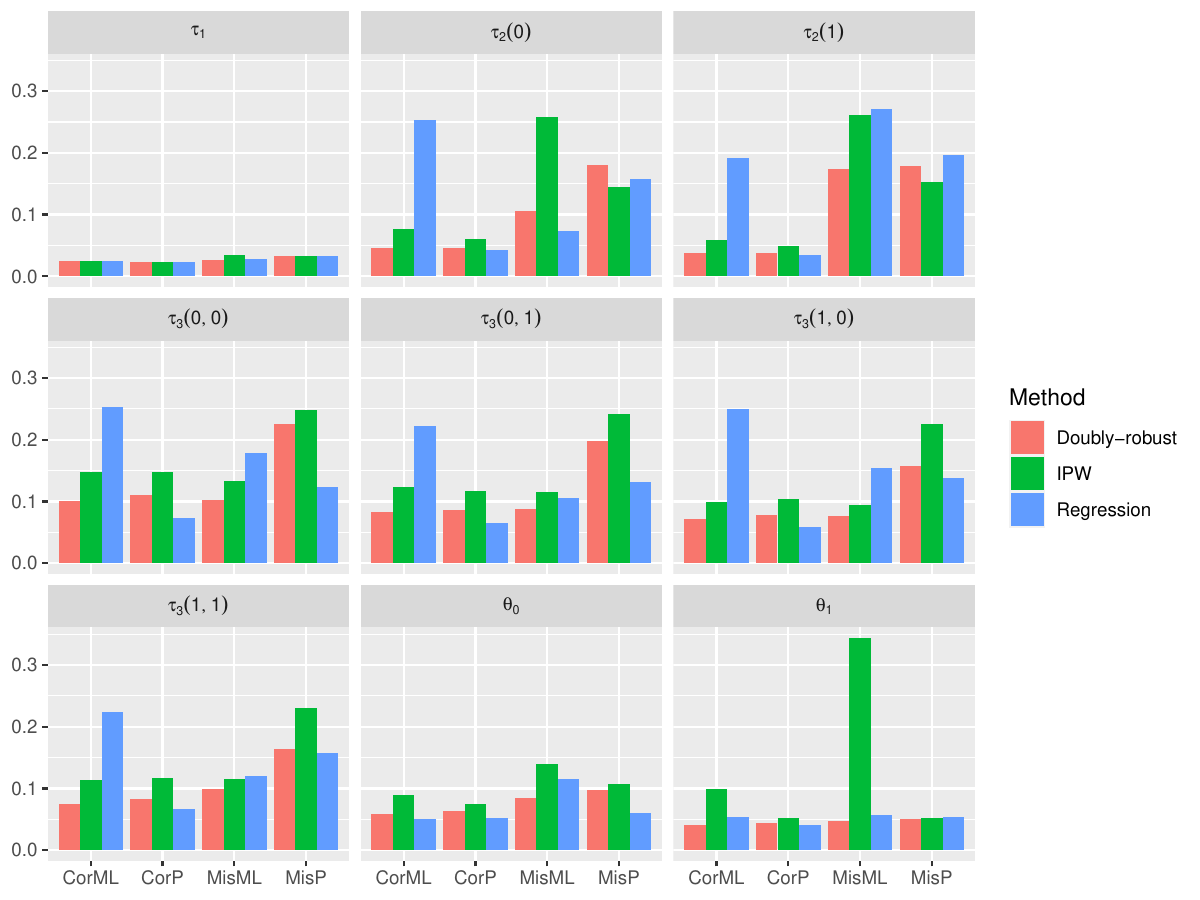}
 \caption{Biases of the estimators across $500$ replications with continuous outcomes, $\delta=0$, and $n=5,000$. Each row presents the results for one target causal quantity. The column labels ``CorML,'' ``CorP,'' ``MisML,'' and ``MisP'' stand for the nonparametric estimation with correctly specified covariates, parametric estimation with correctly specified covariates,  nonparametric estimation with  mis-specified covariates, and parametric estimation with  mis-specified covariates, respectively.}
\label{fig::sim_inv_n2_con_bias}
\end{figure}

\begin{figure}[htbp]
 \centering \spacingset{1}
 \includegraphics[width=\textwidth]{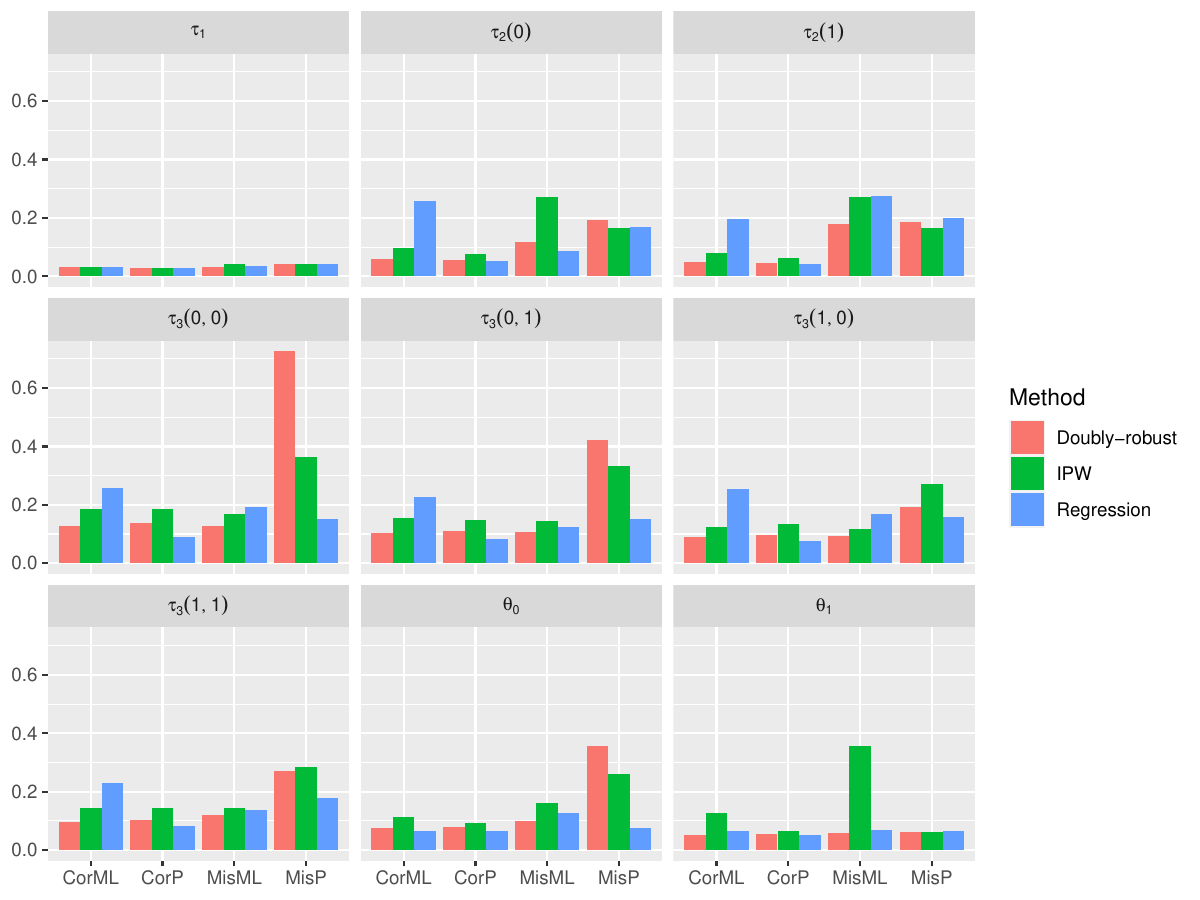}
 \caption{Root mean squared errors for the estimators with continuous outcomes, $\delta=0$, and $n=5,000$. Each subplot presents the results for one target causal quantity. Within each subplot, the $x$-axis labels `CorML,'' ``CorP,'' ``MisML,'' and ``MisP'' stand for the nonparametric estimation with correctly specified covariates, parametric estimation with correctly specified covariates,  nonparametric estimation with  mis-specified covariates, and parametric estimation with  mis-specified covariates, respectively.  The heights of the bars represent the root mean squared errors of the regression (blue), inverse probability weighting (red), and doubly robust estimators (green).}
\label{fig::sim_inv_n2_con_rmse}
\end{figure}

Figure~\ref{fig::sim_inv_n2_con_bias} shows the absolute biases of the estimators with time invariant confounders and sample size $n=5,000$ and Figure~\ref{fig::sim_inv_n2_con_rmse} presents the corresponding RMSEs. 

\begin{figure}[htbp]
 \centering \spacingset{1}
 \includegraphics[width=\textwidth]{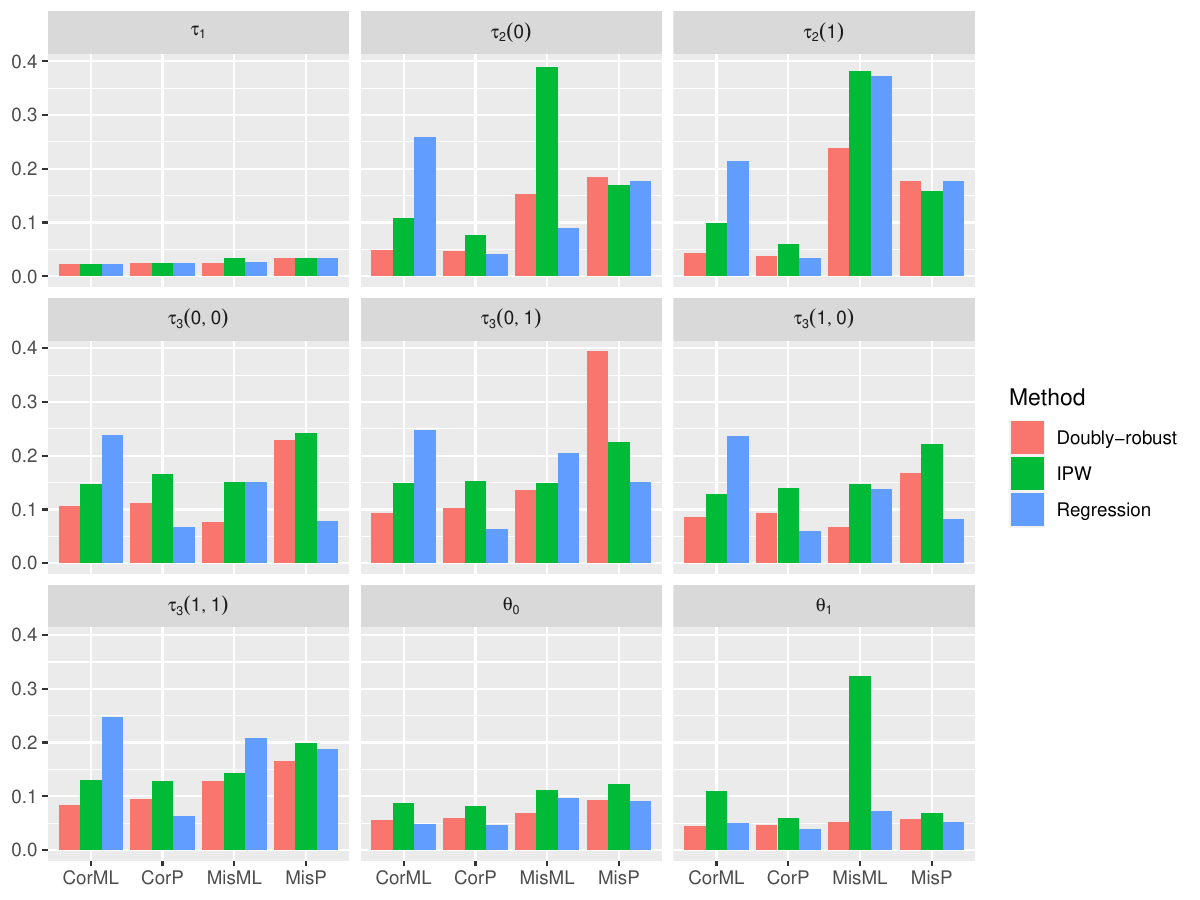}
 \caption{Biases of the estimators across $500$ replications with continuous outcomes, $\delta=0.5$, and $n=5,000$. Each row presents the results for one target causal quantity. The column labels ``CorML,'' ``CorP,'' ``MisML,'' and ``MisP'' stand for the nonparametric estimation with correctly specified covariates, parametric estimation with correctly specified covariates,  nonparametric estimation with  mis-specified covariates, and parametric estimation with  mis-specified covariates, respectively.}
\label{fig::sim_var_n2_con_bias}
\end{figure}

\begin{figure}[htbp]
 \centering \spacingset{1}
 \includegraphics[width=\textwidth]{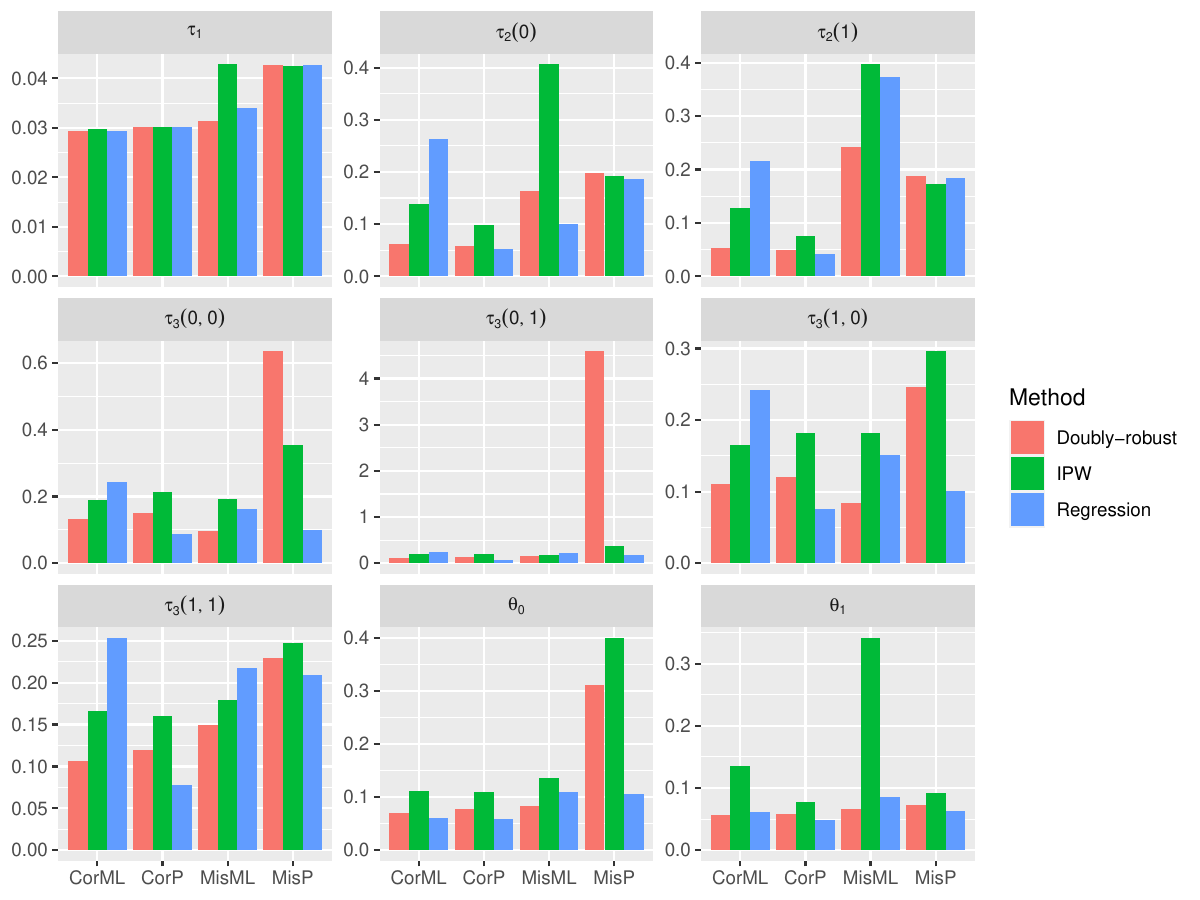}
 \caption{Root mean squared errors for the estimators with continuous outcomes, $\delta=0.5$, and $n=5,000$. Each subplot presents the results for one target causal quantity. Within each subplot, the $x$-axis labels `CorML,'' ``CorP,'' ``MisML,'' and ``MisP'' stand for the nonparametric estimation with correctly specified covariates, parametric estimation with correctly specified covariates,  nonparametric estimation with  mis-specified covariates, and parametric estimation with  mis-specified covariates, respectively.  The heights of the bars represent the root mean squared errors of the regression (blue), inverse probability weighting (red), and doubly robust estimators (green).}
\label{fig::sim_var_n2_con_rmse}
\end{figure}

Figure~\ref{fig::sim_var_n2_con_bias} shows the absolute biases of the estimators with time invariant confounders and sample size $n=5,000$ and Figure~\ref{fig::sim_var_n2_con_rmse} presents the corresponding RMSEs. 

\subsection{Results with binary outcomes}
We maintain the same data generating mechanism as in the scenarios with continuous outcomes except that the outcomes are now generated from logistic models with the same predictors and coefficients. 

\begin{figure}[htbp]
 \centering \spacingset{1}
 \includegraphics[width=\textwidth]{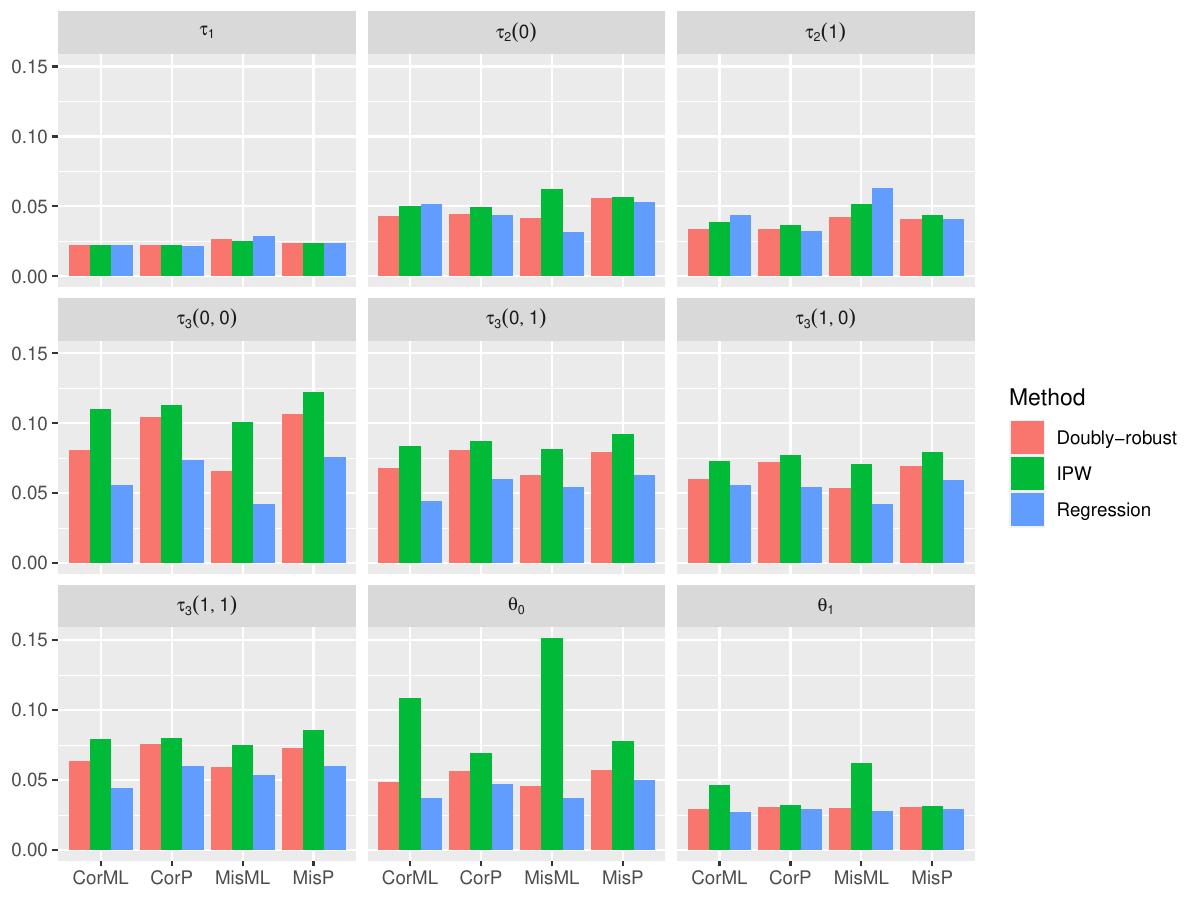}
 \caption{Biases of the estimators across $500$ replications with binary outcomes, $\delta=0$, and $n=1,000$. Each row presents the results for one target causal quantity. The column labels ``CorML,'' ``CorP,'' ``MisML,'' and ``MisP'' stand for the nonparametric estimation with correctly specified covariates, parametric estimation with correctly specified covariates,  nonparametric estimation with  mis-specified covariates, and parametric estimation with  mis-specified covariates, respectively.}
\label{fig::sim_inv_n1_bin_bias}
\end{figure}

\begin{figure}[htbp]
 \centering \spacingset{1}
 \includegraphics[width=\textwidth]{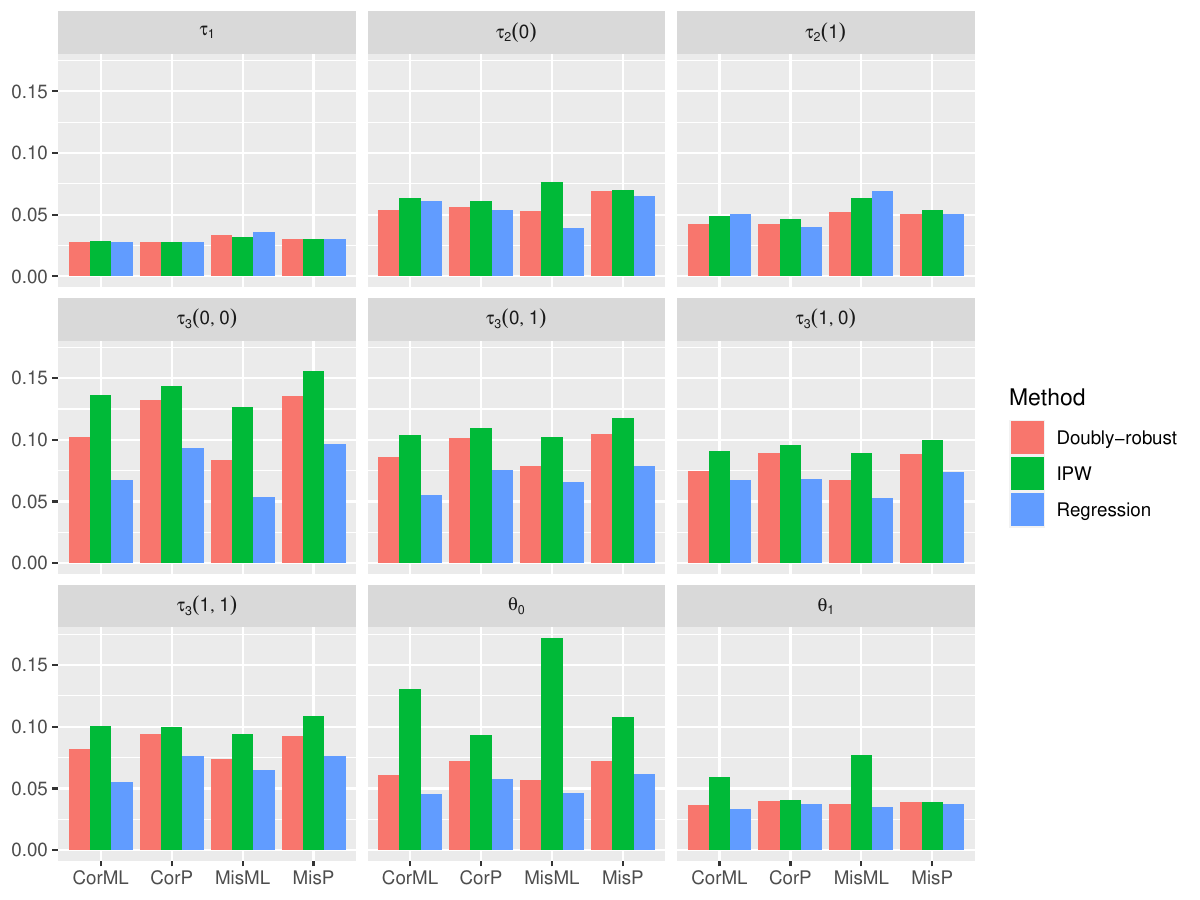}
 \caption{Root mean squared errors for the estimators with binary outcomes, $\delta=0$, and $n=1,000$. Each subplot presents the results for one target causal quantity. Within each subplot, the $x$-axis labels `CorML,'' ``CorP,'' ``MisML,'' and ``MisP'' stand for the nonparametric estimation with correctly specified covariates, parametric estimation with correctly specified covariates,  nonparametric estimation with  mis-specified covariates, and parametric estimation with  mis-specified covariates, respectively.  The heights of the bars represent the root mean squared errors of the regression (blue), inverse probability weighting (red), and doubly robust estimators (green).}
\label{fig::sim_inv_n1_bin_rmse}
\end{figure}

Figure~\ref{fig::sim_inv_n1_bin_bias} shows the absolute biases of the estimators with time invariant confounders and sample size $n=1,000$ and Figure~\ref{fig::sim_inv_n1_bin_rmse} presents the corresponding RMSEs. 

\begin{figure}[htbp]
 \centering \spacingset{1}
 \includegraphics[width=\textwidth]{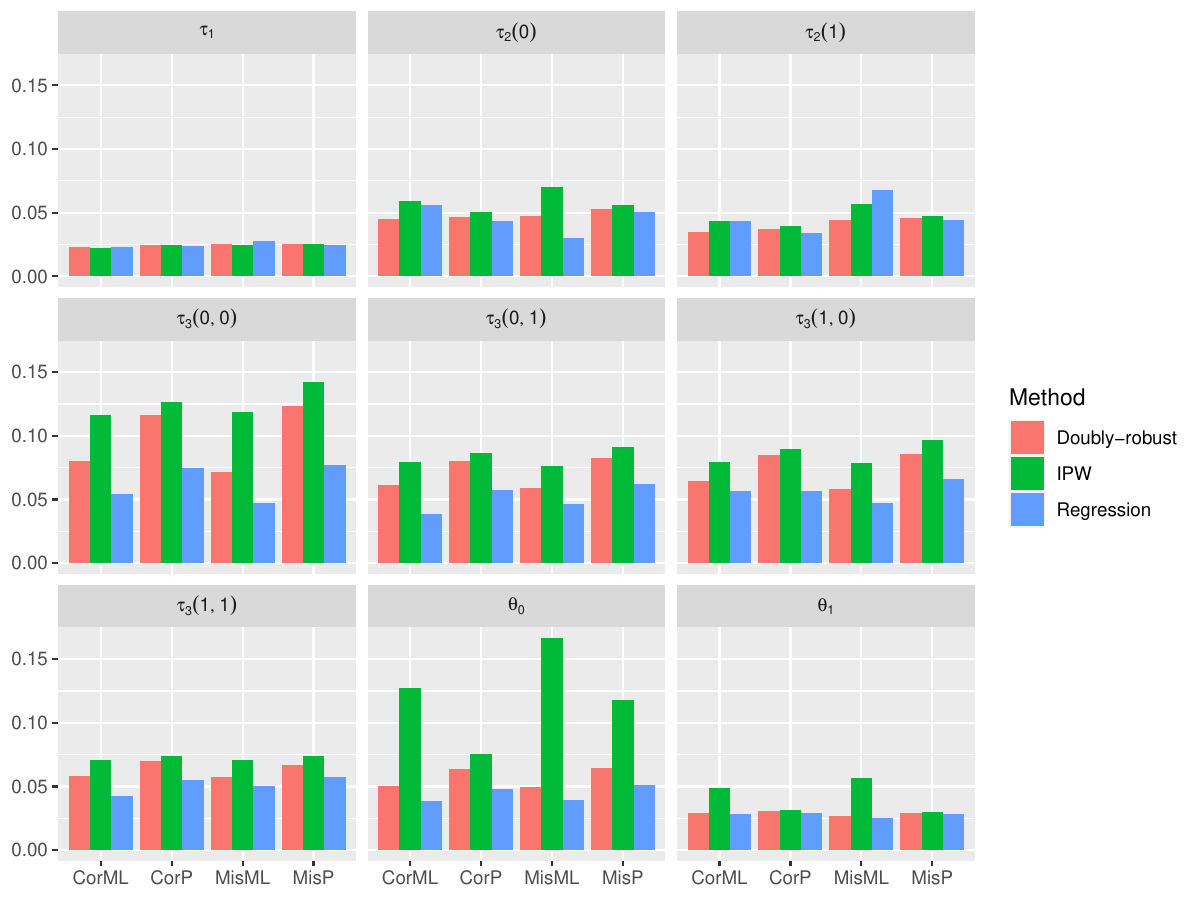}
 \caption{Biases of the estimators across $500$ replications with binary outcomes, $\delta=0.5$, and $n=1,000$. Each row presents the results for one target causal quantity. The column labels ``CorML,'' ``CorP,'' ``MisML,'' and ``MisP'' stand for the nonparametric estimation with correctly specified covariates, parametric estimation with correctly specified covariates,  nonparametric estimation with  mis-specified covariates, and parametric estimation with  mis-specified covariates, respectively.}
\label{fig::sim_var_n1_bin_bias}
\end{figure}

\begin{figure}[htbp]
 \centering \spacingset{1}
 \includegraphics[width=\textwidth]{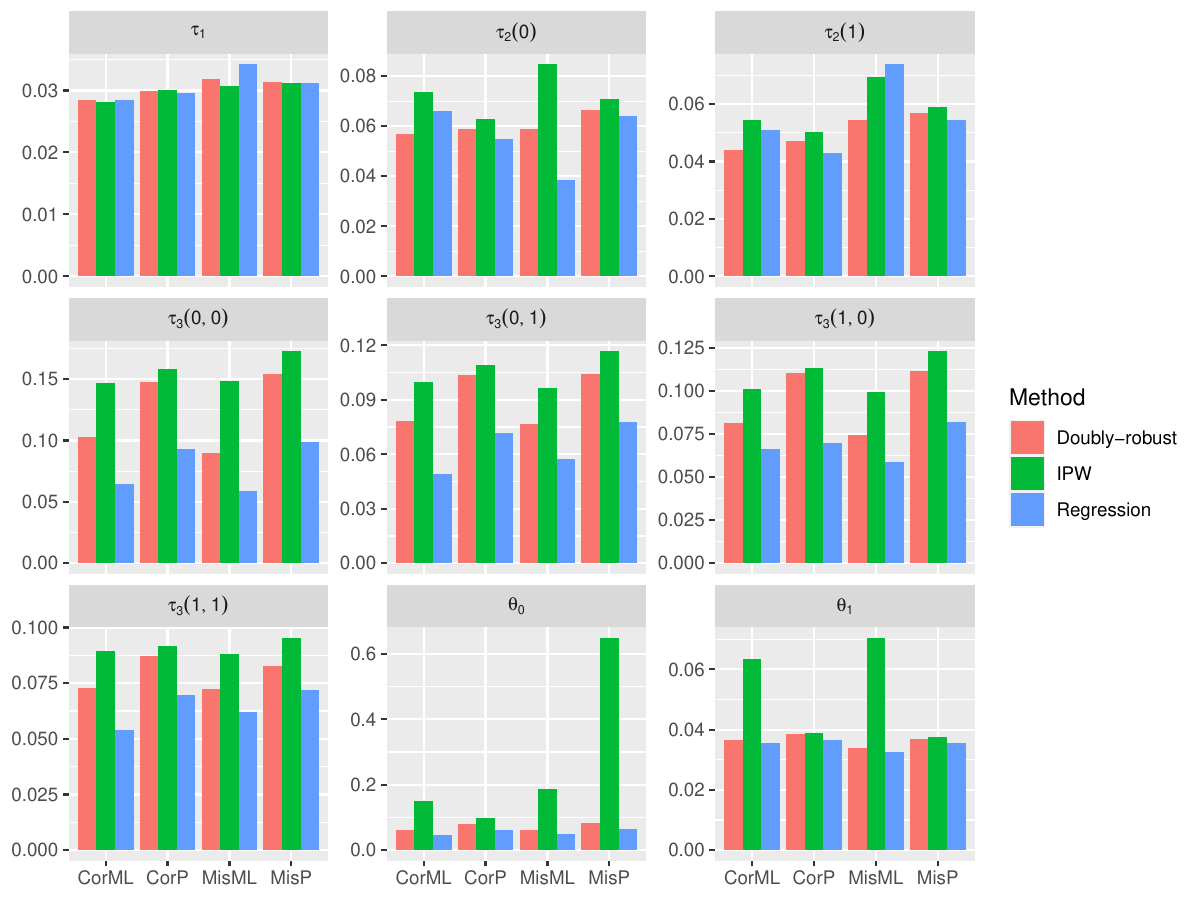}
 \caption{Root mean squared errors for the estimators with binary outcomes, $\delta=0.5$, and $n=1,000$. Each subplot presents the results for one target causal quantity. Within each subplot, the $x$-axis labels `CorML,'' ``CorP,'' ``MisML,'' and ``MisP'' stand for the nonparametric estimation with correctly specified covariates, parametric estimation with correctly specified covariates,  nonparametric estimation with  mis-specified covariates, and parametric estimation with  mis-specified covariates, respectively.  The heights of the bars represent the root mean squared errors of the regression (blue), inverse probability weighting (red), and doubly robust estimators (green).}
\label{fig::sim_var_n1_bin_rmse}
\end{figure}

Figure~\ref{fig::sim_var_n1_bin_bias} shows the absolute biases of the estimators with time varying confounders and sample size $n=1,000$ and Figure~\ref{fig::sim_var_n1_bin_rmse} presents the corresponding RMSEs. 

\begin{figure}[htbp]
 \centering \spacingset{1}
 \includegraphics[width=\textwidth]{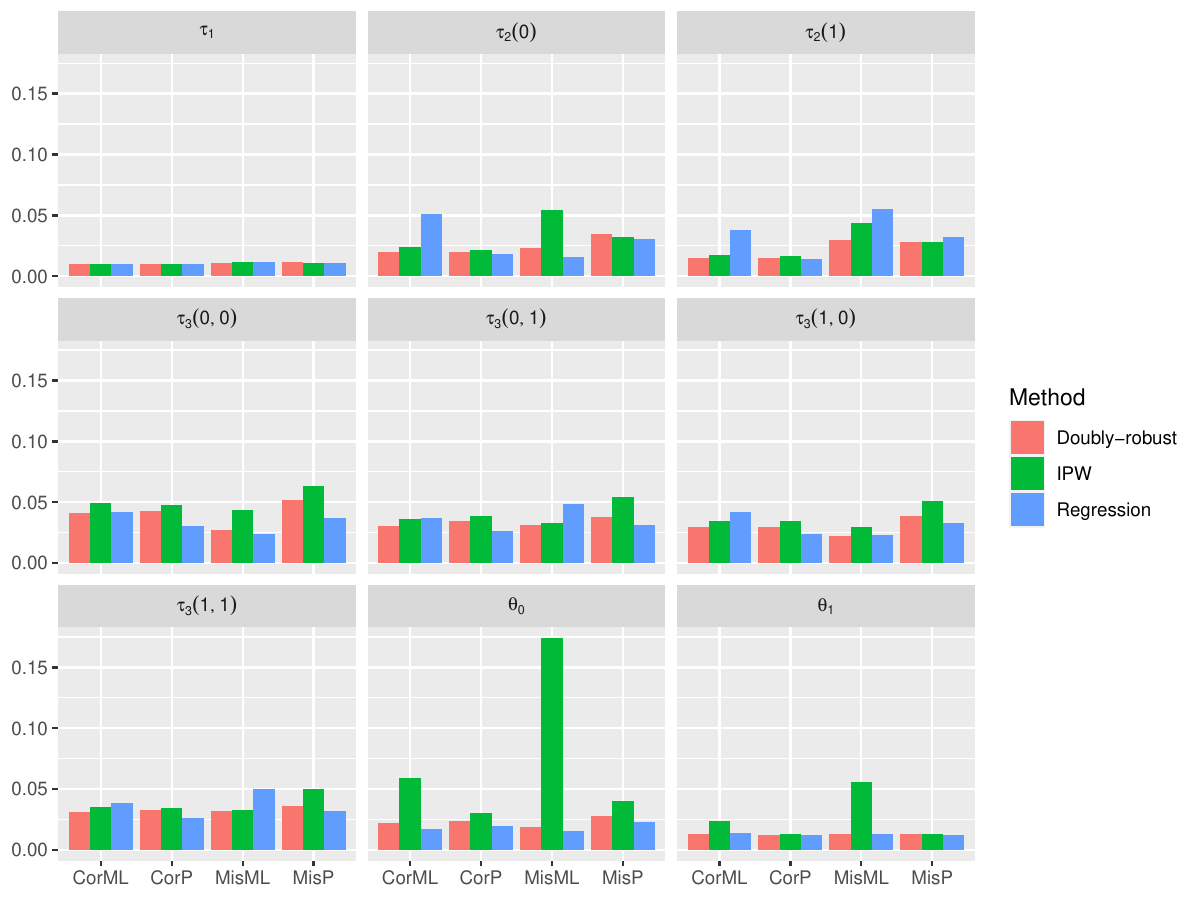}
 \caption{Biases of the estimators across $500$ replications with binary outcomes, $\delta=0$, and $n=5,000$. Each row presents the results for one target causal quantity. The column labels ``CorML,'' ``CorP,'' ``MisML,'' and ``MisP'' stand for the nonparametric estimation with correctly specified covariates, parametric estimation with correctly specified covariates,  nonparametric estimation with  mis-specified covariates, and parametric estimation with  mis-specified covariates, respectively.}
\label{fig::sim_inv_n2_bin_bias}
\end{figure}

\begin{figure}[htbp]
 \centering \spacingset{1}
 \includegraphics[width=\textwidth]{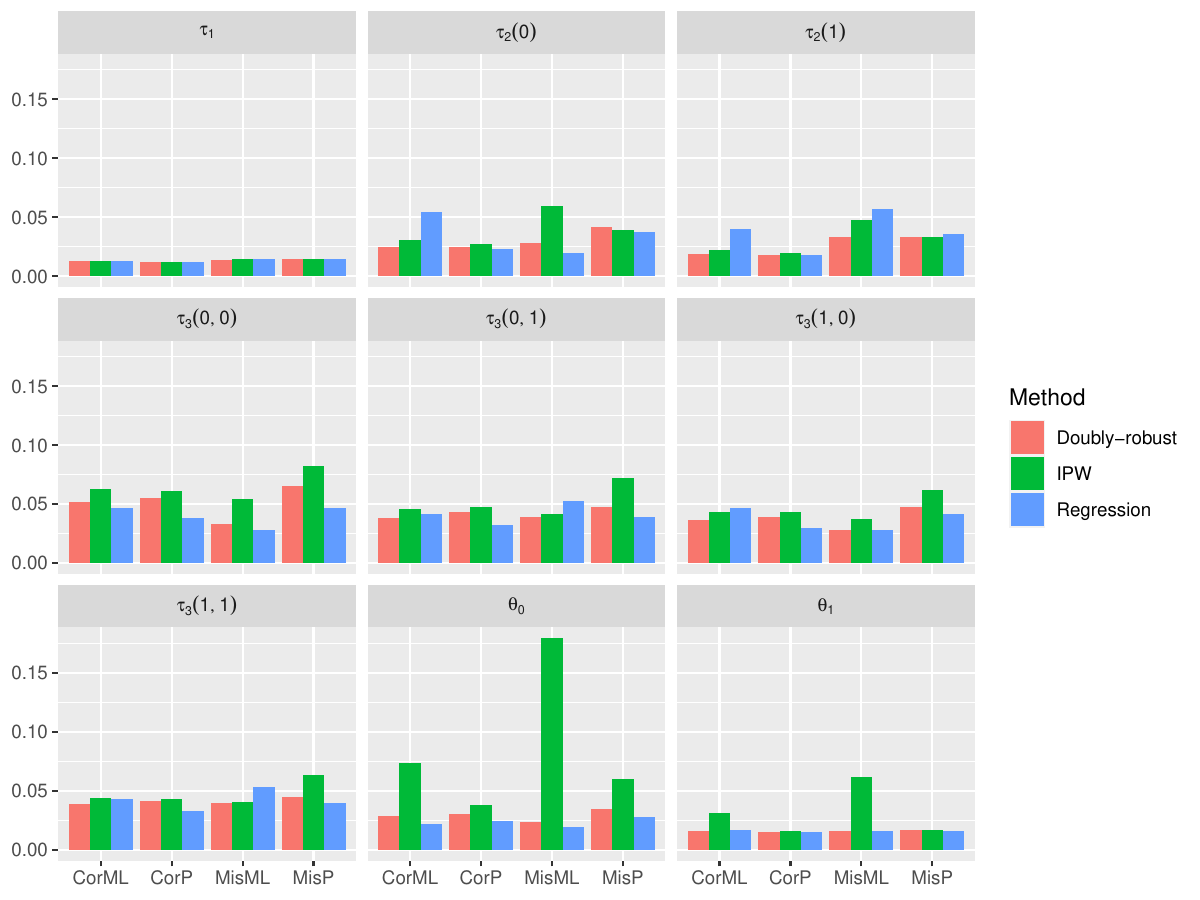}
 \caption{Root mean squared errors for the estimators with binary outcomes, $\delta=0$, and $n=5,000$. Each subplot presents the results for one target causal quantity. Within each subplot, the $x$-axis labels `CorML,'' ``CorP,'' ``MisML,'' and ``MisP'' stand for the nonparametric estimation with correctly specified covariates, parametric estimation with correctly specified covariates,  nonparametric estimation with  mis-specified covariates, and parametric estimation with  mis-specified covariates, respectively.  The heights of the bars represent the root mean squared errors of the regression (blue), inverse probability weighting (red), and doubly robust estimators (green).}
\label{fig::sim_inv_n2_bin_rmse}
\end{figure}

Figure~\ref{fig::sim_inv_n2_bin_bias} shows the absolute biases of the estimators with time invariant confounders and sample size $n=5,000$ and Figure~\ref{fig::sim_inv_n2_bin_rmse} presents the corresponding RMSEs. 

\begin{figure}[htbp]
 \centering \spacingset{1}
 \includegraphics[width=\textwidth]{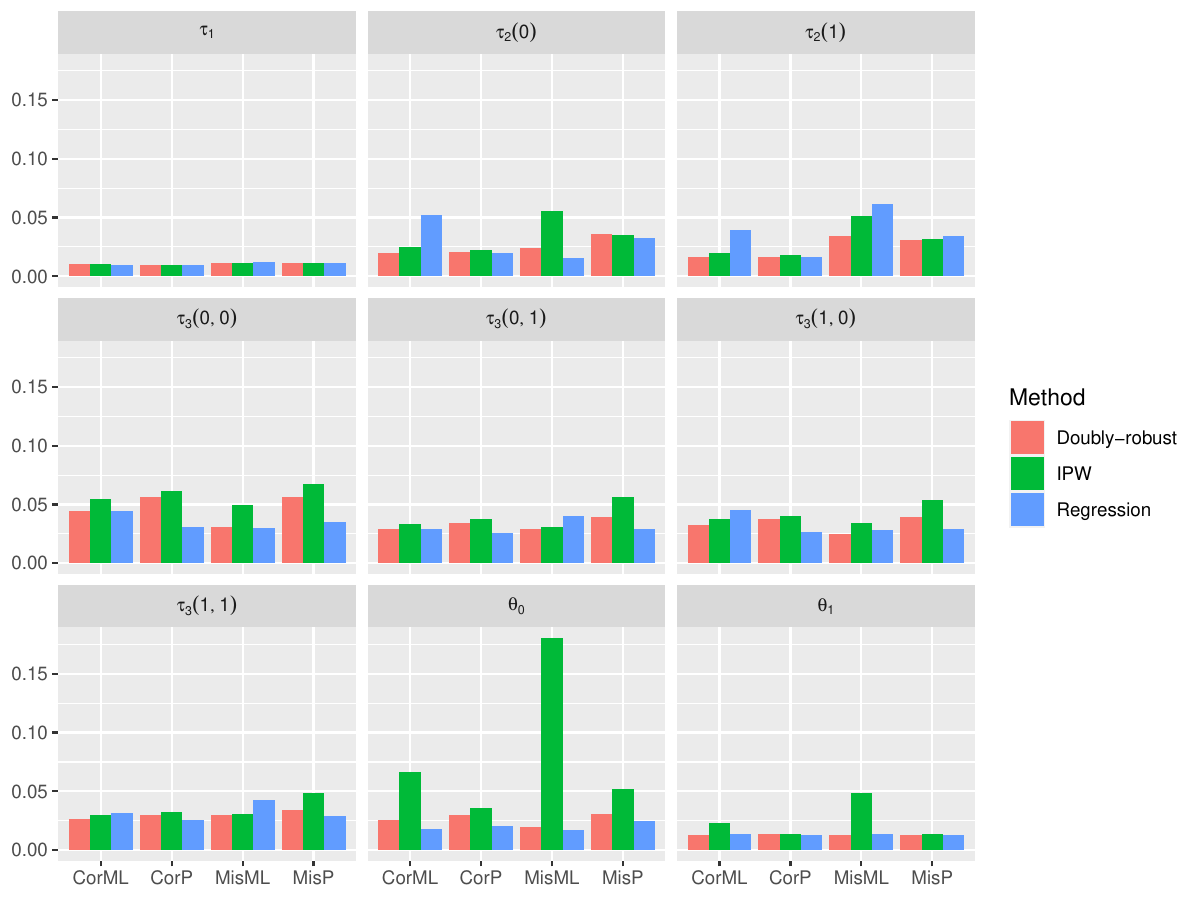}
 \caption{Biases of the estimators across $500$ replications with
   binary outcomes, $\delta=0.5$, and $n=5,000$. Each row presents the
   results for one target causal quantity. The column labels
   ``CorML,'' ``CorP,'' ``MisML,'' and ``MisP'' stand for the
   nonparametric estimation with correctly specified covariates,
   parametric estimation with correctly specified covariates,
   nonparametric estimation with  mis-specified covariates, and
   parametric estimation with  mis-specified covariates,
   respectively. The heights of the bars represent the bias of the regression (blue), inverse probability weighting (green), and doubly robust estimators (red).}
\label{fig::sim_var_n2_bin_bias}
\end{figure}

\begin{figure}[htbp]
 \centering \spacingset{1}
 \includegraphics[width=\textwidth]{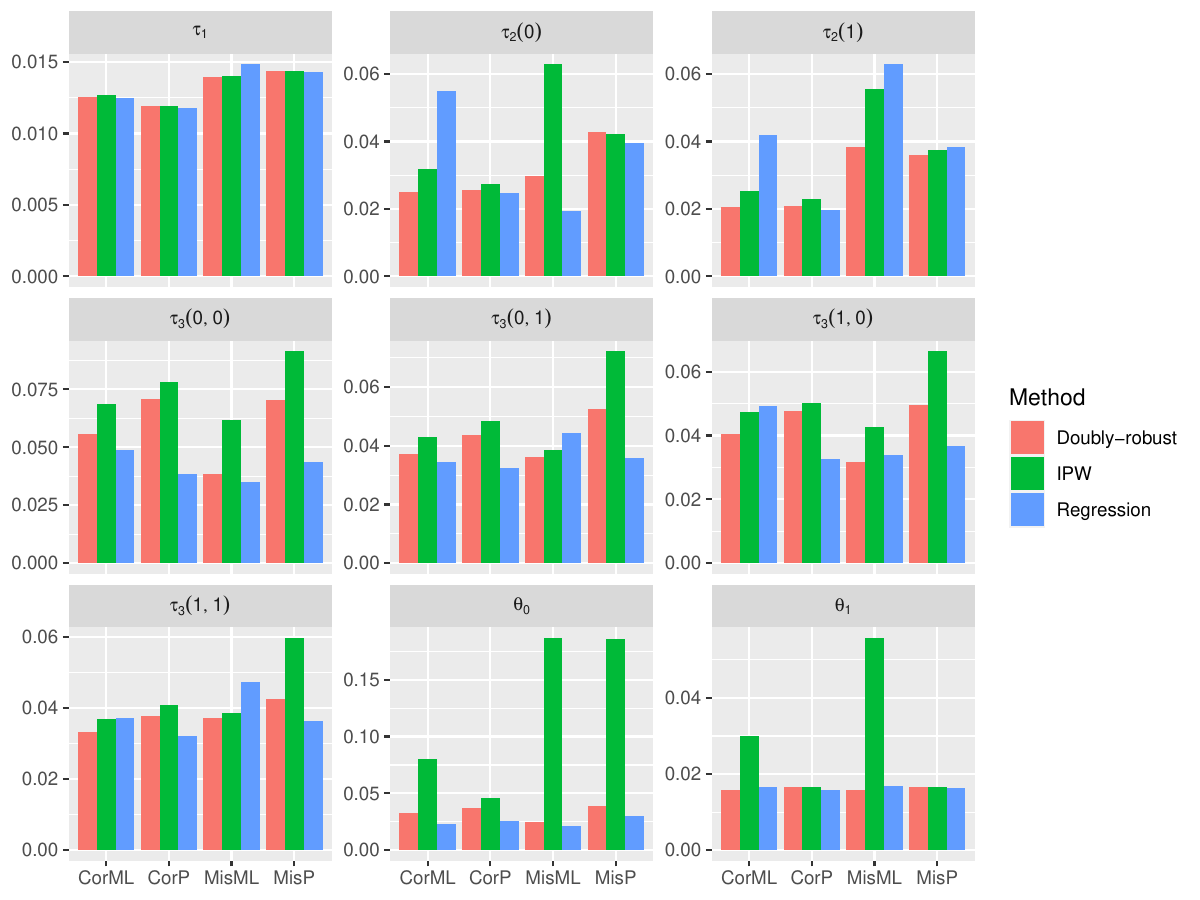}
 \caption{Root mean squared errors for the estimators with binary outcomes, $\delta=0.5$, and $n=5,000$. Each subplot presents the results for one target causal quantity. Within each subplot, the $x$-axis labels `CorML,'' ``CorP,'' ``MisML,'' and ``MisP'' stand for the nonparametric estimation with correctly specified covariates, parametric estimation with correctly specified covariates,  nonparametric estimation with  mis-specified covariates, and parametric estimation with  mis-specified covariates, respectively.  The heights of the bars represent the root mean squared errors of the regression (blue), inverse probability weighting (green), and doubly robust estimators (red).}
\label{fig::sim_var_n2_bin_rmse}
\end{figure}

Figure~\ref{fig::sim_var_n2_bin_bias} shows the absolute biases of the estimators with time varying confounders and sample size $n=5,000$ and Figure~\ref{fig::sim_var_n2_bin_rmse} presents the corresponding RMSEs. 

\end{document}